\documentclass{aa}

\usepackage[colorlinks=true,allcolors=blue]{hyperref}
\usepackage{graphicx}
\usepackage{txfonts}
\usepackage{lipsum}
\usepackage{caption}
\usepackage{subcaption}         
\usepackage{lscape}             
\usepackage{placeins}           
                                
\PassOptionsToPackage{colorlinks=true,allcolors=blue}{hyperref}
\usepackage{mathtools}

\usepackage{xcolor}

\definecolor{forestgreen}{rgb}{0.13, 0.55, 0.13}

\begin{document}

   \title{From residuals to warps: Probing disk geometry with gas kinematics\\
   }



\author{
    C. N. Kimmig\inst{1}
    \and G. P. Rosotti\inst{1}
    \and A. J. Winter\inst{2}
    \and G. Lodato\inst{1}
    \and E. Ragusa\inst{1}
    \and S. Facchini\inst{1}
    \and A. F. Izquierdo\inst{3}
    \and F. Zagaria\inst{4}
    \and M. Benisty\inst{4}
    \and T. Birnstiel\inst{5,6}
    }

\institute{
    Dipartimento di Fisica, Università degli Studi di Milano, via Celoria 16, 20133 Milano, Italy
    \and
    Astronomy Unit, School of Physics and Astronomy, Queen Mary University of London, London E1 4NS, UK
    \and
    Department of Astronomy, University of Florida, Gainesville, FL 32611, USA
    \and
    Max-Planck Institute for Astronomy (MPIA), Königstuhl 17, 69117 Heidelberg, Germany.
    \and
    University Observatory, Faculty of Physics, Ludwig-Maximilians-Universität München, Scheinerstr. 1, 81679 Munich, Germany
    \and
    Exzellenzcluster ORIGINS, Boltzmannstr. 2, D-85748 Garching, Germany
}

\date{Received September 30, 20XX}


\abstract
{Evidence of shadowing among protoplanetary disks suggests that warped protoplanetary disks are widespread, making it crucial to constrain their three-dimensional shape in order to understand their dynamical evolution and origin. Yet, inferring this warp shape from observations is challenging, as the typical signatures of a warp, especially shadows, are often degenerate with the warp shape. Kinematics of disks may provide a method to recover the shape by interpreting the residuals from a Keplerian disk model as local changes in inclination and position angle. In this work, we test its applicability and limitations. We perform radiative transfer simulations of disks with a parametrized warp shape to create synthetic maps of $^{12}\mathrm{CO}$ rotational line emission at ALMA wavelengths and compare the fitted warp with the true warp shape.
We find that the warp shape can be recovered for nearly face-on disks with small warp misalignments of a few degrees, even without knowledge about disk properties such as emission height. The overall warp shape remains traceable even at moderate spectral resolutions up to $0.8\,\mathrm{km/s}$.
However, we identify regimes where the method breaks down, including disks with an inclination higher than $40^\circ$, a high emission height of the tracer, and sharp warp transitions. In addition, radial and vertical motions can introduce apparent warps and twists, motivating extensions of the method to account for these effects.} 


\maketitle
\nolinenumbers.

\section{Introduction}

Shadowing in scattered light observations is common among protoplanetary disks, which has been interpreted as indirect evidence that these disks frequently have a warped or broken geometry \citep[e.g.,][]{Marino2015, Benisty2017, Benisty2018, Garufi2022, Pinte2023}.
This idea was investigated directly by \cite{Bohn2022}, who probed the orientation of different disk regions independently with VLTI/GRAVITY observations for the inner regions and ALMA velocity maps for the outer disk orientation. Asymmetries in edge-on disks \citep{Villenave2024} and frequent moderate warps proposed as explanation for kinematic residuals in the recent exoALMA survey \citep{Teague2025, Winter2025} further support the high prevalence of warps.

Constraining the parameters of the misalignment from observations may reveal the dynamical origin and evolution of the warp. However, recovering the warp shape is challenging.
While shadows can weakly constrain maximum misalignments \citep[see, e.g.,][]{Min2017, Facchini2018, Kraus2020, Muro-Arena2020, Nealon2020, Kimmig2025}, the underlying warp shape is often ambiguous.

\begin{figure*}[ht!]
    \centering
    \includegraphics[width=\linewidth]{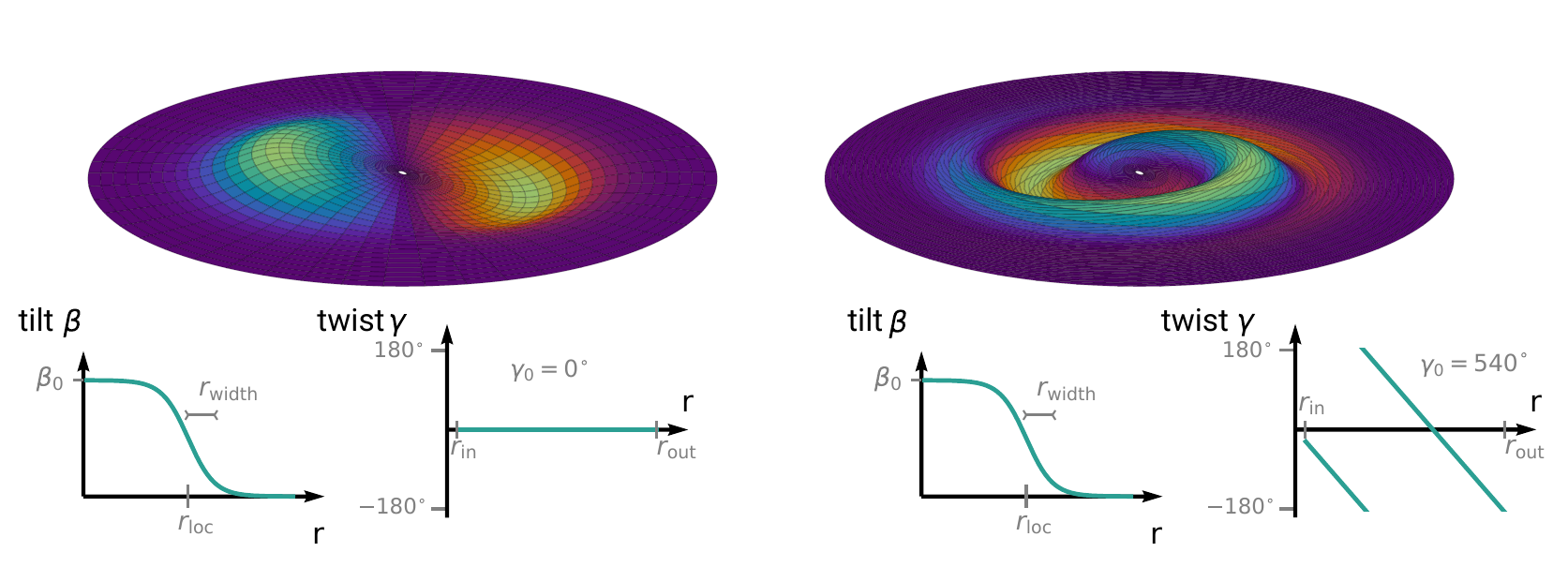}
    \caption{Warped disk without a twist (left) and twisted disk (right) in a schematic view. The colors indicate the height above (blue) or below (orange) the $x$-$y$-plane (indicated in purple). The bottom panels show the profiles of the angles $\beta$ and $\gamma$ for both cases as given in Eqs.~\ref{eq:incl-ini}~\&~\ref{eq:twist}.}
    \label{fig:twisteddisk-scematics}
\end{figure*}

Theoretical models show that the shape of a warp is strongly dependent on both the origin and evolution. Different formation mechanisms can produce distinct radial profiles and amplitudes of warps, while the subsequent dynamical evolution can further modify its shape. There are multiple proposed formation scenarios for warped disks, including gravitational interaction with an inclined binary companion \citep{Papaloizou1995, Foucart2013, Lodato2013, Facchini2013, Deng2022, Rabago2023, Young2023}, an inclined stellar fly-by \citep{Clarke1993, Cuello2019, Kimmig2026}, late-infall events \citep{Dullemond2019, Kuffmeier2021, Kuffmeier2024}, stellar magnetic fields \citep{Foucart2011, Romanova2021}, or shadow-induced warps \citep{Zhang2025}. It is yet to be determined, which of these scenarios is most common and which characteristic warp profiles are associated with each formation mechanism.

The presence of a warp fundamentally alters the dynamical evolution of a disk. In the wave-like regime relevant to protoplanetary disks, where the viscosity often parametrized with $\alpha$ \citep{Shakura1973} is much smaller than the disk's aspect ratio $h = H_\mathrm{p}/r$ (where $H_\mathrm{p}$ is the pressure scale height),  warps propagate as bending waves \citep{Papaloizou1995, Lubow2000, Martin2019}.
In the other limit, $\alpha \gtrsim h$, warp dissipation becomes relevant and the evolution transitions to the diffusive regime \citep{Papaloizou1983, Pringle1992, Ogilvie1999}. This regime becomes relevant for thin, highly viscous disks, for example accretion disks around black holes.
The warp evolution drives a variety of fluid motions \citep{Ogilvie2013-2, Ogilvie2013, Dullemond2022} and can excite the parametric instability, a hydrodynamic instability arising from resonances of inertial waves in oscillating shear flows \citep{Gammie2000, Deng2021, Fairbairn2023, Held2025}.
These dynamical effects play a significant role for the disk's morphology.

As observed warp shapes may grant insight into the system's dynamical past and connect to the warp origin, it is crucial to find a reliable method for retrieving the warp parameters from observed disks. \citet[][hereafter W25]{Winter2025} recently developed a method to determine the warp shape using the kinematics derived from emission line observations. This method interprets structures in the kinematic residuals as radial changes in observed inclination and position angle across the projected disk structure. A similar method was previously suggested by \cite{Casassus2019}.
These changes can then be translated into a warp shape. W25 applied this method to the molecular line observations from the exoALMA sample \citep{Teague2025}, finding that small warps of a few degrees may be common and could naturally explain a wide range of kinematic substructures.

It is well known that warps can leave characteristic signatures in kinematic observations. \cite{Juhasz2017} were the first to investigate these signatures and showed that the integrated flux (Moment~1 maps) show deviations of the projected zero-velocity contour from a straight line at the disk systemic velocity for warped disks. This is because the orbital planes vary with radius, shifting the projected zero-velocity location for different radial regions.
Such twisted contours are indeed observed in some disks, for example in HD~100546 \citep{Walsh2017}, J16042165-J130284 \citep{Mayama2018, Fukagawa2026}, HD~143006 \citep{Perez2018}, and HD~142527 \citep{Garg2021}. We refer to \cite{Pinte2023} for a review.
However, such wiggled zero-velocity contours can be degenerate with other physical effects such as planet-driven perturbations and fluid instabilities \cite{Pinte2023}. In addition, 
\cite{Zuleta2024} showed that a radial flow (following \citealp{Rosenfeld2014}) can also lead to a bent zero-velocity contour, which appears very similar to the signature of a warped disk.

\cite{Juhasz2017} showed that the detectability of warps depends on the observed disk inclination and azimuthal orientation, implying that some warps may remain undetected. These results were confirmed by \cite{Young2022}, who found that warps can generate spiral patterns in kinematic residuals relative to an unwarped, Keplerian model. Such spiral structures have been observed in the disk around MWC~758 \citep{Teague2025, Fukagawa2026, Izquierdo2026}.

In this work, we test the applicability and limitations of the W25 method using synthetic observations of parametrized warped disks. Fitting these data with the W25 kinematic framework allows us to compare recovered and input warp profiles. Our methodology is detailed in Section~\ref{sec:methods} and we present the results in Section~\ref{sec:results}. Section~\ref{sec:othereffects} explores additional dynamical processes, and we discuss the implications in Section~\ref{sec:discussion}. We conclude in Section~\ref{sec:conclusions}.

\section{Methods} \label{sec:methods}

We set up a disk model with a parametrized warp, including tilts and twists, in order to control the warp shape input. We then run radiative transfer simulations of $^{12}\mathrm{CO}$ line emission to compute kinematic maps, which we then fit using the method developed by W25.

\subsection{Disk model} \label{sec:methods-diskmodel}

For the disk, we assume a gas distribution with a surface density that follows a power law ${\Sigma \propto r^{-p}}$ with exponential cut-offs both at the inner and at the outer edge, and we scale it such that the disk to stellar mass ratio is ${M_\mathrm{disk}/M_* = 0.01}$ \citep[for details, we refer to][their Eq.~1]{Kimmig2026}. In all our simulations, the disk extends from ${r_\mathrm{in}=10\,\mathrm{au}}$ to ${r_\mathrm{out}=250\,\mathrm{au}}$ and we adopt a surface density slope of ${p=1}$.
For the vertical structure, we assume a locally isothermal model, resulting in an aspect ratio ${h = H_\mathrm{p} /r \propto r^{f}}$ with ${h_0=0.05}$ at a reference radius ${r_0 = 5.2\,\mathrm{au}}$.
We choose a slope, often called flaring index, of ${f=0.25}$, motivated by the temperature structure in a passively irradiated disk with $T(r) \propto r^{-1/2}$ \citep{Chiang1997}.

We create this density structure on a spherical grid that extends from ${5 - 300\,\mathrm{au}}$ with 300 logarithmically spaced cells in the radial direction, 200 azimuthal and 250 vertical cells, where the vertical coordinate includes the full sphere, i.e. ${\theta \in [0,2\pi]}$.
For the velocity structure, we set up a Keplerian disk with a correction due to pressure gradients according to the vertical disk structure, leading to a sub-Keplerian velocity toward the surfaces, and hence a vertical shear (see \cite{Kimmig2026}, Appendix~A)
\begin{equation}
    {v_\phi^2 (r) = v_\mathrm{k}^2 \left(\frac{r_\mathrm{cyl}}{r}\right)^{2f} + \frac{r_\mathrm{cyl}}{\rho} \frac{\partial\ P(r_\mathrm{cyl},z)}{\partial r_\mathrm{cyl}},}
\end{equation}
where $v_\mathrm{k} = \sqrt{G M_* / r}$ is the Keplerian velocity, $r$ the spherical radius and $r_\mathrm{cyl}$ the cylindrical radius in the coordinate system where the central star is at the origin, $\rho$ the density, and $P$ the pressure.
We do not include any vertical or radial velocities in our models.

\begin{figure*}
    \sidecaption
    \includegraphics[width=12cm]{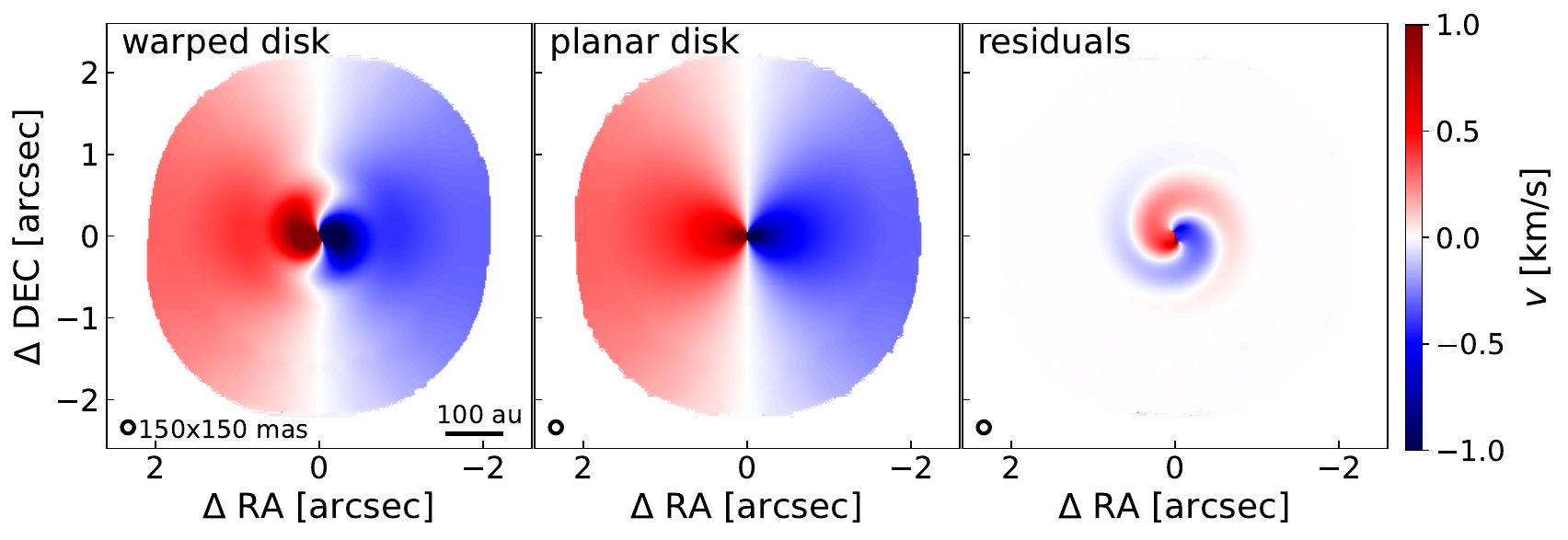}
    \caption{Kinematic $^{12}$CO-emission maps computed using the quadratic method for a warped disk with $\beta_\mathrm{0}=5^\circ$ containing a twist of $\gamma_\mathrm{0} = 540^\circ$ (left panel) and a planar disk (middle panel). The residuals between warped and planar disks are shown in the right panel and correspond to the best case model described in Sect.~\ref{sec:methods-warpfitter}. The (outer) disks are inclined by $10^\circ$, and the maps are already deprojected.
    }
    \label{fig:kinematic-maps-wa5}
\end{figure*}

The warp shape can be characterized with two angles, the tilt $\beta$ and the twist $\gamma$, that are derived from the local angular momentum axis of each disk region. The tilt $\beta$ is defined as the angle between the local angular momentum unit vector $\vec{l}$ and a reference axis $\hat{z}$, and the twist $\gamma$ is defined as the angle in the $x$-$y$-plane with the $x$-axis that is orthogonal to the $z$-axis. A commonly used quantity for the warp is the warp amplitude $\psi$, defined as the derivative of the unit angular momentum vector $l$ in each radial disk region, leading to
\begin{equation} \label{eq:definitions}
    \beta = \arccos(\vec{l} \cdot \hat{z}), \\
    \gamma = \mathrm{atan2}(\vec{l}\cdot\hat{y}, \vec{l}\cdot\hat{x}), \\
    \psi(r) = \left| \frac{\mathrm{d} \vec{l}}{\mathrm{d} \ln r} \right|,
\end{equation}
where $\hat{x}$, $\hat{y}$, $\hat{z}$ are unit vectors in the direction of the $x$, $y$, and $z$-axis, respectively. The dot denotes the scalar product between the two vectors.
Consequently, the definitions of $\beta$ and $\gamma$ require the choice a reference coordinate system, which is simple in the case of our models, but may be ambiguous for observations. 
We set up the warp shape in similar way as \cite{Kimmig2024} with
\begin{equation}\label{eq:incl-ini}
    \beta(r) = \frac{\beta_\mathrm{0}}{2} \left[1 - \tanh \left( \frac{r - r_\mathrm{loc}}{r_\mathrm{width}} \right)\right],
\end{equation}
where $\beta_\mathrm{0}$ is the maximum tilt between inner and outer disk, $r_\mathrm{loc}$ is the location of the sharpest tilt transition, and $r_\mathrm{width}$ is the width of the tilt transition. Fig.~\ref{fig:twisteddisk-scematics} shows an example of the warp structure. As fiducial values for our models, we adopt ${r_\mathrm{loc} = 100\,\mathrm{au}}$ and ${r_\mathrm{width}=50\,\mathrm{au}}$.  This is set up such that the outer disk edge aligns with the $x$-$y$ plane of the chosen coordinate system, while the inner region is tilted.
In disks that are additionally twisted, we assume a linear twist motivated by the result of W25 for the disk of MWC~758
\begin{equation} \label{eq:twist}
    \gamma(r) = \gamma_\mathrm{0} \left( \frac{r}{r_\mathrm{in} - r_\mathrm{out}} \right)
\end{equation}
with the maximum twist angle of $\gamma_\mathrm{0}$ that determines the strength by how much the disk is wound up.
To achieve the warp shape, we perform rotations of each radial grid shell for both the density and the velocities around the $y$-axis according to Eq.~\ref{eq:incl-ini}, and subsequently around the $z$-axis according to Eq.~\ref{eq:twist}.
Our fiducial model is warped by ${\beta_\mathrm{0} = 5^\circ}$ and twisted by ${\gamma_\mathrm{0} = 540^\circ}$, which means that the disk winds up for one and a half circles as indicated in Fig.~\ref{fig:twisteddisk-scematics}, right panel. We vary both parameters throughout this work.

\subsection{Radiative transfer models}

To create the kinematic maps, we run radiative transfer simulations using the code RADMC-3D \citep{radmc3d}. For that, we assume a constant dust-to-gas ratio of 0.01 and compute the dust temperature using the \texttt{mctherm} function of RADMC-3D (see Fig.~\ref{fig:temperature}). For each disk, we use $10^8$~photon packages for a central Sun-like star with ${M_* = 1 M_\odot}$ and ${R_* = R_\odot}$, and enable the modified random walk option that improves the performance when rays would be trapped in optically thick regions.
As we assume the gas to be in thermal equilibrium with the dust, we set the gas and dust temperature equal.
For the dust opacity, we assume the DIANA standard composition with carbon and silicate (pyroxene with 70\% magnesium and 30\% iron) and a mass ratio of carbon to pyroxene of ${0.87/0.13}$, and a porosity of 25\% \citep{Dorschner1995, Zubko1996, Woitke2016}. We adopt a dust size distribution from ${0.1\,\mathrm{\mu m}}$ to ${10\,\mathrm{\mu m}}$ with a power law ${\mathrm{d}n/\mathrm{d}a \propto a^{3.5}}$.
The opacity is computed using \texttt{optool} \citep{optool}.

We use the ray-tracing function of RADMC-3D to compute the line emission of the $^{12}\mathrm{CO}$ $J=2-1$ rotational transition at ${230.538\,\mathrm{GHz}}$ ${(\lambda = 1.3\,\mathrm{mm})}$ and assume a molecular abundance of ${^{12}\mathrm{CO/H_2} = 10^{-4}}$. We take into account the freeze-out at the condensation temperature of $20\,\mathrm{K}$ \citep{Williams2014} and the photodissociation for column number densities lower than ${N_\mathrm{dissoc} = 5 \times 10^{19}\,\mathrm{H_2/cm^2}}$ \citep[see][]{VanDishoeck1988, Visser2009, Trapman2023, Rosotti2025} and set the molecular abundance to zero in these regions.
We compute 200 channels in a window of $\pm 8\,\mathrm{km/s}$ around the molecular line, resulting in a spectral resolution of $0.08\,\mathrm{km/s}$. We checked that the spectral window encompasses all line emission, and we subtract the continuum, which we take to be the first, line-free channel from all channels.

In order to approximate an observation, we convolve the data cube with a two-dimensional Gaussian beam of ${150\times150\,\mathrm{mas}}^2$. For the distance to the source, we assume $150\,\mathrm{pc}$. The inclination varies throughout the paper, but for simplicity, we fix the position angle to $-90^\circ$. Each image is computed with ${800 \times 800}$ pixels with $0.8\,\mathrm{au/pix}$. We do not assume any thermal noise for simplicity. Since the warp shape is obtained by fitting an azimuthal profile, we expect the noise to be averaged out and therefore have only a minor impact on the results. We will, however, relax this assumption in future work (Spaan et al., in prep.).

In order to save computation time, we ignore the emission that results from scattering in the ray-tracing step by activating the option \texttt{noscat} in RADMC-3D. To validate this assumption, we ran a comparison simulation with exactly the same parameters, but with full scattering treatment. The residuals between the two simulations are at least two orders of magnitude lower than the line emission and do not influence the qualitative outcome of the kinematic evaluation. We can thus safely ignore scattering.

\subsection{Fitting the warp with \texttt{warpfitter}} \label{sec:methods-warpfitter}

We evaluate the radiative transfer simulations to fit a warp using the method developed by W25. This means that the residuals of the kinematic maps are interpreted as radially varying inclination and position angle. To compute the velocity map of a disk, we use the quadratic method to fit the line profile, where the line centroid is fitted with a quadratic function, rather than using the intensity-weighted first moment of the entire line profile, enabling sub-channel precision \citep{Teague2018}. \citet{Young2022} demonstrated for warped disks specifically that this method provides more accurate velocity maps than the first-moment approach. Afterwards, we deproject the map and transform it to polar coordinates.

\begin{table}[ht!]
    \centering
    \caption{{Overview of the two methods used to compute the kinematic residuals}}
    \begin{tabular}{p{0.45\linewidth}|p{0.45\linewidth}}
        best case model & worst case model \\ \hline \vspace{0.05cm}
        \vspace{0.05cm} subtracting the velocity map of a planar disk with the same parameters (except for the warp) that is computed with the same radiative transfer method & \vspace{0.05cm} simply subtracting the Keplerian velocity at the midplane of the disk \newline (\textbf{not} accounting for emission height and pressure correction for simplicity)
    \end{tabular}
    \label{tab:models}
\end{table}

To compute the residuals, we use two different methods, corresponding to the best- and worst-case scenario, summarized in Tab.~\ref{tab:models}. For the best-case scenario, we set up a planar, unwarped disk as described in the previous section and compute the $^{12}\mathrm{CO}$-emission the same way as for the warped disks. This scenario describes the case where the disk parameters such as thickness and emission height are perfectly known. Typically, this is not the case for observed disks, which is why we compare to the worst-case scenario, where we simply subtract the Keplerian velocity at the midplane of the disk and do not account for emission height and pressure correction in the azimuthal velocity. 
In observations, the vertical structure and the corresponding pressure correction may be constrained, but are generally not known precisely. In Appendix~\ref{sec:appendix-aspectratio}, we investigate the effect of an incorrectly assumed aspect ratio by using planar reference models that are either thinner or thicker than the warped disk. For the nearly face-on models, the recovered tilt profiles retain the qualitative shape of the input warp, whereas larger discrepancies occur at higher inclination. Our best and worst case scenarios therefore provide useful guidance on the recovery achievable in observations, although they do not constitute strict bounds when disk parameters are uncertain.
We stress that we do not fit for disk inclination or position angle, instead assuming they are known exactly, and investigate the impact of uncertainties in Section~\ref{sec:uncertainties}.

We then use the \texttt{warpfitter} tool by W25 on kinematic residuals to fit deviations in inclination $\delta i(r)$ and in position angle $\delta \mathrm{PA}(r)$. For details on the fitting process, we refer to W25. From these fits, we can compute the warp tilt
\begin{equation}
    \beta(r) \approx \sqrt{\delta i (r)^2 + \delta \mathrm{PA(r)^2} \sin^2i_0},
\end{equation}
the twist
\begin{equation}
    \gamma(r) \approx \arctan2 (\delta i(r), -\sin i_0 \delta \mathrm{PA(r))},
\end{equation}
and the warp amplitude as derivative of the unit angular momentum vector $\vec{l}$
\begin{equation} \label{eq:psi}
    \psi(r) = r \sqrt{\left( \frac{\partial i}{\partial r} \right)^2 + \sin^2 i \left( \frac{\partial \mathrm{PA}}{\partial r} \right)^2}.
\end{equation}
Here, $i_0$ is the inclination of the observed disk assuming a planar disk, which we take to be the inclination of the outer disk in our work. The inclination and position angles are given as $i = i_0 + \delta i$ and $\mathrm{PA} = \mathrm{PA}_0 +\delta \mathrm{PA}$, where we set $\mathrm{PA_0}=0$ without loss of generality \citep[see][]{Winter2025}.
This process then enables us to compare the fitted warp shape parameters with the model setup.

\section{Method robustness} \label{sec:results}

In this section, we assess the robustness of W25's method and constrain limiting application cases. We start with parametrized warps according to Eq.~\ref{eq:incl-ini}, and motivated by the twisted shape of MWC~758 found in W25, we set a strong twist of $\gamma_\mathrm{0} = 540^\circ$ according to Eq.~\ref{eq:twist}.

\begin{figure}
    \centering
    \includegraphics[width=0.95\linewidth]{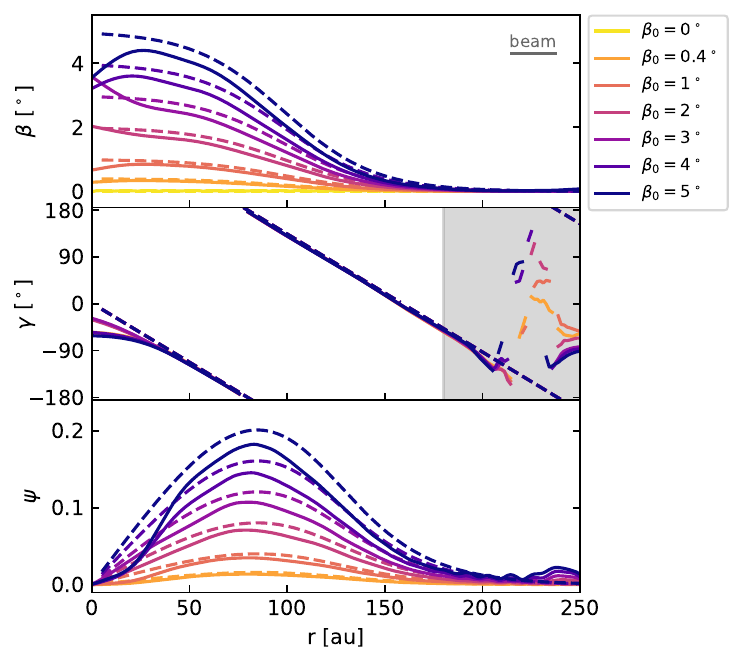}
    \caption{The disk tilt $\beta$ (top), twist $\gamma$ (middle), and warp amplitude $\psi$ (bottom) for simulations of varying $\beta_\mathrm{0}$. The dashed lines show the input profiles, while the solid lines show the fitted results from \texttt{warpfitter} in the best case scenario (i.e. disk thickness, emission height, etc. are known). The inclination of the outer edge of the disk is $10^\circ$. The gray shaded area indicates the region where the twist is undefined for all models, as $\beta \approx 0$. We do not plot the twist $\gamma$, as well as $\psi$, in the case of the $\beta_\mathrm{0}=0^\circ$ simulation (yellow) in the lower panels, as the twist is undefined everywhere in this case.}
    \label{fig:warpfitter-wa}
\end{figure}

\subsection{Nearly face-on disks} \label{sec:faceondisks}

We first investigated disks that are nearly face-on, specifically the outer edge of the disk is inclined by $i=10^\circ$ with respect to the line of sight. Fig.~\ref{fig:kinematic-maps-wa5} shows an example of the synthetically observed kinematic maps of a warped (left) and planar disk (middle) as well as the residuals in the best case scenario (right).

First, we vary the maximum tilt of the warp $\beta_\mathrm{0}$ between $0^\circ$ and $5^\circ$ (where $0^\circ$ corresponds to a planar, unwarped disk).
Fig.~\ref{fig:warpfitter-wa} shows the comparison of warp tilt $\beta$, twist $\gamma$, and warp amplitude $\psi$ between the input model (dashed lines) and the fitted data using the kinematic pipeline including \texttt{warpfitter}. The solid lines correspond to the best case scenario, where disk parameters such as vertical thickness, density, emission height, inclination, and position angle are perfectly known (see Sect.~\ref{sec:methods-warpfitter} for details).

We find that in this case, the method to fit warps from kinematic data works well, even down to a tilt of only ${\beta_\mathrm{0} = 0.4^\circ}$ between inner and outer disk region (orange line).
Generally, the method seems to slightly underestimate the maximum tilt of the warp, especially for higher $\beta_\mathrm{0}$.
The twist $\gamma$ is traced perfectly for all models up to $200\,\mathrm{au}$. Further out than that, the twist is undefined, since the tilt $\beta$ is zero, so we can ignore the deviations in this regime. For the same reason, we do not show the yellow line for a planar disk with $\beta_\mathrm{0} = 0^\circ$ in the middle panel.
The warp amplitude $\psi$ (Eq.~\ref{eq:definitions}), which accounts for both the tilt and the twist, is traced reasonably well. Also here, the warp amplitude is slightly underestimated for larger $\beta_\mathrm{0}$.

\begin{figure}
    \centering
    \includegraphics[width=\linewidth]{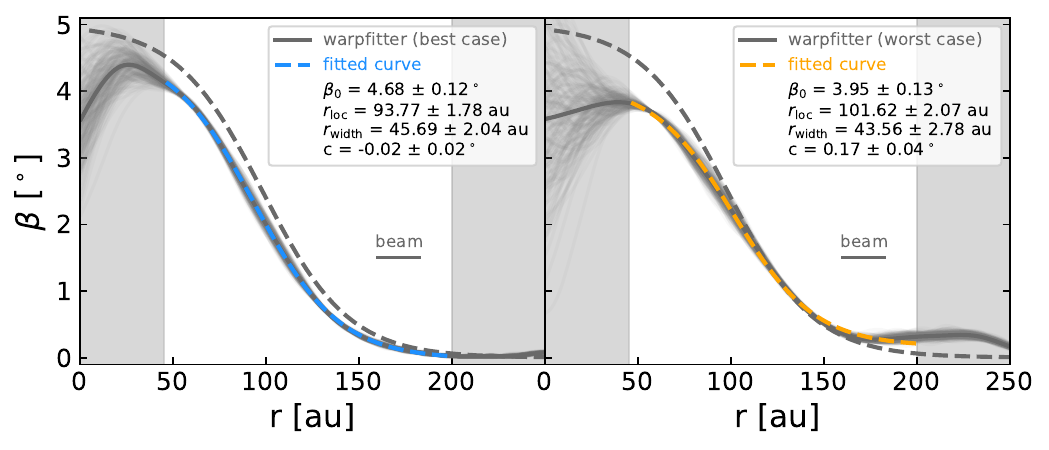}
    \caption{Warp tilt profiles of the simulation with $\beta_\mathrm{0}=5^\circ$, with the residuals calculated using the best case (left panel) and worst case model (right panel). Gray dashed lines show the model input, gray solid lines the best fit from \texttt{warpfitter} and the light grey lines show the posterior distribution from the Gausian process fit by \texttt{warpfitter}. The orange and blue dashed lines indicate the fit according to Eq.~\ref{eq:incl-ini} (allowing for a constant offset $c$) of best and worst case model, respectively. The gray shaded region shows regions we exclude from the fit.}
    \label{fig:fits-10deg}
\end{figure}

In order to quantify how well the warp profile can be traced with \texttt{warpfitter}, we fit the resulting warp tilt profile with the function we use to set up the warp shape in the gas density structure (Eq.~\ref{eq:incl-ini}) using \texttt{scipy.optimize.curve\_fit} \citep{SciPy}. At this point, we only focus on the warp tilt $\beta$ and ignore the twist $\gamma$. We allow for a constant offset $c$, as it is required for a good fit in some cases. This offset would correspond to an apparent difference in inclination of the outer disk. However, $c$ is expected to be close to zero for most models.

Two examples of such fits are shown in Fig.~\ref{fig:fits-10deg}.
We estimate the errors using the posterior distributions from the Gaussian process fit by \texttt{warpfitter}, indicated with the solid light grey lines. For this, we fit each of the lines with the same function and compute the standard deviation with ${\sqrt{\sum_i(a_i - \bar{a})^2/(N-1)}}$, with $a_i$ the fitted values of each curve, $\bar{a}$ the mean value, and $N$ the number of fitted curves.
However, the errors we get from this approach are likely underestimated during the \texttt{warpfitter} fit, because the fitting process only takes the deviation across an annulus into account and neglects systematic errors inherent in the observations.
All fits we show throughout the paper are fitted in a constrained radial regime from $45\,\mathrm{au}$ up to $200\,\mathrm{au}$, excluding the inner two beam sizes and the outer part where the emission of the back side is significant, especially for the higher inclined disks.

\begin{figure}
    \centering
    \includegraphics[width=0.95\linewidth]{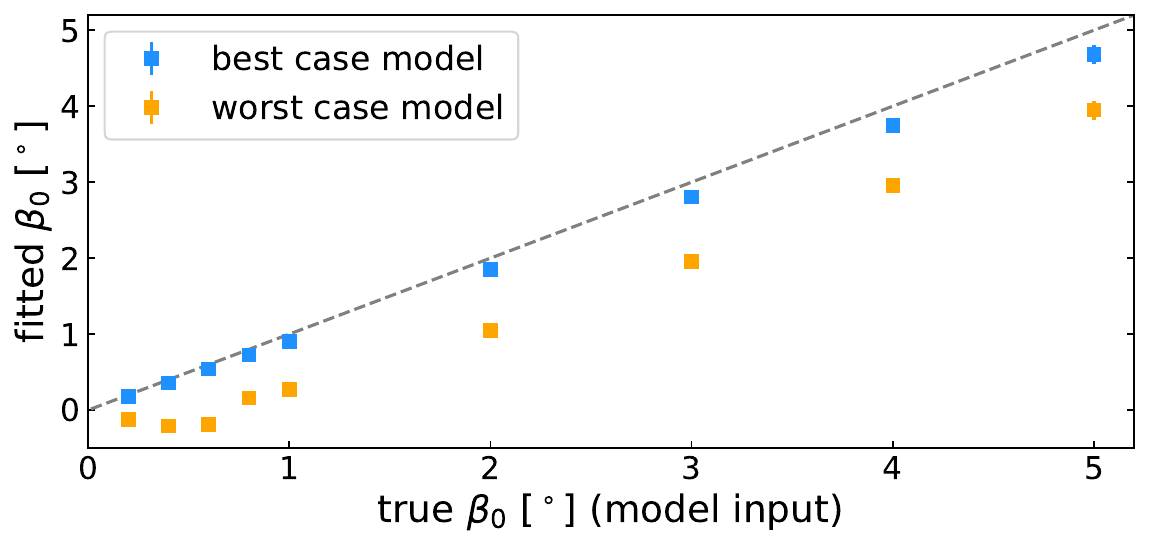}
    \caption{Fitted values of $\beta_\mathrm{0}$ in comparison to the true $\beta_\mathrm{0}$-values of the model input. Blue squares represent the fits on the best case model, orange squares from the worst case model. The dashed line indicates the $y=x$-relation.}
    \label{fig:fits-betamax}
\end{figure}

Fig.~\ref{fig:fits-betamax} shows the comparison of the fitted values of $\beta_\mathrm{0}$ to the input values for different models. Here, the inclination and position angle are known perfectly, meaning that there are no uncertainties associated to these parameters. We investigate possible uncertainties in this regard in Sect.~\ref{sec:uncertainties}. Both the best case and worst case models are shown in Fig.~\ref{fig:fits-betamax} (blue and orange squares, respectively).
The best case model traces the tilt of the warp shape very well, although it slightly underestimates it for higher values. The worst case model underestimates the warp tilt roughly by one degree for most models. This case fails for the models with ${\beta_\mathrm{0} < 1^\circ}$, as it results in a negative $\beta_\mathrm{0}$, which results from the fact that $\beta_\mathrm{0}$ is defined as a function parameter in our work. We show an example of the fit for ${\beta_\mathrm{0}=0.2^\circ}$ in Appendix~\ref{sec:appendix-limits}, Fig.~\ref{fig:fits-betamax02}. There, we can see that the the wrong shape is traced in the worst case model, resulting in a flipped curve explaining the negative fit for $\beta_\mathrm{0}$. In the context of the evaluation in this part, we ignore the negative $\beta_\mathrm{0}$-values and simply note that the method of \texttt{warpfitter} fails for these models.
The worst case model often shows an offset in inclination (larger than zero) at the outer disk as seen in Fig.~\ref{fig:fits-10deg}, left panel (also for example in Fig.~\ref{fig:fits-betamax02} and \ref{fig:fits-30and60deg}, left column), which is the reason why the parameter $\beta_0$ is systematically smaller by roughly a degree in this case.

We note that the best case scenario is unrealistic for observations of real disks. Nevertheless, when the emission height is well constrained by observations, we expect the values recovered in realistic cases to typically lie between the blue and the orange squares in Fig.~\ref{fig:fits-betamax}.

\subsection{Effect of disk inclination} \label{sec:observerincl}

\begin{figure}
    \centering
    \includegraphics[width=0.95\linewidth]{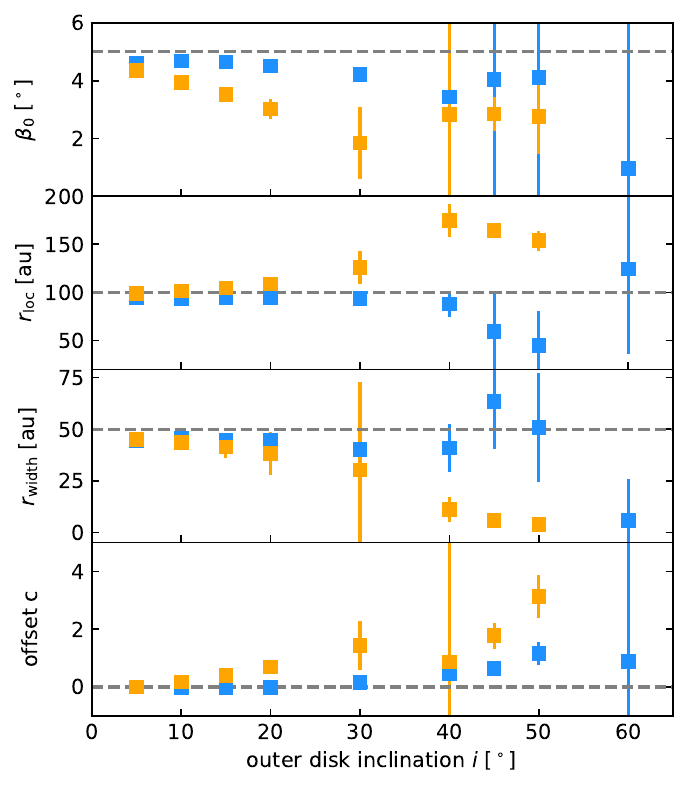}
    \caption{Fitted values of the tilt tilt $\beta_\mathrm{0}$, the location of steepest transition $r_\mathrm{loc}$, the warp width $r_\mathrm{width}$, and an offset $c$ for the fit, depending on inclination of the outer disk region. Blue squares represent the best case, orange squares the worst case models. Grey dashed lines indicate the input parameter.}
    \label{fig:observer-inclination}
\end{figure}

We now explore how the result from \texttt{warpfitter} changes for different inclinations of the outer disk edge. For this, we focus on the model with $\beta_\mathrm{0}=5^\circ$ with a twist of $\gamma_\mathrm{0}= 540^\circ$.
Fig.~\ref{fig:observer-inclination} shows the result from the fits of the four fitting parameters warp tilt $\beta_\mathrm{0}$, location of the warp $r_\mathrm{loc}$, warp width $r_\mathrm{width}$, and offset $c$ depending on the observed inclination of the outer disk.

We find that the method works best for nearly face-on disks. The warp shape for disks up to an inclination of $30^\circ$ can be traced well, especially in the best case model.
Even the worst case model is able to retrieve the warp shape well up to an inclination of $20^\circ$, but starts to fail for $\gtrsim 30^\circ$. We show the fit for $i = 30^\circ$ in Fig.~\ref{fig:fits-30and60deg}. For disks with even higher observed inclination, it becomes very difficult to trace the warp shape in both the best and worst case models.
This may be related to the deprojection of the data. In our models, we do not take the emission surface into account for the deprojection, and simply deproject using the fixed inclination of the outer disk. This leads to stronger residuals for higher inclinations.
Additionally, the back side of the disk contributes to the emission at higher inclinations, which complicates fitting the warp shape \textit{in our best and worst case scenarios}, particularly in the outer disk region. This back side emission can be seen in the residual maps of the disks with higher inclination, see Fig.~\ref{fig:kinematic-maps-inclination}.
We note that, although our approach performs best for nearly face-on disks, the inferred disk scale height may be more uncertain in this regime. This is because the backside of the disk is not visible and therefore cannot be accounted for when fitting the scale height. We will investigate this effect in more detail in follow-up work (Spaan et al., in prep.).

In summary, the higher the observed inclination, the more difficult it is to trace the true warp structure.
This result is highly relevant for evaluating potential warps in kinematic observations of protoplanetary disks.

\subsection{Robustness to underlying disk structure}
\subsubsection{Vertical thickness}
In this section, we investigate how well the warp can be traced in disks with different physical disk structures. In a first comparison, we set up a thicker disk using $h_0 = 0.1$ at $r_0 = 5.2\,\mathrm{au}$, but keep the flaring the same, $f=0.25$ in the parametrization ${h = H_\mathrm{p}/r \propto r^f}$, with the pressure scale height $H_\mathrm{p}$.
As statistically more disks have a higher inclination, we focus on disks inclined by $30^\circ$ in this and the following subsections.

\begin{figure}
    \centering
    \includegraphics[width=\linewidth]{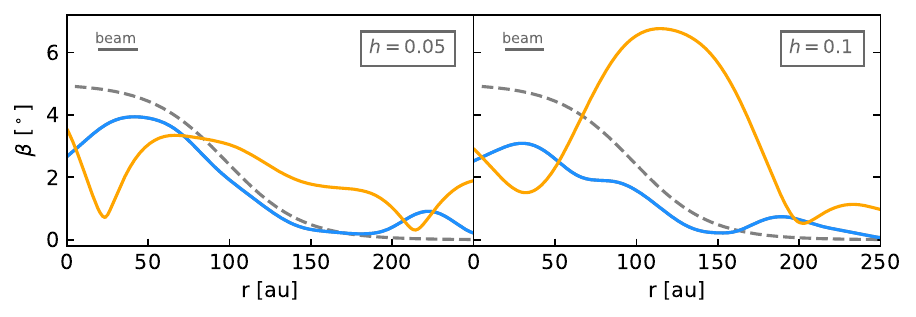}
    \caption{Fitted warp shapes $\beta(r)$ for disks with different thicknesses of $h=0.05$ (first panel) and $h=0.1$ (second panel). Both cases are shown for an outer disk inclination of $i=30^\circ$. The blue and orange lines in each plot correspond to the best and worst case residual model, respectively. Grey dashed lines indicate the underlying model.}
    \label{fig:aspectratio}
\end{figure}

Fig.~\ref{fig:aspectratio} shows that the retrieved warp shape matches less well for a thicker disk, even in the best case scenario where we know the thickness of the disk. In the worst case scenario (orange lines), the warp shape can not be retrieved for the thicker disk. We obtain qualitatively similar results for a disk with the same aspect ratio as our fiducial case $h_0=0.05$, but stronger flaring of $f=0.5$, which effectively leads to a thicker disk as well, especially in the outer part of the disk.
This may be explained by geometric effects due to the vertical height of the emission surface, which we discuss in more detail in Sect.~\ref{sec:discussion-vertical-extent}.

\subsubsection{Broken disk structure} \label{sec:brokendisks}

\begin{figure}
    \centering
    \includegraphics[width=0.95\linewidth]{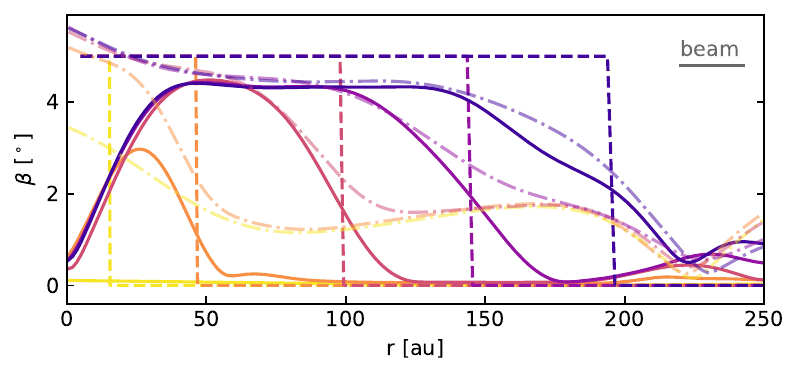}
    \caption{Disk tilt $\beta(r)$ for the models of broken disks, where the break is at different locations (different colors). The dashed lines correspond to the model input, the solid lines are the fitted results using the best case model, and the light dash-dotted lines are fitted using the worst case model (simple Keplerian velocity at the midplane). The outer disks are observed at $30^\circ$ inclination.}
    \label{fig:brokendisk}
\end{figure}

To investigate a different warp shape, we set up an extreme case with a very sharp warp transition of $r_\mathrm{width}=1\,\mathrm{au}$ in Eq.~\ref{eq:incl-ini}.
This corresponds to a broken disk with an inner disk region inclined by $5^\circ$ with respect to the outer region. We note that there is no radial gap between inner and outer disk in our models.
In order to investigate how well the warp tracing method performs for warps at different locations in the disk, we set up five different models with ${r_\mathrm{loc} \in \{20, 50, 100, 150, 200\}\,\mathrm{au}}$.
Fig.~\ref{fig:brokendisk} shows the result of these models for an outer disk inclined by $30^\circ$ to the line of sight. We see that the warp cannot be traced for the model with a break at $20\,\mathrm{au}$. This occurs due to the Gaussian convolution of the synthetic observations and our choice of a beam size (FWHM) of $0.15\arcsec$ that corresponds to a size of $22.5\,\mathrm{au}$ at the chosen distance of $150\,\mathrm{pc}$. This means, of course, that warps at the inner regions may be hidden by the beam. These results are also relevant for future high angular resolution Integral Field Unit (IFU) facilities such as Mid-infrared ELT Imager and Spectrograph (METIS), which will be able to probe molecular line emission from the innermost disk regions at substantially finer spatial scales \citep[see, e.g.,][]{Oeberg2023}.

\begin{figure}
    \centering
    \includegraphics[width=0.95\linewidth]{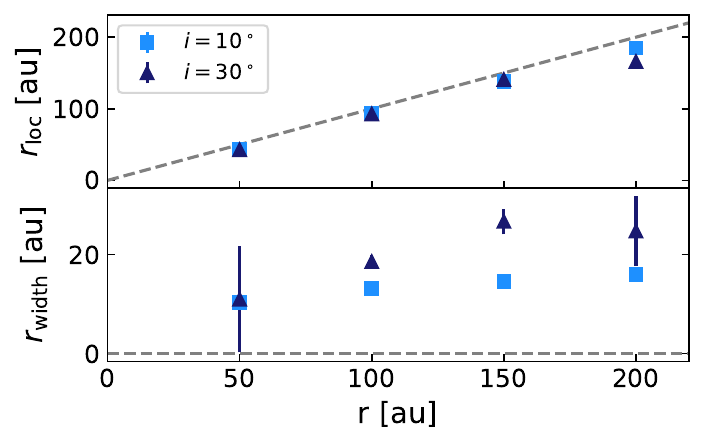}
    \caption{Fit parameters of the warp location $r_\mathrm{loc}$ and warp width $r_\mathrm{width}$ for the models of a broken disk with the break at different locations. The grey dashed line in the top panel indicates the $x=y$-relation. These data points were obtained using the best case model.}
    \label{fig:brokendisk-fits}
\end{figure}

In the other models, the transition between inner and outer region can be traced, although the results from \texttt{warpfitter} show a much shallower transition than we set in the models. This is also linked to the convolution with a Gaussian beam.
To investigate this more quantitatively, we fit the results of the best case scenario with the model input of the warp shape in the same way as described in Sect.~\ref{sec:faceondisks}, both for outer disks inclined by $10^\circ$ and $30^\circ$, and show the results in Fig.~\ref{fig:brokendisk-fits}. We were not able to fit the model with $r_\mathrm{loc}=20\,\mathrm{au}$ and thus cannot include the data points of this model.
The location of the warp can be retrieved accurately for all models, with a slight deviation for the model of $r_\mathrm{loc}=200\,\mathrm{au}$, particularly for higher inclination. This is likely due to the emission from the back side, as discussed in Sect.~\ref{sec:observerincl}.
As mentioned, we are not able to retrieve the steep warp profile. However, the retrieved warp width does not seem to strongly depend on the location of the warp in the disk.

In summary, we find that a warp is more difficult to detect in vertically thicker disks.
Warps in the very inner region of the disk can also be hidden due the beam smearing, and warps with a sharp transition between differently tilted regions may not be traced accurately, depending on the beam size.
For higher inclined disks, a warp close to the outer disk edge may also be contaminated by emission from the back side of the disk.

\subsection{Spectral resolution}

\begin{figure}
    \centering
    \includegraphics[width=\linewidth]{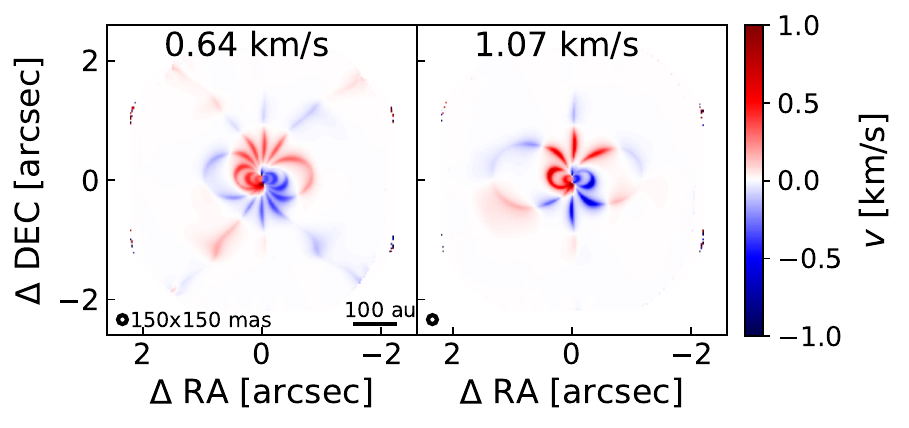}
    \caption{Kinematic residuals obtained using the best case model of a warped and twisted disk for lower spectral resolutions. The outer disk is inclined by $30^\circ$ with respect to the observer.}
    \label{fig:maps-spectralres}
\end{figure}

In this section, we investigate the spectral resolution necessary for the warp tracing method. For that, we use the existing fiducial model and compute the moment maps by binning the channels into larger spectral intervals.
Fig.~\ref{fig:maps-spectralres} shows two examples of the residual moment maps computed using the quadratic method. Even though the lower resolution maps do not look smooth, the fitting procedure can still return a warp profile. While some details may be lost at lower spectral resolutions, the overall warp shape can still be recovered, as the method averages over the azimuthal region.
The fitted warp shapes for different spectral resolutions can be found in Fig.~\ref{fig:spectralres-incl30}.

\begin{figure}
    \centering
    \includegraphics[width=0.95\linewidth]{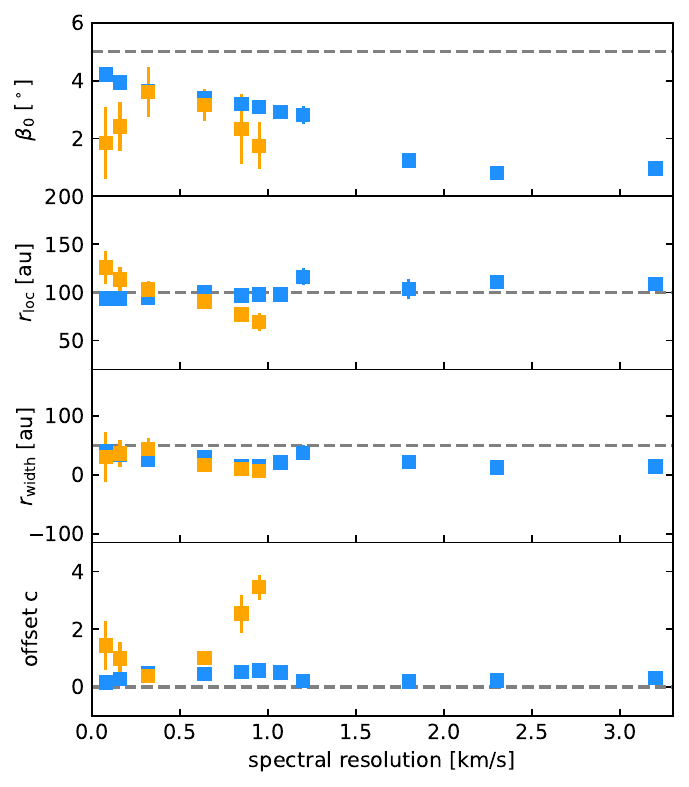}
    \caption{Fit parameters of the warp profile obtained for a warped and twisted disk in different spectral resolutions. Blue correspond to the best case model, orange to the worst case model, and the grey dashed lines indicate the true parameter of the input model. The outer disk is observed at an inclination of $30^\circ$. The fitting procedure fails for resolutions larger than ${1\,\mathrm{km/s}}$ in the worst case model.}
    \label{fig:spectralres-fits}
\end{figure}

Again, we fit the warp shape with the model input function. Fig.~\ref{fig:spectralres-fits} shows the fit parameters for different spectral resolutions, while we keep all parameters of the model fixed. For the best case scenario (blue squares), the warp shape could be traced relatively accurately for spectral resolutions ${\lesssim 1.0\,\mathrm{km/s}}$, although the maximum tilt $\beta_\mathrm{0}$ becomes less accurate with decreasing spectral resolution.
For the warp fits obtained in the worst case scenario, the warp shape can be traced for spectral resolutions ${\lesssim 0.8\,\mathrm{km/s}}$. In this case, the fit by \texttt{warpfitter} failed for resolutions ${> 1.8\,\mathrm{km/s}}$.
For resolutions between ${1.0\,\mathrm{km/s}}$ and ${1.8\,\mathrm{km/s}}$, despite \texttt{warpfitter} returning a profile, we were not able to fit the warp profile with the model input function, as the warp profile in the outer disk seems to obtain a steep increase in $\beta$ (see Fig.~\ref{fig:spectralres-incl30}, lower panel).
We find that in the worst-case scenario, the recovery of $r_\mathrm{loc}$ and $r_\mathrm{width}$ degrades at spectral resolutions higher than $0.5\,\mathrm{km/s}$. This is caused by a less accurate recovery of the inclination in the outer disk regions (see Fig.~\ref{fig:spectralres-incl30}, resulting in a less steep warp transition.

Based on this result, we expect it to be possible to extract the warp shape for observations with a spectral resolution of ${\lesssim 0.8\,\mathrm{km/s}}$. We recall that we investigated disks inclined by $30^\circ$ in this section, and note that more face-on disks may require a higher spectral resolution, as the absolute values of the projected velocities are lower. This result enables a reevaluation of archival kinematic data in future work, since many kinematic observations of protoplanetary disks fall in this category \citep[see, e.g.,][]{Kurtovic2026}.

\subsection{Uncertainty in disk geometry and stellar mass} \label{sec:uncertainties}

In the previous sections, we have assumed that the observational parameters inclination and position angle of the outer disk are perfectly known. Here, we aim to investigate the sensitivity of the method with respect to the orientation of the disk on the sky. For this, we analyzed our fiducial model of a warped and twisted disk with $\beta_\mathrm{0}=5^\circ$ and $\gamma_\mathrm{0}=540^\circ$ for slight deviations from the true inclination and position angle. Specifically, we used an incorrect inclination or position angle to deproject the synthetic kinematic data\footnote{We also used the incorrect inclination or position angle for the values of $i_0$ or $\mathrm{PA}_0$ in the fitting procedure of \texttt{warpfitter}}. In other words, we investigated what the warp profile would look like, if we inferred the inclination or position angle incorrectly.

\begin{figure}
    \centering
    \includegraphics[width=\linewidth]{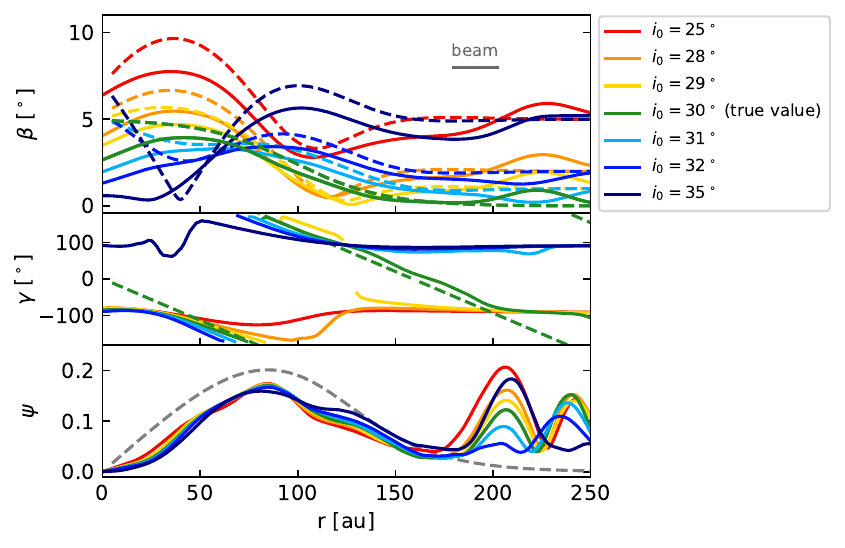}
    \caption{Warp profiles $\beta(r)$, twist $\gamma(r)$, and warp amplitudes $\psi(r)$ for evaluations with different inclinations (solid lines). For better comparison in the top panel, the dashed lines indicate the model input as it would appear using reference frames with different disk inclinations $i_0$ (i.e., translated into the new reference frame). We note that $\psi$ does not depend on the reference system, thus the gray dashed line is the same for all cases. However, the recovered profiles can differ, as the assumed inclination affects the deprojection and fitting of the kinematic data.}
    \label{fig:incl-err}
\end{figure}

Fig.~\ref{fig:incl-err} shows the fitted warp profiles $\beta(r)$ in the top panel. However, these profiles of $\beta$ extracted from the synthetic observation are now given in a different reference frame, because a different disk inclination was assumed. Thus, in order to compare the fitted profiles with the model input, we need to transform the input warp shape to the new (incorrectly assumed) reference frame. The dashed lines in the top panel of Fig.~\ref{fig:incl-err} show the new input $\beta$ profile after this transformation, corresponding to the new rotated reference axis defined by the assumed inclination. The green lines correspond to the correct inclination and thus shows the input profile in the original reference frame. The fact that the same profile, when transformed to different reference frames (dashed lines of different colors), can differ substantially in shape emphasizes the dependence of $\beta$ on the choice of reference frame. This has also been pointed out by \cite{Young2022}. As expected, the outer part of the disk appears tilted (i.e., with a non-zero, roughly constant $\beta$ in the outer regime) if the inclination is incorrectly assumed.
When comparing the fitted profiles with the corresponding transformed input profiles (solid and dashed lines of the same color), we find that the fitted warp generally reproduces the shape of the input profile, even when the assumed inclination differs from the true inclination by $5^\circ$. Although the deviations between the fitted and transformed input profiles become larger for the largest inclination offsets, the overall shape remains similar.

While both the warp $\beta$ and the twist $\gamma$ depend on the choice of reference system, the warp amplitude $\psi$ is invariant under the change of the reference frame, as it is defined from the local derivative of the angular momentum vectors (see Eq.~\ref{eq:definitions}), also found in \cite{Juhasz2017}.
Although Eq.~\ref{eq:psi} contains a dependence on inclination $i$, a change of reference frame changes both $i(r)$ and $\mathrm{PA}(r)$, and their radial derivatives together, leaving $\psi$ unchanged.
However, this does not mean that the measurement of $\psi$ is independent
from inclination. An incorrect inclination affects the deprojection of the kinematic data and the fitted profiles of $\beta$ and $\gamma$, from which $\psi$ is then calculated.
We find that the recovered profiles show a similar qualitative shape in $\psi$ over the main warp region. Larger deviations occur in the outer disk, where the backside emission contributes to the residuals ({$\gtrsim 200\,\mathrm{au}$}).
We note that the warp amplitude depends on both $\beta(r)$ and $\gamma(r)$ and is therefore degenerate in these angles. Consequently, $\psi$ is not sufficient on its own to fully characterize the warp shape. However, its invariance to the choice of reference system provides a useful way to compare the recovered warp profiles.
Thus, for the inclination offsets considered here, we find that the qualitative shape of $\psi$ is not strongly affected, although care must be taken in the choice of reference system when interpreting $\beta$ and $\gamma$.

\begin{figure}
    \centering
    \includegraphics[width=0.95\linewidth]{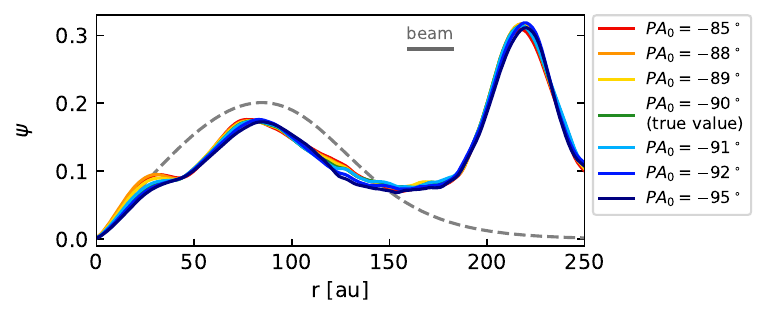}
    \caption{Warp amplitude $\psi$ for evaluations of different position angles, evaluated in the worst case model. The gray dashed line is directly computed from the input model.}
    \label{fig:pa-err}
\end{figure}

We perform the same analysis for errors in the position angle of the disk.
Analogously, we find that the warp amplitude profiles $\psi$ are very similar for the different assumed position angles (Fig.~\ref{fig:pa-err}). Here, we evaluated the worst case model in order to gauge the applicability of this worst case scenario.
Taking into account the uncertainties in inclination, position angle, or a combination of both would likely give a better estimate on the errors in the warp shape fitting process.

\begin{figure}
    \centering
    \includegraphics[width=\linewidth]{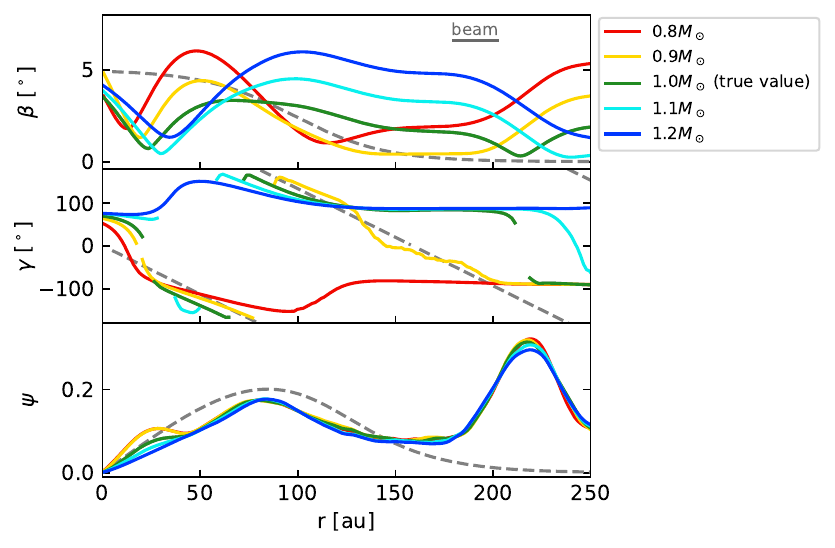}
    \caption{Warp $\beta(r)$, twist $\gamma(r)$, and warp amplitude $\psi(r)$ in the worst case scenario, evaluated for incorrectly assumed stellar masses. The grey dashed line indicate the input model.}
    \label{fig:mstar-err}
\end{figure}

Further, we have assumed that we know the stellar mass perfectly for the models in the worst case models. Usually, the stellar mass is subject to many uncertainties. We therefore analyzed the twisted disk also for errors in the stellar mass, analogously to the analysis above. Fig.~\ref{fig:mstar-err} shows that guessing the stellar mass incorrectly may lead to an apparent tilt of the outer disk. This can be explained by the kinematic residual, that remains axisymmetric when subtracting the incorrect Keplerian velocity. This axisymmetric residual then gets misinterpreted as an overall tilt in the \texttt{warpfitter} process. Again, evaluating $\psi(r)$, we find that the recovered profiles show a similar qualitative shape for the different assumed stellar masses.

However, we want to note that unfortunate combinations of incorrectly guessed stellar mass, position angle, inclination, and other parameters might lead to signatures that may be misinterpreted as a disk warp. Constraining some of these parameters from observations is difficult, as they may be degenerate. Therefore, caution must be taken when interpreting actual observations.

Overall, we find that fitting a warp to the kinematic residuals can recover the qualitative warp shape for disks with low to moderate inclination.
However, the agreement with the input profile depends in part on the uncertainties in observable disk properties, especially the inclination and position angle, as well as on assumptions about the three-dimensional disk geometry, particularly its thickness.
Additionally, we cannot retrieve warp features that are smaller than the beam resolution such as sharp transitions or warps in the very inner region of the disk, but the beam size provides an upper limit on the extent of the warp.
A spectral resolution of up to $0.8\,\mathrm{km/s}$ seems to be sufficient.
For the tested errors in inclination, position angle, and stellar mass, the recovered warp amplitude profiles show a similar qualitative shape over the main warp region.

\section{Impact of other physical effects} \label{sec:othereffects}

\begin{figure}
    \centering
    \includegraphics[width=0.95\linewidth]{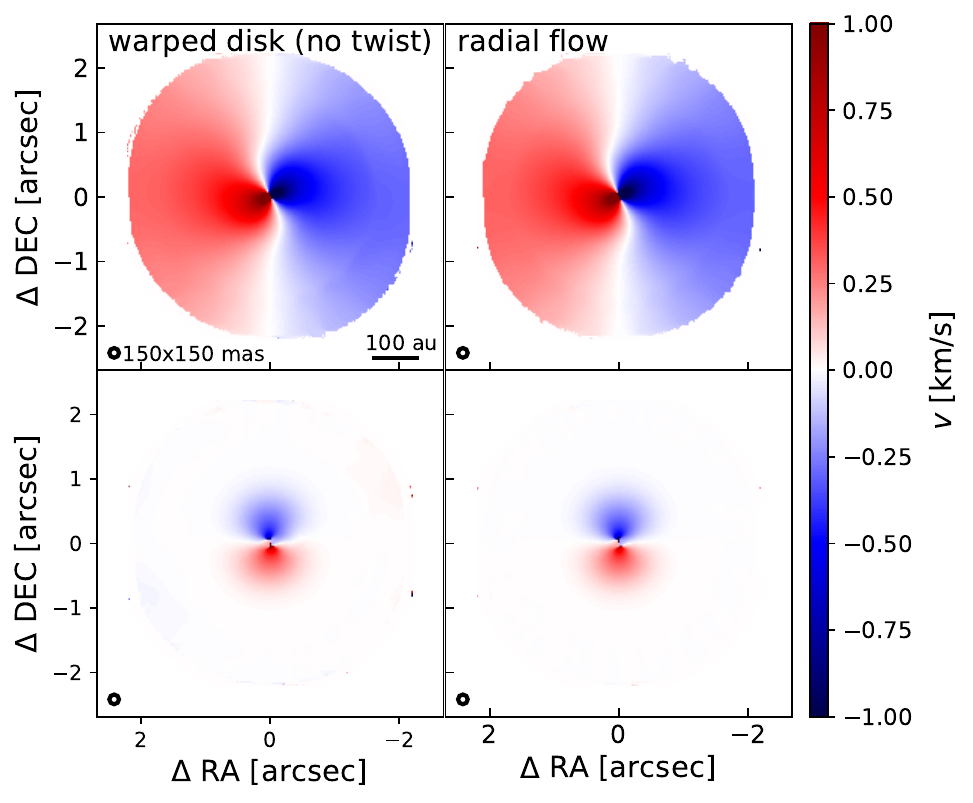}
    \caption{Kinematic maps computed using the quadratic method (top row) of a warped, but untwisted disk (left column) and a disk containing a radial flow (right column). The (outer) disk is viewed at an inclination of $10^\circ$. The bottom row shows the residuals from an unwarped disk (best-case scenario) without a radial flow for both cases.}
    \label{fig:maps-radialflow}
\end{figure}

\begin{figure}
    \centering
    \includegraphics[width=\linewidth]{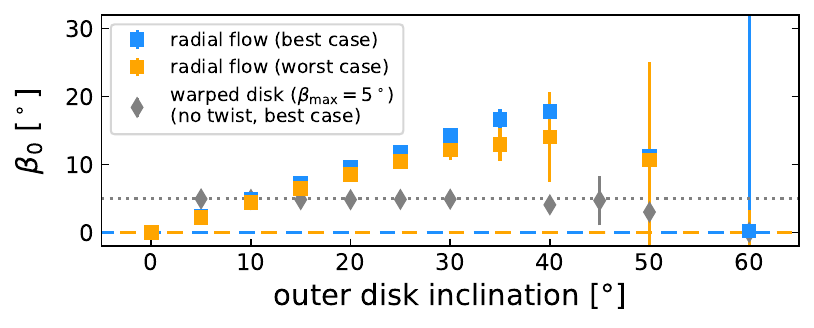}
    \caption{Fits of the obtained warp strength $\beta_\mathrm{0}$ for the warped disk without a twist (grey diamonds) and the unwarped disk containing a radial flow in the best and worst case scenario (blue and orange squares, respectively).}
    \label{fig:fits-radialflow}
\end{figure}

One caveat of interpreting the kinematic residuals as warps is that other physical effects can also lead to deviations from a smooth Keplerian model.
Such deviations are commonly observed \citep{Fukagawa2026} and can be caused by dynamical interactions of the disk with planets \citep{Perez2015, Pinte2018, Teague2018, Izquierdo2026}, fluid instabilities \citep{Barraza-Alfaro2025}, vortices \citep{Boehler2021, Wolfer2025}, and other disk perturbations. Observations also indicate that vertical flows \citep{Teague2018, Benisty2026} and gas pressure maxima \citep{Teague2018, Rosotti2020, Stadler2025} leave distinctive signatures in the kinematics.
This is why we aim to investigate other dynamical processes that might be misinterpreted as warps.

As radial flows can resemble the kinematic signatures of warped disks \citep{Rosenfeld2014, Zuleta2024}, we investigate such radial flows in Sect.~\ref{sec:radialflow}.
Further, the sloshing and breathing motions resulting from warp dynamics \citep{Lubow2000, Ogilvie2013, Dullemond2022} also influence the kinematic observations, thus we investigate these effects in Sect.~\ref{sec:sloshing}.

\subsection{Radial flow} \label{sec:radialflow}

\begin{figure*}
    \sidecaption
    \includegraphics[width=12cm]{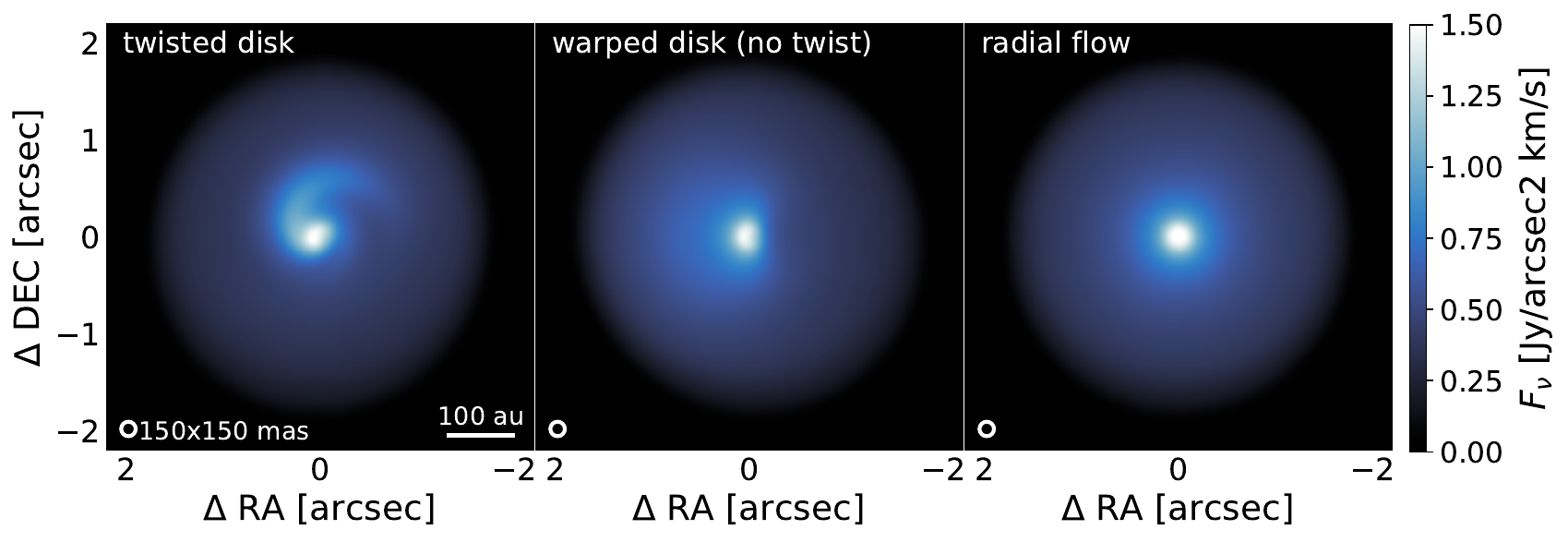}
    \caption{Integrated intensity maps along the spectral axis (``Moment~0'') of a warped and twisted disk (first panel), a warped, but untwisted disk (second panel), and an unwarped disk that contains a radial flow (third panel). In all panels, the (outer) disk is synthetically observed at $10^\circ$ inclination and the images are deprojected. The peak intensities can be seen in Fig.~\ref{fig:mom0-peak}.}
    \label{fig:mom0-maps}
\end{figure*}

To investigate the risk of misinterpreting a radial flow as a warp, we set up an unwarped disk including an axisymmetric radial flow that intentionally resembles the signatures of a warped disk, similar to \cite{Zuleta2024}.
This axisymmetric radial flow can not resemble a twisted disk, which is why we set up a comparison simulation of a warped, but untwisted disk, specifically ${\beta_\mathrm{0}=5^\circ}$ and ${\gamma_\mathrm{0}=0^\circ}$. The kinematic map of this simulation is shown in Fig.~\ref{fig:maps-radialflow}, left column.

The disk with a radial flow does not contain a warp. Instead, we added a radial velocity that is proportional to the Keplerian velocity $v_\mathrm{k}$ as follows:
\begin{equation} \label{eq:raidal-flow}
    v_r = -\chi(r) \ v_\mathrm{k}, \hspace{0.2cm} \mathrm{with} \hspace{0.2cm}
    \chi(r) = \frac{\chi_0}{2} \left[1 - \tanh\left(\frac{r-\chi_\mathrm{loc}}{\chi_\mathrm{width}}\right)\right],
\end{equation}
where $\chi(r)$ is a function to scale the strength of the radial flow with radius, where we adopt the same shape as for the warp profile, Eq.~\ref{eq:incl-ini}.
Here, analogously, $\chi_0$ is the scaling parameter at the inner edge of the disk, and $\chi_\mathrm{loc}$ and $\chi_\mathrm{width}$ determine the steepness of the function.
We chose these parameters such that the kinematic maps are visually indistinguishable for the warp and the radial flow model at an (outer) disk inclination of $10^\circ$, as shown in Fig.~\ref{fig:maps-radialflow}.
In particular, we set ${\chi_0=0.5}$, ${\chi_\mathrm{loc} = r_\mathrm{loc} = 100\,\mathrm{au}}$, and ${\chi_\mathrm{width} = r_\mathrm{width} = 50\,\mathrm{au}}$. The parameter ${\chi_0=0.5}$ indicates a very strong radial flow, which is necessary to visually match the two models in the kinematics. This also means that even a modest warp of only $5^\circ$ requires a very large, likely unphysical, radial flow to have a comparable kinematic signature.
We note that this specific shape of the radial flow function is motivated by the aim to create kinematic maps that exactly resemble the warp morphology in our models. Physically, this would mean that the radial flow is enhanced in the inner regions of the disk.
Although such a radial flow profile may have a plausible physical origin, such as magnetic disk winds \citep{Martel2022}, exploring its physical interpretation is not the aim of this work and is therefore beyond its scope.

The residuals from an unwarped, Keplerian disk without a radial flow are shown in the bottom row of Fig.~\ref{fig:maps-radialflow}.
We again fit the warp shape $\beta(r)$ as described in Sect.~\ref{sec:faceondisks} and show the fitted warp strength $\beta_\mathrm{0}$ as function of inclination in Fig.~\ref{fig:fits-radialflow}.
We emphasize that the true value of $\beta_\mathrm{0}$ in the radial flow model is $0^\circ$ (blue and orange dashed lines), since the disk does not contain a warp.

We find that the kinematic residuals of the radial flow model are misinterpreted as a warp with the applied method. For an inclination of ${i=10^\circ}$, we indeed obtain ${\beta_\mathrm{0} = 5^\circ}$, as expected with our setup. However, the strength of the apparent warp highly depends on the inclination, strongly increasing with inclination, up to almost $20^\circ$. This can have consequences for the interpretation of observed protoplanetary disks, which we discuss in Sect.~\ref{sec:discussion-inclination}.

The kinematic map and the residuals from a Keplerian model at an inclination of $10^\circ$ are visually indistinguishable in the model of a warped disk versus a radial flow. However, the integrated flux along the spectral axis, called Moment~0 maps (see Fig.~\ref{fig:mom0-maps}), show an asymmetry for the warped disk both with and without the twist, whereas the radial flow model is perfectly axisymmetric.
We note that is not necessarily the case for a realistic disk, where a combination of perturbations in radial and azimuthal flows can mimic the signatures of a twist. However, a warp inevitably leads to an azimuthally asymmetric Moment~0 map, which provides an additional diagnostic.
This has also been stated by \citet[see their Fig.~9]{Juhasz2017}. 

Additionally, a radial flow of this kind can only resemble a specific orientation of the warped disk with respect to the observer. If the tilt direction of the inclined inner region is not perfectly perpendicular to the line of sight, the kinematic signature of the warp looks different. For the radial flow, however, the appearance does not change, as it is axisymmetric. An example of this can be seen in Fig.~\ref{fig:maps-radialflow-phi}. This shows that in most cases, it is possible to distinguish the two (idealized) scenarios. A statistical approach on multiple sources can also reveal insight to the origin, as for warp as cause, an isotropic distribution of the line of nodes is expected in contrast to radial flows.
Furthermore, the radial flow can only resemble a warp for specific radial dependencies of the flow strength and the physical mechanism to induce this specific shape is dubious.
Assuming a simpler radial dependence (e.g., constant $\chi(r) = \chi_0$), the kinematic signature does not resemble the that of a warp and thus cannot be misinterpreted as such.

\subsection{Evolution of warps} \label{sec:sloshing}

\begin{figure}
    \centering
    \includegraphics[width=0.9\linewidth]{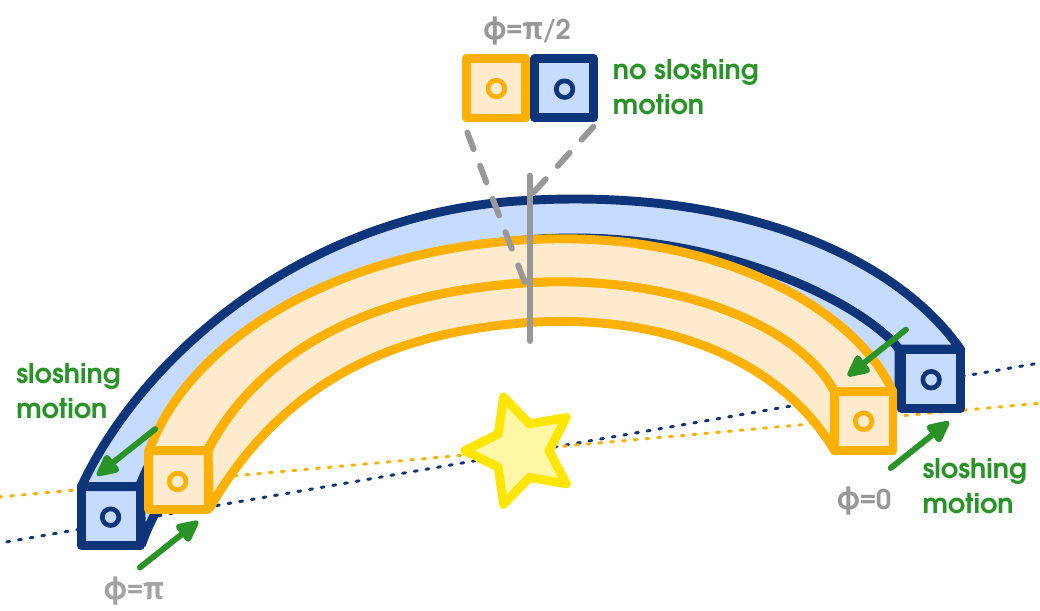}
    \caption{Schematic representation of the radial sloshing motion $v_{r,\mathrm{slosh}}$ (green arrows) caused by a misalignment of neighboring regions within a warped disk.}
    \label{fig:sloshing-explanation}
\end{figure}

\begin{figure*}
    \sidecaption
    \includegraphics[width=12cm]{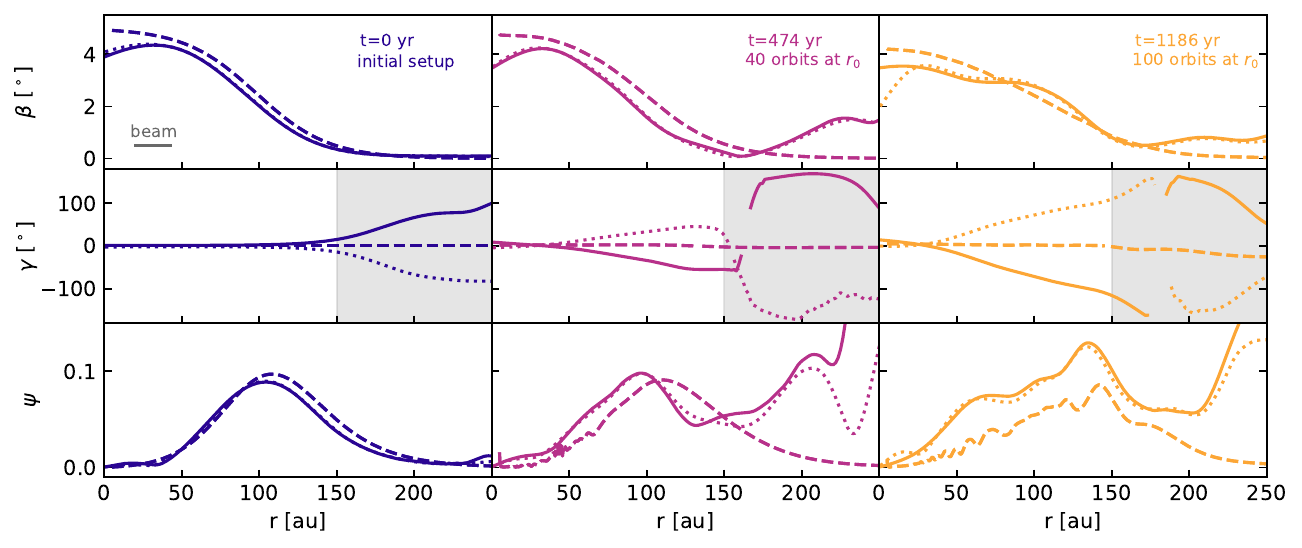}
    \caption{Fitted warp profile $\beta(r)$ , twist $\gamma(r)$, and warp amplitude $\psi(r)$ for different times in a hydrodynamic simulation of a warped, but untwisted disk. We define $r_0=5.2\,\mathrm{au}$. The solid lines are obtained with \texttt{warpfitter} at an outer disk inclination of $i_0=10^\circ$. The dotted lines show the \texttt{warpfitter} fit of the opposite side of the disk (flipped case).
    The dashed lines show the true profiles extracted from the 3D hydrodynamic simulations.
    The gray shaded regions indicate where the true $\beta$ is close to zero, hence the twist $\gamma$ is undefined.}
    \label{fig:hydro-simulation}
\end{figure*}

Warped protoplanetary disks are dynamic. The warp shape changes due to internal and external torques acting on the local orbital planes in the disk \citep{Papaloizou1983, Papaloizou1995}. In typical conditions of protoplanetary disks, where the aspect ratio $h$ is higher than the viscosity parameter $\alpha$, the warp travels in a wave through the disk \citep{Lubow2000}. The warp evolution is driven by internal torques that are caused by pressure gradients that arise due to the misplacement of inclined neighboring disk regions. This gives rise to epicyclic resonant motions called sloshing (in radial and azimuthal direction) and breathing (in vertical direction) \citep{Ogilvie2013, Dullemond2022}
These motions, especially the sloshing that occurs in radial and azimuthal direction, are stronger toward the disk surfaces and can even become supersonic \citep[see for example][their Fig.~25]{Kimmig2024}. In this work so far, we have ignored the effects of the sloshing and breathing motions due to the warp shape.

\begin{figure}
    \centering
    \includegraphics[width=\linewidth]{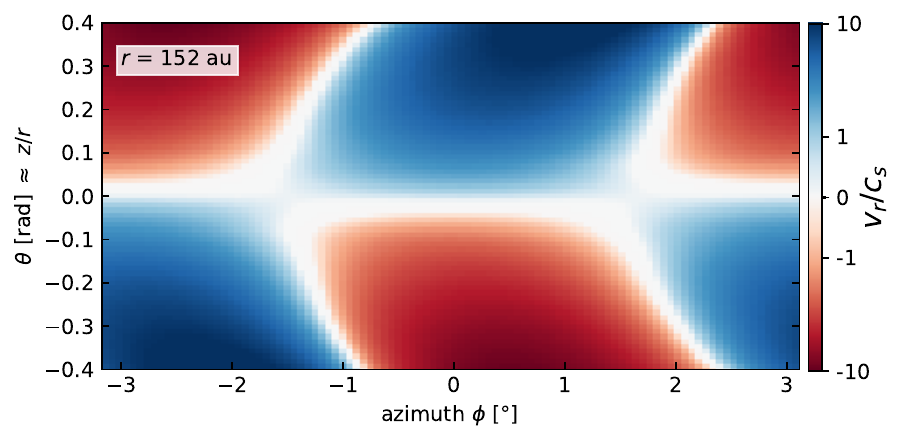}
    \caption{Sloshing motions along an annulus at $r=152\,\mathrm{au}$ at the latest stage in our simulation, $t=1,186\,\mathrm{yr}$. The radial velocity is given in the local warped coordinate frame.}
    \label{fig:sloshing-radvel}
\end{figure}

In linear theory, it is possible to estimate the velocity of the sloshing, depending on the warp amplitude $\psi$ \citep{Lodato2007, Dullemond2022, Winter2025}. The sloshing velocity varies azimuthally, as indicated in Fig.~\ref{fig:sloshing-explanation}, and the local maximum velocity during one orbit can be given by
\begin{equation}
    v_{r,\mathrm{slosh}} \approx \frac{\psi}{2\alpha} \Omega z, \hspace{1cm} v_{\phi,\mathrm{slosh}} \approx \frac{\psi}{4 \alpha} \Omega z,
\end{equation}
where $\Omega$ is the angular frequency and $z$ the vertical height above the local midplane. This means that at one pressure scale height, $z_\mathrm{surf} = H$, the maximum sloshing velocity can become roughly $5 c_\mathrm{s}$ for ${\psi = 10^{-2}}$ and ${\alpha = 10 ^{-3}}$, where $c_\mathrm{s}$ denotes the sound speed. We note that this is a linear approximation and a full treatment should consider shocks produced by supersonic motions, leading to effectively lower sloshing velocities.
We thus investigate the sloshing motions in our hydrodynamic simulations. Fig.~\ref{fig:sloshing-radvel} shows the radial velocities in the warped coordinate frame transformed to align with the local disk geometry. The radial velocity pattern is characteristic of warp evolution, with alternating regions of inward and outward motions along the annulus, corresponding to the sloshing motion \citep[see][Sect.~5.2]{Kimmig2024}.

In this section, we investigate the effect of sloshing on the resulting fit of the warp. Because the linear analysis only provides a rough estimation of the sloshing motion and does not account for shocks and other non-linear effects, we set up a 3D hydrodynamic simulation using FARGO3D \citep{FARGO3D} of the same disk setup described in Sect.~\ref{sec:methods}.
For the disk viscosity, we choose ${\alpha = 10^{-3}}$ and recall that we set the aspect ratio to ${h_0 = 0.05}$ at ${r_0 = 5.2\,\mathrm{au}}$, which ensures that the warp evolution falls into the wave-like regime.
We simulate the first few orbits at the inner disk region, so that the disk can dynamically adjust to the warp shape, giving rise to the sloshing and breathing motions, but the shape itself does not change significantly. Fig.~\ref{fig:sloshing-radvel} shows that this evolution time is sufficient for the sloshing motions to develop.
Because the dynamics of a warped disk without a twist in an isolated environment is generally understood better than the dynamics of a twisted disk, we set up a warp without twist with ${\beta_\mathrm{0}=5^\circ}$ and ${\gamma_\mathrm{0}=0^\circ}$.
We can then create synthetic observations at different stages of the hydrodynamic simulation and evaluate the disk kinematics in the same way as in the previous sections using \texttt{warpfitter}.

We present the results of three different times of the simulation in Fig.~\ref{fig:hydro-simulation}. The first column shows the initial setup of the simulation. The setup, as well as the results, are similar to the disk investigated in Sect.~\ref{sec:results}, but the disk does not contain a twist. We note that the twist $\gamma$ is undefined if $\beta=0^\circ$. This is why we see some deviations in the twist profile in the outer part of the disk, and we do not consider these regions (gray shaded regions in the second row) in the analysis. The dashed lines in this figure are the profiles extracted from the three-dimensional density of the hydrodynamic simulation.

As the hydrodynamical simulation evolves, we find that the extracted warp profiles $\beta$ (solid lines in the top row) remain close to the profile extracted directly from the hydrodynamical simulation (dashed lines).
In contrast, the twist $\gamma$ (second row) develops strong deviations as the simulation advances in time. Particularly, the disk looks twisted (non-zero $\gamma$), even though the true disk is not twisted, as the dashed lines, which are extracted from the density structure of the hydrodynamic simulation, remain zero.
We note that we use a sufficiently high resolution in our simulation such that the numerical precession of the disk found by \cite{Kimmig2024} is negligible. \cite{Kimmig2024} also find that an undriven warp develops a twist due to physical internal precession. In our model, this twist has not yet developed, as shown by the dashed lines in Fig.~\ref{fig:hydro-simulation}. If we evolved the simulation for a longer time, we would expect this physical twist to develop.
As the profile $\psi$ depends on both $\beta$ and $\gamma$, it also shows significant deviations from the warp amplitude in the simulation.

We suspect that this apparent twist is a result of the sloshing and breathing motions that develop at the later stages of the simulation.
This interpretation is supported by a direct comparison between models with and without sloshing motions (first and third column in Fig.~\ref{fig:hydro-simulation}). Both models are viewed from the same angle, which means that effects such as line-of-sight projection and emission height as found in \cite{Young2022} are the same in both cases and do not contribute to the difference in the observed twist profile. Since the apparent twist is only present in the model including sloshing motions, this twist cannot be attributed to projection effects alone.
In this case, the sloshing motion would be antisymmetric between the top and bottom disk surfaces, as indicated in Fig.~\ref{fig:sloshing-explanation}.
Therefore, we computed additional images using ray-tracing of the same simulation snapshots in time, but flipped by $180^\circ$ around the disk major axis. For reference, an example of the kinematic maps of both disk sides is shown in Fig.~\ref{fig:kinematic-maps-hydro-flipped}.
Since the two disk sides are otherwise subject to the same viewing geometry, this comparison allows us to isolate the effect of antisymmetric motions due to the warp evolution.
The results are shown with the dotted lines in Fig.~\ref{fig:hydro-simulation}. Both the extracted warp profile $\beta$ and amplitude $\psi$ are almost the same for the two sides of the disk, indicating that the breathing motions do not have a large effect on the kinematics. The twist $\gamma$, however, is flipped, indicating that the sloshing motions have a significant effect on the apparent warp morphology.
For reference, the kinematic maps of the last time step (right column) are shown in Fig.~\ref{fig:kinematic-maps-hydro-flipped}.

\section{Discussion} \label{sec:discussion}

\subsection{Warp parametrization} \label{sec:discussion-warp-parametrization}

The warp profile adopted in this work, described by Eq.~\ref{eq:incl-ini} in Sect.~\ref{sec:methods-diskmodel}, provides a controlled and simple representation of a continuously warped disk.
However, it does not capture the full range of possible warp profiles that can arise from hydrodynamical evolution. In particular, the warp evolves with time, which changes the shape of the disk.
Nevertheless, we expect the general conclusions of this work to hold for different continuously warped configurations, as our parametrization serves primarily as controlled example for investigating how a warp is traced in kinematic observations.

To assess how representative our parametrized warp shape is of hydrodynamically produced warps, we compare it with simulations of warps created by an inclined binary. \cite{Young2023} present such simulations. In the cases where the warp remains continuously connected, the warp profile can be qualitatively similar (see, e.g., their Figs.~8 and 9), if the profiles of the tilt and twist angles are transformed to a reference frame aligned with the outer disk geometry.
Additionally, fly-by events as presented in \cite{Kimmig2026} can lead to comparable warp profiles.
Therefore, the parametrization adopted here can provide a reasonable representation of physically produced continuously warped disks.

More extreme warp transitions can arise when the disk breaks into two or more separate disk planes \citep{Lodato2010, Facchini2013, Facchini2018, Rabago2023, Young2023}.
Such sharp transitions are more difficult to trace observationally in disk kinematics, because beam convolution can smear out the break, as demonstrated in Sect.~\ref{sec:brokendisks}. Our results should therefore primarily be interpreted in the context of continuously warped disks.

\subsection{Effect of disk thickness} \label{sec:discussion-vertical-extent}

\begin{figure}
    \centering
    \includegraphics[width=0.85\linewidth]{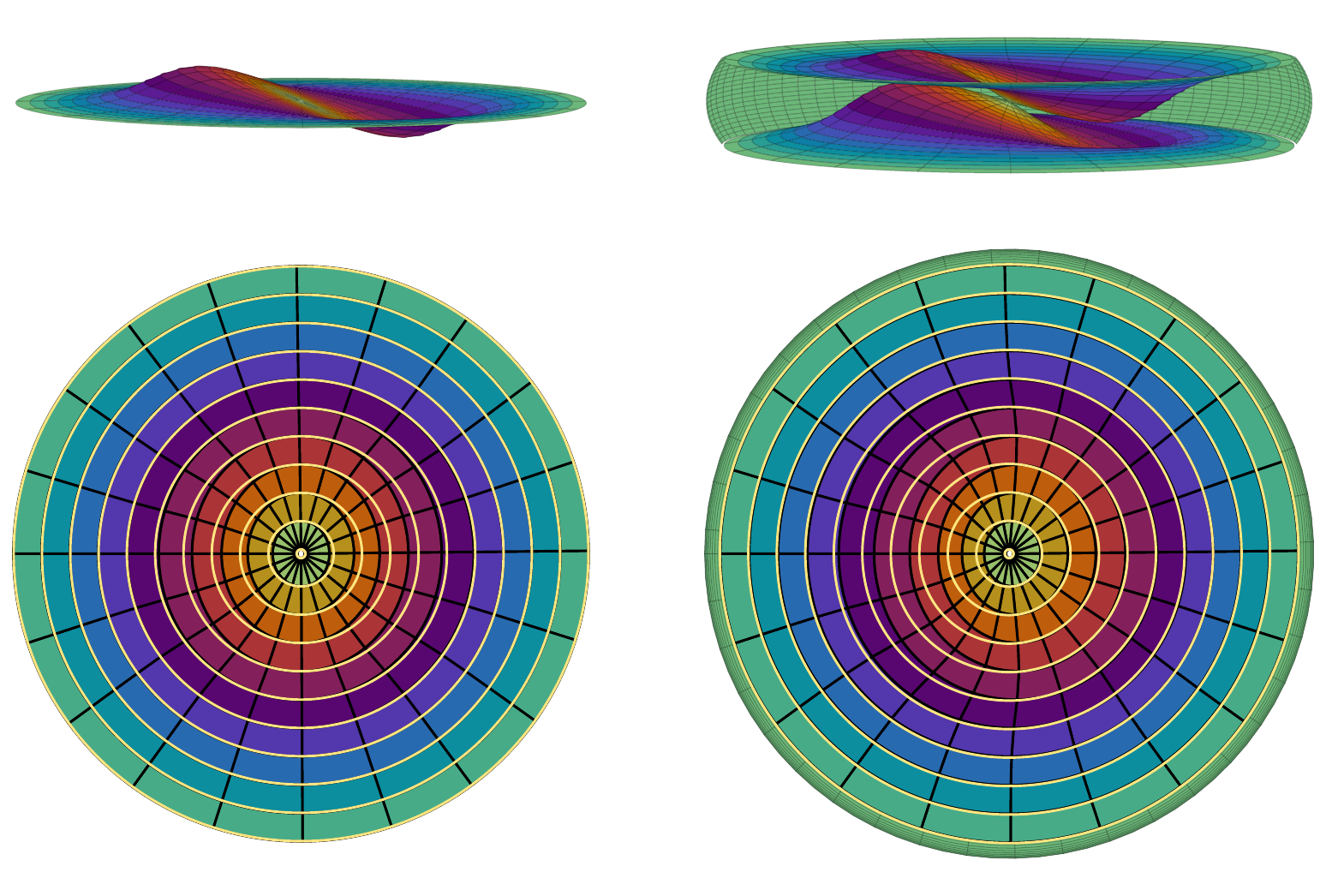}
    \caption{Schematics of projection effects in a infinitesimally thin disk (left column) and a vertically extended disk (right), in this example with a constant aspect ratio. The bottom row shows a face-on view of the outer edge in both cases. The yellow overplotted circles are perfectly concentric in both cases.}
    \label{fig:schematics-emission-height}
\end{figure}

With the method of fitting a warp to the kinematic residuals by W25, we assume an infinitesimally thin disk. Realistic protoplanetary disks, however, have a significant vertical extent, often described with the aspect ratio ${h = H_\mathrm{p}/r}$, where $H_\mathrm{p}$ is the pressure scale height of the disk.
This vertical extent shifts the disk surface, causing distortions in the projected disk plane. We visualize this effect in Fig.~\ref{fig:schematics-emission-height}, where we show a perfectly thin disk in the left column, and a vertically extended disk with a constant aspect ratio in the right column.
The constant aspect ratio leads to a thinner disk at the inner regions (yellow), becoming thicker toward the outer regions (purple to green).

The bottom row in the figure shows the projected radial regions (colors and black lines) when the outer disk is viewed perfectly face-on. We overplot yellow concentric circles, which are the lines along which the velocity residuals are evaluated in the method by W25.
It becomes clear specifically in the inner (orange) regions that the inclined radial circles are symmetric in the perfectly thin case on the left, while they are slightly shifted to the right in a thick disk (as seen in the comparison between black ellipses and yellow concentric circles).
The current warp evaluation of the residuals does not account for this shift caused by the vertical thickness of the disk, which also explains our results that the method is not able to accurately recover the warp shape in thicker disks (see Fig.~\ref{fig:aspectratio}).

We note that creating a disk surface with a constant aspect ratio is not the same transformation as simply shifting the infinitesimally thin disk up and down. In the thin case, the orbital regions are purely rotated, while there is a translation depending on $r$ in addition to the rotation in the thick disk. This means that the warped shape of the surface is fundamentally different, even though the warp shape at the midplane is identical to the infinitesimally thin disk.
It is possible to extend the method to include this vertical offset of the emitting layer in the disk.
A framework of how to include these effects has been suggested by \cite{Aizawa2026}.
However, this means that the vertical thickness of an observed disk needs to be known, or simultaneously fitted when fitting the model for computing the residuals. We will explore this direction in more detail in future work.

\subsection{Implications for observations} \label{sec:discussion-inclination}

\begin{figure}
    \centering
    \includegraphics[width=0.9\linewidth]{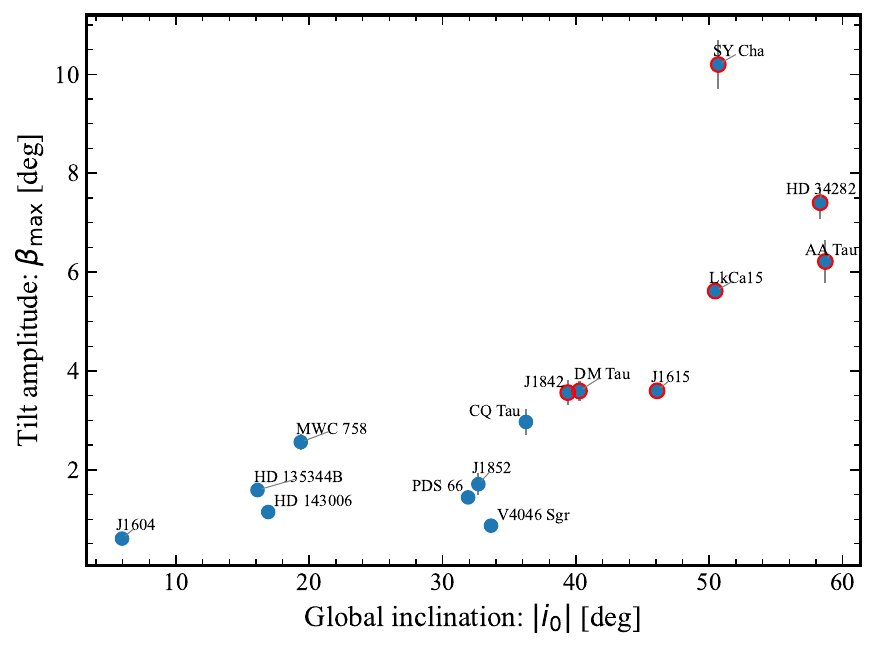}
    \caption{Observed maximum tilt amplitude of the tentative warps in the exoALMA sample over global inclination. The tilt amplitude $\beta_\mathrm{max}$ is defined as the difference between maximum and minimum $\beta(r)$-value in each disk. Red circles are disks where the backside is visible, which needed to be accounted for in to rotation curve modelling using a double bell line profile \citep[see][]{Winter2025}.}
    \label{fig:exoalma-disks}
\end{figure}

In this section, we take a closer look at the tentative warps in the exoALMA sample \citep{Teague2025}. Fig.~\ref{fig:exoalma-disks} shows the observed maximum tilts extracted by W25 in all exoALMA disks.
We define this maximum tilt $\beta_\mathrm{max}$ as the maximum value of the tilt profile in each disk, which means that it is not a fitting parameter as opposed to the $\beta_\mathrm{0}$ values we recovered from our synthetic simulations. The comparison is challenging because there is no uniquely defined disk plane that can be used as a reference. In our analysis, we use the orientation of the outer disk as a reference, which is not necessarily the same as what is inferred from observations, especially if a warp is located in the outer regions.
However, because of our chosen warp shape (Eq.~\ref{eq:incl-ini}), $\beta_0$ is the same as $\beta_\mathrm{max}$ in our models with the outer disk as reference plane.

The maximum tilt values in Fig.~\ref{fig:exoalma-disks} show no clear correlation between inclination and maximum tilt for nearly face-on disks. In contrast, higher inclined disks, particularly $i\geq 40\deg$ show a strong correlation with inclination (red circles). These disks are also fitted with double bell profiles when computing the kinematic residuals, because they have visible backsides in the line profiles (see W25).
When looking at our result in Fig.~\ref{fig:observer-inclination}, we can understand this observation: the method interpreting kinematic residuals as warps is not reliable for disks at higher inclinations.

Looking at the results of Fig.~\ref{fig:fits-radialflow}, such a correlation between maximum tilt amplitude with observed inclination may hint at radial flows as an underlying physical process, as in our work, we find a correlation with inclination for radial flows, but not for warps. However, we recall that an axisymmetric radial flow can only resemble a very specific warp shape in a very specific orientation of the warp with respect to the observer (see Fig.~\ref{fig:maps-radialflow-phi}).
Nevertheless, we can point out that radial flows may become progressively more important at higher inclinations, providing a possible origin for the relationship between inclination and maximum warp tilt for higher observed inclination, as the amplitude of the velocity due to the disk rotation grows with inclination.

\subsection{Further discussion}

An important takeaway of this work is that kinematics can indeed give insight into the overall warp shape. However, the interpretation of kinematic residuals as warps should be taken with care, as other physical processes can lead to kinematic signatures that can be misinterpreted as warps. Thus, the kinematics should not be taken as the only indication of a warp. Indications of warps from other observations, such as shadows in scattered light or deviations in the continuum emission, could serve as additional indication to validate the interpretation of kinematic residuals as warps \citep[see][]{Bensity2023}. In the example of the exoALMA sample, \cite{Curone2025} find indications of a warp in the continuum emission of some disks. 

Further, this work shows that hydrodynamic effects that arise due to the warp shape are significant and should not be ignored in the evaluation of kinematic residuals. As a first step, the linear approximation of the sloshing and breathing motions could be taken into account. However, this will require an iterative fitting, as the strength of these motions depend on the local warp amplitude. Additionally, the analytic equation used for the residual evaluation in W25 only allows interpretation for small warp amplitudes, as it includes the small angle approximation. Replacing the analytic equation with a numerical approach for the fit of the kinematics would enable a generalization to arbitrarily large warps. This would also facilitate to include the effects of sloshing and breathing motions, vertical displacement of the emission surface, and further effects such as eccentricities \citep[][Ragusa et al., in prep.]{Ragusa2024}. 

Similar methods to investigate the warp in kinematics have been studied in other works. \cite{Young2022} present the characteristics of small warps in CO-line emission and compare their synthetic observations to the underlying simulation in a similar way. However, instead of describing the warp shape using $\beta$ and $\gamma$, they decompose the residual velocity along the azimuth as a sinusoid ${v_\mathrm{los} = v_\mathrm{max} \sin(\phi - \theta_\mathrm{p})}$, where $v_\mathrm{max}$ is the maximum velocity and $\theta_\mathrm{p}$ the phase angle of the peak velocity.
On the other hand, W25 describe this azimuthal residual velocity as ${v_\mathrm{los}^\mathrm{warp} = v_\phi \sin(i_0 + \delta i) \cos(\phi - \mathrm{PA}_0 - \delta \mathrm{PA})}$, where $v_\phi$ is the azimuthal velocity, $i_0$, $\mathrm{PA}_0$ are the overall inclination and position angles of the disk, and $\delta i$, $\delta \mathrm{PA}$ are small deviations of them, respectively.
To compare the methods, we can transform the two descriptions to find the respective equivalents.
The analysis of \cite{Young2022} considers the face-on case of a disk, meaning ${i_0 = 0^\circ}$ and hence arbitrarily $\mathrm{PA}_0 = 0^\circ$.
They also investigate the effect of viewing inclination by considering inclinations up to $35^\circ$.
In the assumption of small warps, $\sin(\delta i) \approx \delta i$. This way, we can relate the parameters ${v_\mathrm{max} = v_\phi \delta i}$ and ${\theta_\mathrm{p} = \delta \mathrm{PA} - \pi/2}$, showing that both methods are equivalent. However, for arbitrary inclinations of the disk $i_0$, the relation of the parameter $v_\mathrm{max}$ by \cite{Young2022} to $\delta i$ becomes more complicated, in particular ${v_\mathrm{max} = v_\phi \sin i_0 + v_\phi \delta i \cos i_0}$. Because both $v_\phi$ and $\delta i$ depend on $r$, it is more challenging to isolate the contribution of the warp in the frame of the work by \cite{Young2022}.

An interesting point to note is the effect of optical depth on the kinematic residuals. \cite{Young2022} find that the optical depth especially influences the appearance of the twist, which may look stronger than it actually is in optically thick disks.
As their underlying models are hydrodynamic simulations of a disk around a binary system, it is possible that sloshing and breathing motions contribute to the observed effect. As these motions are antisymmetric for the two disk surfaces, we would expect the motions of the opposite surfaces to cancel out in a face-on view for an optically thin tracer. However, these motions are likely unresolved in the simulations by \cite{Young2022}, and therefore we cannot say whether they play a significant role.

\section{Conclusions} \label{sec:conclusions}

In this work, we investigated the method to trace warps in kinematic maps of protoplanetary disks, developed by \cite{Winter2025}, which interprets kinematic residuals as local changes in inclination and position angle of different disk regions.
The aim of this work was to assess the applicability of the method and explore its limitations.
We list our main conclusions in the following.

\begin{itemize}
    \item{We find that the qualitative warp shape can be recovered from the kinematic residuals, particularly for small to moderate warp misalignments and nearly face-on disks.
    Including information on emission height and pressure correction generally improves the agreement with the input warp shape. For nearly face-on disks, the overall warp shape can also be traced without precise knowledge of the three-dimensional disk properties.
    The accuracy decreases for disks with higher inclination, where the vertical structure becomes more important.}
    \item{Vertically extended disks introduce additional geometric and kinematic effects that are not accounted for by the fitting method. For this reason, the fitting procedure fails or returns spurious features in the warp for thicker disks. This highlights the importance of accounting for the three-dimensional structure of the disk when applying the method to observations.}
    \item{We find that the overall warp shape can still be recovered for spectral resolutions of up to ${0.8\,\mathrm{km/s}}$ in the models considered here. The reason for this is that discontinuities in the kinematic map do not necessarily strongly affect the fitting result, since the local inclination of a disk region is determined by fitting sine and cosine curves along an azimuthal annulus. This suggests that existing archival observations with moderate spectral resolution may be suitable for tracing disk warps. The spatial resolution on the other hand is more important, as sharp warp transitions and warps close to the inner disk region cannot be reliably recovered when their spatial extent is smaller than approximately two beam sizes.}
    \item{A specific radial flow pattern can produce a kinematc signature that resemples a warp, which means that an unwarped disk containing a radial flow may be misinterpreted as a warp.
    For small inclination angles, reproducing the kinematic signature of a warp requires relatively large radial velocities; conversely, smaller radial velocities are sufficient at larger inclinations.
    Further, the integrated intensity (Moment~0) map is axisymmetric for an axisymmetric radial flow, but never perfectly axisymmetric for a warp. Additionally, such a radial flow can only resemble a warp in a particular geometric configuration with respect to the observer.}
    \item{The dynamics of warped disks can affect the observed residuals. In a dynamic warped disk, radial and vertical motions develop at the disk surfaces. By comparing the initial model without these motions to the evolved hydrodynamical model that includes the radial and vertical motions, we find
    that they can induce a spurious twist in the warp shape fit.}
    
\end{itemize}

All in all, we conclude that obtaining the warp shape from kinematic residuals is possible. The current method by \cite{Winter2025} reproduces the overall warp shape in our nearly face-on disk models with moderate warp amplitudes. However, several improvements are possible to make the fitting method applicable to more general cases.

\begin{acknowledgements}
We thank Attila Juhasz for providing python scripts for the creation and evaluation of \texttt{.fits} files.

GR, ER, and CK acknowledge support from the European Union (ERC Starting Grant DiscEvol, project number 101039651) and from Fondazione Cariplo, grant No. 2022-1217.
AJW has been supported by a Royal Society University Research Fellowship, grant number URF\textbackslash R1\textbackslash 241791.
We acknowledge financial support from the project ``Space Observations of Proto-planetary Accretion Disc Evolution (SPADE)'' in the framework of the ``PROgramma di RIcerca Spaziale di base (PRORIS)'', financed by the Italian Ministry of University and Research (Ministero dell’Università e della Ricerca; MUR).
SF acknowledges financial contribution from the European Union (ERC, UNVEIL, 101076613) and from PRIN-MUR 2022YP5ACE.
Support for AFI was provided by NASA through the NASA Hubble Fellowship grant No. HST-HF2-51532.001-A awarded by the Space Telescope Science Institute, which is operated by the Association of Universities for Research in Astronomy, Inc., for NASA, under contract NAS5-26555.
MB has received funding from the European Research Council (ERC) under the European Union’s Horizon 2020 research and innovation programme (PROTOPLANETS, grant agreement No. 101002188).
TB acknowledges funding from the European Union under the European Union's Horizon Europe Research and Innovation Programme 101124282 (EARLYBIRD) and funding by the Deutsche Forschungsgemeinschaft (DFG, German Research Foundation) under grant 325594231, and Germany's Excellence Strategy - EXC-2094 - 390783311.
Views and opinions expressed are, however, those of the author(s) only and do not necessarily reflect those of the European Union or the European Research Council. Neither the European Union nor the granting authority can be held responsible for them.
Computational resources were provided by the INDACO Platform, which is a project of High Performance Computing at the University of MILAN (\url{http://www.unimi.it}).
We acknowledge the use of the the Python \citep{Python3} libraries \texttt{matplotlib} \citep{matplotlib}, \texttt{numpy} \citep{NumPy}, \texttt{scipy} \citep{SciPy}, and \texttt{astropy} \citep{AstroPy}.


\end{acknowledgements}

\bibliographystyle{aa} 
\bibliography{kinematics} 

@ARTICLE{Nealon2020,
       author = {{Nealon}, Rebecca and {Price}, Daniel J. and {Pinte}, Christophe},
        title = "{Rocking shadows in broken circumbinary discs}",
      journal = {\mnras},
         year = 2020,
        month = mar,
       volume = {493},
       number = {1},
        pages = {L143-L147},
          doi = {10.1093/mnrasl/slaa026},
archivePrefix = {arXiv},
       eprint = {2002.02983},
 primaryClass = {astro-ph.EP},
       adsurl = {https://ui.adsabs.harvard.edu/abs/2020MNRAS.493L.143N}
}

@ARTICLE{Lodato2010,
       author = {{Lodato}, Giuseppe and {Price}, Daniel J.},
        title = "{On the diffusive propagation of warps in thin accretion discs}",
      journal = {\mnras},
         year = 2010,
        month = jun,
       volume = {405},
       number = {2},
        pages = {1212-1226},
          doi = {10.1111/j.1365-2966.2010.16526.x},
archivePrefix = {arXiv},
       eprint = {1002.2973},
 primaryClass = {astro-ph.HE},
       adsurl = {https://ui.adsabs.harvard.edu/abs/2010MNRAS.405.1212L}
}

@ARTICLE{Young2023,
       author = {{Young}, Alison K. and {Stevenson}, Struan and {Nixon}, C.~J. and {Rice}, Ken},
        title = "{On the conditions for warping and breaking protoplanetary discs}",
      journal = {\mnras},
         year = 2023,
        month = oct,
       volume = {525},
       number = {2},
        pages = {2616-2631},
          doi = {10.1093/mnras/stad2451},
archivePrefix = {arXiv},
       eprint = {2306.11809},
 primaryClass = {astro-ph.SR},
       adsurl = {https://ui.adsabs.harvard.edu/abs/2023MNRAS.525.2616Y}
}

@ARTICLE{Martel2022,
       author = {{Martel}, {\'E}tienne and {Lesur}, Geoffroy},
        title = "{Magnetised winds in transition discs. I. 2.5D global simulations}",
      journal = {\aap},
         year = 2022,
        month = nov,
       volume = {667},
          eid = {A17},
        pages = {A17},
          doi = {10.1051/0004-6361/202142946},
archivePrefix = {arXiv},
       eprint = {2205.02126},
 primaryClass = {astro-ph.EP},
       adsurl = {https://ui.adsabs.harvard.edu/abs/2022A&A...667A..17M}
}

@ARTICLE{matplotlib,  author={Hunter, John D.},  journal={Computing in Science   Engineering},   title={Matplotlib: A 2D Graphics Environment},   year={2007},  volume={9},  number={3},  pages={90-95},  doi={10.1109/MCSE.2007.55}}

@ARTICLE{NumPy,
  author  = {Harris, Charles R. and Millman, K. Jarrod and
            van der Walt, Stéfan J and Gommers, Ralf and
            Virtanen, Pauli and Cournapeau, David and
            Wieser, Eric and Taylor, Julian and Berg, Sebastian and
            Smith, Nathaniel J. and Kern, Robert and Picus and others},
  title   = {Array programming with {NumPy}},
  journal = {Nature},
  year    = {2020},
  volume  = {585},
  pages   = {357–362},
  doi     = {10.1038/s41586-020-2649-2}
}

@ARTICLE{AstroPy,
       author = {{Astropy Collaboration} and {Price-Whelan}, Adrian M. and {Lim}, Pey Lian and {Earl}, Nicholas and {Starkman}, Nathaniel and {Bradley}, Larry and {Shupe}, David L. and {Patil}, Aarya A. and {Corrales}, Lia and {Brasseur}, C.~E. and {N{\"o}the}, Maximilian and {Donath}, Axel and {Tollerud}, Erik and {Morris}, Brett M. and {Ginsburg}, Adam and {Vaher}, Eero and {Weaver}, Benjamin A. and {Tocknell}, James and {Jamieson}, William and {van Kerkwijk}, Marten H. and {Robitaille}, Thomas P. and {Merry}, Bruce and {Bachetti}, Matteo and {G{\"u}nther}, H. Moritz and {Aldcroft}, Thomas L. and {Alvarado-Montes}, Jaime A. and {Archibald}, Anne M. and {B{\'o}di}, Attila and {Bapat}, Shreyas and {Barentsen}, Geert and {Baz{\'a}n}, Juanjo and {Biswas}, Manish and {Boquien}, M{\'e}d{\'e}ric and {Burke}, D.~J. and {Cara}, Daria and {Cara}, Mihai and {Conroy}, Kyle E. and {Conseil}, Simon and {Craig}, Matthew W. and {Cross}, Robert M. and {Cruz}, Kelle L. and {D'Eugenio}, Francesco and {Dencheva}, Nadia and {Devillepoix}, Hadrien A.~R. and {Dietrich}, J{\"o}rg P. and {Eigenbrot}, Arthur Davis and {Erben}, Thomas and {Ferreira}, Leonardo and {Foreman-Mackey}, Daniel and {Fox}, Ryan and {Freij}, Nabil and {Garg}, Suyog and {Geda}, Robel and {Glattly}, Lauren and {Gondhalekar}, Yash and {Gordon}, Karl D. and {Grant}, David and {Greenfield}, Perry and {Groener}, Austen M. and {Guest}, Steve and {Gurovich}, Sebastian and {Handberg}, Rasmus and {Hart}, Akeem and {Hatfield-Dodds}, Zac and {Homeier}, Derek and {Hosseinzadeh}, Griffin and {Jenness}, Tim and {Jones}, Craig K. and {Joseph}, Prajwel and {Kalmbach}, J. Bryce and {Karamehmetoglu}, Emir and {Ka{\l}uszy{\'n}ski}, Miko{\l}aj and {Kelley}, Michael S.~P. and {Kern}, Nicholas and {Kerzendorf}, Wolfgang E. and {Koch}, Eric W. and {Kulumani}, Shankar and {Lee}, Antony and {Ly}, Chun and {Ma}, Zhiyuan and {MacBride}, Conor and {Maljaars}, Jakob M. and {Muna}, Demitri and {Murphy}, N.~A. and {Norman}, Henrik and {O'Steen}, Richard and {Oman}, Kyle A. and {Pacifici}, Camilla and {Pascual}, Sergio and {Pascual-Granado}, J. and {Patil}, Rohit R. and {Perren}, Gabriel I. and {Pickering}, Timothy E. and {Rastogi}, Tanuj and {Roulston}, Benjamin R. and {Ryan}, Daniel F. and {Rykoff}, Eli S. and {Sabater}, Jose and {Sakurikar}, Parikshit and {Salgado}, Jes{\'u}s and {Sanghi}, Aniket and {Saunders}, Nicholas and {Savchenko}, Volodymyr and {Schwardt}, Ludwig and {Seifert-Eckert}, Michael and {Shih}, Albert Y. and {Jain}, Anany Shrey and {Shukla}, Gyanendra and {Sick}, Jonathan and {Simpson}, Chris and {Singanamalla}, Sudheesh and {Singer}, Leo P. and {Singhal}, Jaladh and {Sinha}, Manodeep and {Sip{\H{o}}cz}, Brigitta M. and {Spitler}, Lee R. and {Stansby}, David and {Streicher}, Ole and {{\v{S}}umak}, Jani and {Swinbank}, John D. and {Taranu}, Dan S. and {Tewary}, Nikita and {Tremblay}, Grant R. and {de Val-Borro}, Miguel and {Van Kooten}, Samuel J. and {Vasovi{\'c}}, Zlatan and {Verma}, Shresth and {de Miranda Cardoso}, Jos{\'e} Vin{\'\i}cius and {Williams}, Peter K.~G. and {Wilson}, Tom J. and {Winkel}, Benjamin and {Wood-Vasey}, W.~M. and {Xue}, Rui and {Yoachim}, Peter and {Zhang}, Chen and {Zonca}, Andrea and {Astropy Project Contributors}},
        title = "{The Astropy Project: Sustaining and Growing a Community-oriented Open-source Project and the Latest Major Release (v5.0) of the Core Package}",
      journal = {\apj},
         year = 2022,
        month = aug,
       volume = {935},
       number = {2},
          eid = {167},
        pages = {167},
          doi = {10.3847/1538-4357/ac7c74},
archivePrefix = {arXiv},
       eprint = {2206.14220},
 primaryClass = {astro-ph.IM},
       adsurl = {https://ui.adsabs.harvard.edu/abs/2022ApJ...935..167A}
}

@ARTICLE{SciPy,
  author  = {Virtanen, Pauli and Gommers, Ralf and Oliphant, Travis E. and
            Haberland, Matt and Reddy, Tyler and Cournapeau, David and
            Burovski, Evgeni and Peterson, Pearu and Weckesser, Warren and
            Bright, Jonathan and {van der Walt}, St{\'e}fan J. and
            Brett, Matthew and Wilson, Joshua and Millman, K. Jarrod and
            Mayorov, Nikolay and Nelson, Andrew R. J. and Jones, Eric and
            Kern, Robert and Larson, Eric and Carey, C J and
            Polat, {\.I}lhan and Feng, Yu and Moore, Eric W. and
            {VanderPlas}, Jake and Laxalde, Denis and Perktold, Josef and
            Cimrman, Robert and Henriksen, Ian and Quintero, E. A. and
            Harris, Charles R. and Archibald, Anne M. and
            Ribeiro, Ant{\^o}nio H. and Pedregosa, Fabian and
            {van Mulbregt}, Paul and {SciPy 1.0 Contributors}},
  title   = {{{SciPy} 1.0: Fundamental Algorithms for Scientific
            Computing in Python}},
  journal = {Nature Methods},
  year    = {2020},
  volume  = {17},
  pages   = {261--272},
  adsurl  = {https://rdcu.be/b08Wh},
  doi     = {10.1038/s41592-019-0686-2},
}

@book{Python3,
 author = {Van Rossum, Guido and Drake, Fred L.},
 title = {Python 3 Reference Manual},
 year = {2009},
 isbn = {1441412697},
 publisher = {CreateSpace},
 address = {Scotts Valley, CA}
}

@ARTICLE{Oeberg2023,
       author = {{Oberg}, N. and {Kamp}, I. and {Cazaux}, S. and {Rab}, Ch. and {Czoske}, O.},
        title = "{Observing circumplanetary disks with METIS}",
      journal = {\aap},
         year = 2023,
        month = feb,
       volume = {670},
          eid = {A74},
        pages = {A74},
          doi = {10.1051/0004-6361/202244845},
archivePrefix = {arXiv},
       eprint = {2212.03007},
 primaryClass = {astro-ph.EP},
       adsurl = {https://ui.adsabs.harvard.edu/abs/2023A&A...670A..74O}
}

@ARTICLE{Kurtovic2026,
       author = {{Kurtovic}, Nicolas T. and {Flores-Rivera}, Lizxandra and {Perez}, Laura M. and {Vioque}, Miguel and {Benisty}, Myriam and {Alarc{\'o}n}, Felipe and {Barraza-Alfaro}, Marcelo and {Curone}, Pietro and {Doi}, Kiyoaki and {Grant}, Sierra and {Jiang}, Haochang and {Kataoka}, Akimasa and {Long}, Feng and {Ribas}, {\'A}lvaro and {Sierra}, Anibal and {Stapper}, Lucas and {Temmink}, Milou and {Zagar{\'\i}a}, Francesco},
        title = "{An archival summary: 15 years of ALMA observations on disks and planet formation}",
      journal = {arXiv e-prints},
         year = 2026,
        month = may,
          eid = {arXiv:2605.30023},
        pages = {arXiv:2605.30023},
          doi = {10.48550/arXiv.2605.30023},
archivePrefix = {arXiv},
       eprint = {2605.30023},
 primaryClass = {astro-ph.EP},
       adsurl = {https://ui.adsabs.harvard.edu/abs/2026arXiv260530023K}
}

@ARTICLE{Min2017,
       author = {{Min}, M. and {Stolker}, T. and {Dominik}, C. and {Benisty}, M.},
        title = "{Connecting the shadows: probing inner disk geometries using shadows in transitional disks}",
      journal = {\aap},
         year = 2017,
        month = aug,
       volume = {604},
          eid = {L10},
        pages = {L10},
          doi = {10.1051/0004-6361/201730949},
archivePrefix = {arXiv},
       eprint = {1704.01844},
 primaryClass = {astro-ph.EP},
       adsurl = {https://ui.adsabs.harvard.edu/abs/2017A&A...604L..10M}
}

@ARTICLE{Dorschner1995,
       author = {{Dorschner}, J. and {Begemann}, B. and {Henning}, T. and {Jaeger}, C. and {Mutschke}, H.},
        title = "{Steps toward interstellar silicate mineralogy. II. Study of Mg-Fe-silicate glasses of variable composition.}",
      journal = {\aap},
         year = 1995,
        month = aug,
       volume = {300},
        pages = {503},
       adsurl = {https://ui.adsabs.harvard.edu/abs/1995A&A...300..503D}
}

@ARTICLE{Zubko1996,
       author = {{Zubko}, V.~G. and {Mennella}, V. and {Colangeli}, L. and {Bussoletti}, E.},
        title = "{Optical constants of cosmic carbon analogue grains - I. Simulation of clustering by a modified continuous distribution of ellipsoids}",
      journal = {\mnras},
         year = 1996,
        month = oct,
       volume = {282},
       number = {4},
        pages = {1321-1329},
          doi = {10.1093/mnras/282.4.1321},
       adsurl = {https://ui.adsabs.harvard.edu/abs/1996MNRAS.282.1321Z}
}

@ARTICLE{VanDishoeck1988,
       author = {{van Dishoeck}, Ewine F. and {Black}, John H.},
        title = "{The Photodissociation and Chemistry of Interstellar CO}",
      journal = {\apj},
         year = 1988,
        month = nov,
       volume = {334},
        pages = {771},
          doi = {10.1086/166877},
       adsurl = {https://ui.adsabs.harvard.edu/abs/1988ApJ...334..771V}
}

@ARTICLE{Visser2009,
       author = {{Visser}, R. and {van Dishoeck}, E.~F. and {Black}, J.~H.},
        title = "{The photodissociation and chemistry of CO isotopologues: applications to interstellar clouds and circumstellar disks}",
      journal = {\aap},
         year = 2009,
        month = aug,
       volume = {503},
       number = {2},
        pages = {323-343},
          doi = {10.1051/0004-6361/200912129},
archivePrefix = {arXiv},
       eprint = {0906.3699},
 primaryClass = {astro-ph.GA},
       adsurl = {https://ui.adsabs.harvard.edu/abs/2009A&A...503..323V}
}

@ARTICLE{Zhang2025,
       author = {{Zhang}, Shangjia and {Zhu}, Zhaohuan and {Fairbairn}, Callum W.},
        title = "{Shadow-induced Warps in Protoplanetary Disks}",
      journal = {\apjl},
         year = 2025,
        month = dec,
       volume = {995},
       number = {1},
          eid = {L33},
        pages = {L33},
          doi = {10.3847/2041-8213/ae2023},
archivePrefix = {arXiv},
       eprint = {2511.11358},
 primaryClass = {astro-ph.EP},
       adsurl = {https://ui.adsabs.harvard.edu/abs/2025ApJ...995L..33Z}
}

@INPROCEEDINGS{Bensity2023,
       author = {{Benisty}, M. and {Dominik}, C. and {Follette}, K. and {Garufi}, A. and {Ginski}, C. and {Hashimoto}, J. and {Keppler}, M. and {Kley}, W. and {Monnier}, J.},
        title = "{Optical and Near-infrared View of Planet-forming Disks and Protoplanets}",
    booktitle = {Protostars and Planets VII},
         year = 2023,
       editor = {{Inutsuka}, S. and {Aikawa}, Y. and {Muto}, T. and {Tomida}, K. and {Tamura}, M.},
       series = {Astronomical Society of the Pacific Conference Series},
       volume = {534},
        month = jul,
        pages = {605},
          doi = {10.48550/arXiv.2203.09991},
archivePrefix = {arXiv},
       eprint = {2203.09991},
 primaryClass = {astro-ph.EP},
       adsurl = {https://ui.adsabs.harvard.edu/abs/2023ASPC..534..605B}
}

@ARTICLE{Rosotti2020,
       author = {{Rosotti}, G.~P. and {Benisty}, M. and {Juh{\'a}sz}, A. and {Teague}, R. and {Clarke}, C. and {Dominik}, C. and {Dullemond}, C.~P. and {Klaassen}, P.~D. and {Matr{\`a}}, L. and {Stolker}, T.},
        title = "{Spiral arms in the protoplanetary disc HD100453 detected with ALMA: evidence for binary-disc interaction and a vertical temperature gradient}",
      journal = {\mnras},
         year = 2020,
        month = jan,
       volume = {491},
       number = {1},
        pages = {1335-1347},
          doi = {10.1093/mnras/stz3090},
archivePrefix = {arXiv},
       eprint = {1911.00518},
 primaryClass = {astro-ph.EP},
       adsurl = {https://ui.adsabs.harvard.edu/abs/2020MNRAS.491.1335R}
}

@ARTICLE{Boehler2021,
       author = {{Boehler}, Y. and {M{\'e}nard}, F. and {Robert}, C.~M.~T. and {Isella}, A. and {Pinte}, C. and {Gonzalez}, J.-F. and {van der Plas}, G. and {Weaver}, E. and {Teague}, R. and {Garg}, H. and {M{\'e}heut}, H.},
        title = "{Vortex-like kinematic signal, spirals, and beam smearing effect in the HD 142527 disk}",
      journal = {\aap},
         year = 2021,
        month = jun,
       volume = {650},
          eid = {A59},
        pages = {A59},
          doi = {10.1051/0004-6361/202040089},
archivePrefix = {arXiv},
       eprint = {2103.13474},
 primaryClass = {astro-ph.EP},
       adsurl = {https://ui.adsabs.harvard.edu/abs/2021A&A...650A..59B}
}

@ARTICLE{Pinte2018,
       author = {{Pinte}, C. and {Price}, D.~J. and {M{\'e}nard}, F. and {Duch{\^e}ne}, G. and {Dent}, W.~R.~F. and {Hill}, T. and {de Gregorio-Monsalvo}, I. and {Hales}, A. and {Mentiplay}, D.},
        title = "{Kinematic Evidence for an Embedded Protoplanet in a Circumstellar Disk}",
      journal = {\apjl},
         year = 2018,
        month = jun,
       volume = {860},
       number = {1},
          eid = {L13},
        pages = {L13},
          doi = {10.3847/2041-8213/aac6dc},
archivePrefix = {arXiv},
       eprint = {1805.10293},
 primaryClass = {astro-ph.SR},
       adsurl = {https://ui.adsabs.harvard.edu/abs/2018ApJ...860L..13P}
}

@ARTICLE{Perez2015,
       author = {{Perez}, Sebastian and {Dunhill}, A. and {Casassus}, S. and {Roman}, P. and {Szul{\'a}gyi}, J. and {Flores}, C. and {Marino}, S. and {Montesinos}, M.},
        title = "{Planet Formation Signposts: Observability of Circumplanetary Disks via Gas Kinematics}",
      journal = {\apjl},
         year = 2015,
        month = sep,
       volume = {811},
       number = {1},
          eid = {L5},
        pages = {L5},
          doi = {10.1088/2041-8205/811/1/L5},
archivePrefix = {arXiv},
       eprint = {1505.06808},
 primaryClass = {astro-ph.EP},
       adsurl = {https://ui.adsabs.harvard.edu/abs/2015ApJ...811L...5P}
}

@ARTICLE{Ragusa2024,
       author = {{Ragusa}, Enrico and {Lynch}, Elliot and {Laibe}, Guillaume and {Longarini}, Cristiano and {Ceppi}, Simone},
        title = "{Probing the eccentricity in protostellar discs: Modelling kinematics and morphologies}",
      journal = {\aap},
         year = 2024,
        month = jun,
       volume = {686},
          eid = {A264},
        pages = {A264},
          doi = {10.1051/0004-6361/202449583},
archivePrefix = {arXiv},
       eprint = {2404.02958},
 primaryClass = {astro-ph.EP},
       adsurl = {https://ui.adsabs.harvard.edu/abs/2024A&A...686A.264R}
}

@ARTICLE{FARGO3D,
       author = {{Ben{\'\i}tez-Llambay}, Pablo and {Masset}, Fr{\'e}d{\'e}ric S.},
        title = "{FARGO3D: A New GPU-oriented MHD Code}",
      journal = {APJS},
         year = 2016,
        month = mar,
       volume = {223},
       number = {1},
          eid = {11},
        pages = {11},
          doi = {10.3847/0067-0049/223/1/11},
archivePrefix = {arXiv},
       eprint = {1602.02359},
 primaryClass = {astro-ph.IM},
       adsurl = {https://ui.adsabs.harvard.edu/abs/2016ApJS..223...11B}
}

@ARTICLE{Lodato2007,
       author = {{Lodato}, Giuseppe and {Pringle}, J.~E.},
        title = "{Warp diffusion in accretion discs: a numerical investigation}",
      journal = {\mnras},
         year = 2007,
        month = nov,
       volume = {381},
       number = {3},
        pages = {1287-1300},
          doi = {10.1111/j.1365-2966.2007.12332.x},
archivePrefix = {arXiv},
       eprint = {0708.1124},
 primaryClass = {astro-ph},
       adsurl = {https://ui.adsabs.harvard.edu/abs/2007MNRAS.381.1287L}
}

@ARTICLE{Curone2025,
       author = {{Curone}, Pietro and {Facchini}, Stefano and {Andrews}, Sean M. and {Testi}, Leonardo and {Benisty}, Myriam and {Czekala}, Ian and {Huang}, Jane and {Ilee}, John D. and {Isella}, Andrea and {Lodato}, Giuseppe and {Loomis}, Ryan A. and {Stadler}, Jochen and {Winter}, Andrew J. and {Bae}, Jaehan and {Barraza-Alfaro}, Marcelo and {Cataldi}, Gianni and {Cuello}, Nicol{\'a}s and {Fasano}, Daniele and {Flock}, Mario and {Fukagawa}, Misato and {Galloway-Sprietsma}, Maria and {Garg}, Himanshi and {Hall}, Cassandra and {Izquierdo}, Andr{\'e}s F. and {Kanagawa}, Kazuhiro and {Lesur}, Geoffroy and {Longarini}, Cristiano and {Menard}, Francois and {Orihara}, Ryuta and {Pinte}, Christophe and {Price}, Daniel J. and {Rosotti}, Giovanni and {Teague}, Richard and {Wafflard-Fernandez}, Gaylor and {Wilner}, David J. and {W{\"o}lfer}, Lisa and {Yen}, Hsi-Wei and {Yoshida}, Tomohiro C. and {Zawadzki}, Brianna},
        title = "{exoALMA. IV. Substructures, Asymmetries, and the Faint Outer Disk in Continuum Emission}",
      journal = {\apjl},
         year = 2025,
        month = may,
       volume = {984},
       number = {1},
          eid = {L9},
        pages = {L9},
          doi = {10.3847/2041-8213/adc438},
archivePrefix = {arXiv},
       eprint = {2504.18725},
 primaryClass = {astro-ph.EP},
       adsurl = {https://ui.adsabs.harvard.edu/abs/2025ApJ...984L...9C}
}

@ARTICLE{Wolfer2025,
       author = {{W{\"o}lfer}, Lisa and {Barraza-Alfaro}, Marcelo and {Teague}, Richard and {Curone}, Pietro and {Benisty}, Myriam and {Fukagawa}, Misato and {Bae}, Jaehan and {Cataldi}, Gianni and {Czekala}, Ian and {Facchini}, Stefano and {Fasano}, Daniele and {Flock}, Mario and {Galloway-Sprietsma}, Maria and {Garg}, Himanshi and {Hall}, Cassandra and {Huang}, Jane and {Ilee}, John D. and {Izquierdo}, Andr{\'e}s F. and {Kanagawa}, Kazuhiro and {Lesur}, Geoffroy and {Longarini}, Cristiano and {Loomis}, Ryan A. and {Menard}, Francois and {Nath}, Anika and {Orihara}, Ryuta and {Pinte}, Christophe and {Price}, Daniel J. and {Rosotti}, Giovanni and {Stadler}, Jochen and {Wafflard-Fernandez}, Gaylor and {Winter}, Andrew J. and {Yen}, Hsi-Wei and {Yoshida}, Tomohiro C. and {Zawadzki}, Brianna},
        title = "{exoALMA. XVII. Characterizing the Gas Dynamics around Dust Asymmetries}",
      journal = {\apjl},
         year = 2025,
        month = may,
       volume = {984},
       number = {1},
          eid = {L22},
        pages = {L22},
          doi = {10.3847/2041-8213/adc42c},
archivePrefix = {arXiv},
       eprint = {2504.20023},
 primaryClass = {astro-ph.EP},
       adsurl = {https://ui.adsabs.harvard.edu/abs/2025ApJ...984L..22W}
}

@ARTICLE{Barraza-Alfaro2025,
       author = {{Barraza-Alfaro}, Marcelo and {Flock}, Mario and {B{\'e}thune}, William and {Teague}, Richard and {Bae}, Jaehan and {Benisty}, Myriam and {Cataldi}, Gianni and {Curone}, Pietro and {Czekala}, Ian and {Facchini}, Stefano and {Fasano}, Daniele and {Fukagawa}, Misato and {Galloway-Sprietsma}, Maria and {Garg}, Himanshi and {Hall}, Cassandra and {Huang}, Jane and {Ilee}, John D. and {Izquierdo}, Andr{\'e}s F. and {Kanagawa}, Kazuhiro and {Koch}, Eric W. and {Lesur}, Geoffroy and {Longarini}, Cristiano and {Loomis}, Ryan A. and {Orihara}, Ryuta and {Pinte}, Christophe and {Price}, Daniel J. and {Rosotti}, Giovanni and {Stadler}, Jochen and {Wafflard-Fernandez}, Gaylor and {Winter}, Andrew J. and {W{\"o}lfer}, Lisa and {Yen}, Hsi-Wei and {Yoshida}, Tomohiro C. and {Zawadzki}, Brianna},
        title = "{exoALMA. XVI. Predicting Signatures of Large-scale Turbulence in Protoplanetary Disks}",
      journal = {\apjl},
         year = 2025,
        month = may,
       volume = {984},
       number = {1},
          eid = {L21},
        pages = {L21},
          doi = {10.3847/2041-8213/adc42d},
archivePrefix = {arXiv},
       eprint = {2504.19853},
 primaryClass = {astro-ph.EP},
       adsurl = {https://ui.adsabs.harvard.edu/abs/2025ApJ...984L..21B}
}

@ARTICLE{Stadler2025,
       author = {{Stadler}, Jochen and {Benisty}, Myriam and {Winter}, Andrew J. and {Izquierdo}, Andr{\'e}s F. and {Longarini}, Cristiano and {Galloway-Sprietsma}, Maria and {Curone}, Pietro and {Andrews}, Sean M. and {Bae}, Jaehan and {Facchini}, Stefano and {Rosotti}, Giovanni and {Teague}, Richard and {Barraza-Alfaro}, Marcelo and {Cataldi}, Gianni and {Cuello}, Nicol{\'a}s and {Czekala}, Ian and {Fasano}, Daniele and {Flock}, Mario and {Fukagawa}, Misato and {Garg}, Himanshi and {Hall}, Cassandra and {Hammond}, Iain and {Hilder}, Thomas and {Huang}, Jane and {Ilee}, John D. and {Kanagawa}, Kazuhiro and {Lesur}, Geoffroy and {Lodato}, Giuseppe and {Loomis}, Ryan A. and {Menard}, Francois and {Orihara}, Ryuta and {Pinte}, Christophe and {Price}, Daniel J. and {Yen}, Hsi-Wei and {Wafflard-Fernandez}, Gaylor and {Wilner}, David J. and {W{\"o}lfer}, Lisa and {Yoshida}, Tomohiro C. and {Zawadzki}, Brianna},
        title = "{exoALMA. VI. Rotating under Pressure: Rotation Curves, Azimuthal Velocity Substructures, and Gas Pressure Variations}",
      journal = {\apjl},
         year = 2025,
        month = may,
       volume = {984},
       number = {1},
          eid = {L11},
        pages = {L11},
          doi = {10.3847/2041-8213/adb152},
archivePrefix = {arXiv},
       eprint = {2504.20036},
 primaryClass = {astro-ph.EP},
       adsurl = {https://ui.adsabs.harvard.edu/abs/2025ApJ...984L..11S}
}

@ARTICLE{Izquierdo2026,
       author = {{Izquierdo}, Andres F. and {Bae}, Jaehan and {Facchini}, Stefano and {van Dishoeck}, Ewine F. and {Barraza-Alfaro}, Marcelo and {Benisty}, Myriam and {Teague}, Richard and {Stadler}, Jochen and {Andrews}, Sean M. and {Cataldi}, Gianni and {Cuello}, Nicolas and {Curone}, Pietro and {Czekala}, Ian and {Fasano}, Daniele and {Flock}, Mario and {Fukagawa}, Misato and {Galloway-Sprietsma}, Maria and {Hall}, Cassandra and {Huang}, Jane and {Ilee}, John D. and {Isella}, Andrea and {Lawrence}, Jensen and {Lesur}, Geoffroy and {Lodato}, Giuseppe and {Longarini}, Cristiano and {Loomis}, Ryan A. and {Menard}, Fran{\c{c}}ois and {Pinte}, Christophe and {Price}, Daniel J. and {Rosotti}, Giovanni and {Testi}, Leonardo and {Wilner}, David J. and {Winter}, Andrew J. and {W{\"o}lfer}, Lisa and {Zawadzki}, Brianna},
        title = "{exoALMA XX: Tomographic Detection of Embedded Planets in Protoplanetary Disks}",
      journal = {arXiv e-prints},
         year = 2026,
        month = mar,
          eid = {arXiv:2603.13157},
        pages = {arXiv:2603.13157},
          doi = {10.48550/arXiv.2603.13157},
archivePrefix = {arXiv},
       eprint = {2603.13157},
 primaryClass = {astro-ph.EP},
       adsurl = {https://ui.adsabs.harvard.edu/abs/2026arXiv260313157I}
}

@ARTICLE{Benisty2026,
       author = {{Benisty}, Myriam and {Izquierdo}, Andres F. and {Stadler}, Jochen and {Galloway-Sprietsma}, Maria and {Facchini}, Stefano and {Winter}, Andrew J. and {Bae}, Jaehan and {Fukagawa}, Misato and {Teague}, Richard and {Pinte}, Christophe and {Andrews}, Sean M. and {Barraza-Alfaro}, Marcelo and {Cataldi}, Gianni and {Curone}, Pietro and {Czekala}, Ian and {Fasano}, Daniele and {Flock}, Mario and {Garg}, Himanshi and {Huang}, Jane and {Ilee}, John D. and {Kanagawa}, Kazuhiro and {Lawrence}, Jensen and {Lesur}, Geoffroy and {Lodato}, Giuseppe and {Longarini}, Cristiano and {Loomis}, Ryan A. and {Menard}, Francois and {Orihara}, Ryuta and {Price}, Daniel J. and {Rosotti}, Giovanni and {Wafflard-Fernandez}, Gaylor and {Wilner}, David J. and {Wolfer}, Lisa and {Yen}, Hsi-Wei and {Yoshida}, Tomohiro C. and {Zawadzki}, Brianna},
        title = "{exoALMA XXI: The Morphology and Dynamics of Vertical Flows}",
      journal = {arXiv e-prints},
         year = 2026,
        month = mar,
          eid = {arXiv:2603.12861},
        pages = {arXiv:2603.12861},
          doi = {10.48550/arXiv.2603.12861},
archivePrefix = {arXiv},
       eprint = {2603.12861},
 primaryClass = {astro-ph.EP},
       adsurl = {https://ui.adsabs.harvard.edu/abs/2026arXiv260312861B}
}

@ARTICLE{Fukagawa2026,
       author = {{Fukagawa}, Misato and {Izquierdo}, Andr{\'e}s F. and {Stadler}, Jochen and {W{\"o}lfer}, Lisa and {Galloway-Sprietsma}, Maria and {Orihara}, Ryuta and {Aizawa}, Masataka and {Momose}, Munetake and {Fasano}, Daniele and {Benisty}, Myriam and {Teague}, Richard and {Facchini}, Stefano and {Pinte}, Christophe and {Andrews}, Sean M. and {Bae}, Jaehan and {Barraza-Alfaro}, Marcelo and {Cataldi}, Gianni and {Curone}, Pietro and {Czekala}, Ian and {Flock}, Mario and {Garg}, Himanshi and {Hall}, Cassandra and {Huang}, Jane and {Ilee}, John D. and {Lawrence}, Jensen and {Lesur}, Geoffroy and {Lodato}, Giuseppe and {Longarini}, Cristiano and {Loomis}, Ryan A. and {M{\'e}nard}, Francois and {Price}, Daniel J. and {Rosotti}, Giovanni and {Yen}, Hsi-Wei and {Yoshida}, Tomohiro C. and {Wafflard-Fernandez}, Gaylor and {Wilner}, David J. and {Winter}, Andrew J. and {Zawadzki}, Brianna},
        title = "{exoALMA XXII: A Two-dimensional Atlas of Deviations from Keplerian Disks}",
      journal = {arXiv e-prints},
         year = 2026,
        month = mar,
          eid = {arXiv:2603.13015},
        pages = {arXiv:2603.13015},
          doi = {10.48550/arXiv.2603.13015},
archivePrefix = {arXiv},
       eprint = {2603.13015},
 primaryClass = {astro-ph.EP},
       adsurl = {https://ui.adsabs.harvard.edu/abs/2026arXiv260313015F}
}

@ARTICLE{Kimmig2026,
       author = {{Kimmig}, C.~N. and {Weber}, P. and {Rosotti}, G.~P. and {Facchini}, S. and {Dullemond}, C.~P.},
        title = "{Warps survive beyond fly-by encounters in protoplanetary disks: RW Aur A as a case study}",
      journal = {\aap},
         year = 2026,
        month = apr,
       volume = {708},
          eid = {A194},
        pages = {A194},
          doi = {10.1051/0004-6361/202557365},
archivePrefix = {arXiv},
       eprint = {2601.22223},
 primaryClass = {astro-ph.EP},
       adsurl = {https://ui.adsabs.harvard.edu/abs/2026A&A...708A.194K}
}

@ARTICLE{Teague2018,
       author = {{Teague}, Richard and {Foreman-Mackey}, Daniel},
        title = "{A Robust Method to Measure Centroids of Spectral Lines}",
      journal = {Research Notes of the American Astronomical Society},
         year = 2018,
        month = sep,
       volume = {2},
       number = {3},
          eid = {173},
        pages = {173},
          doi = {10.3847/2515-5172/aae265},
archivePrefix = {arXiv},
       eprint = {1809.10295},
 primaryClass = {astro-ph.IM},
       adsurl = {https://ui.adsabs.harvard.edu/abs/2018RNAAS...2..173T}
}

@ARTICLE{Williams2014,
       author = {{Williams}, Jonathan P. and {Best}, William M.~J.},
        title = "{A Parametric Modeling Approach to Measuring the Gas Masses of Circumstellar Disks}",
      journal = {\apj},
         year = 2014,
        month = jun,
       volume = {788},
       number = {1},
          eid = {59},
        pages = {59},
          doi = {10.1088/0004-637X/788/1/59},
archivePrefix = {arXiv},
       eprint = {1312.0151},
 primaryClass = {astro-ph.EP},
       adsurl = {https://ui.adsabs.harvard.edu/abs/2014ApJ...788...59W}
}

@ARTICLE{Rosotti2025,
       author = {{Rosotti}, Giovanni P. and {Longarini}, Cristiano and {Paneque-Carre{\~n}o}, Teresa and {Cataldi}, Gianni and {Galloway-Sprietsma}, Maria and {Andrews}, Sean M. and {Bae}, Jaehan and {Barraza-Alfaro}, Marcelo and {Benisty}, Myriam and {Curone}, Pietro and {Czekala}, Ian and {Facchini}, Stefano and {Fasano}, Daniele and {Flock}, Mario and {Fukagawa}, Misato and {Garg}, Himanshi and {Hall}, Cassandra and {Huang}, Jane and {Ilee}, John D. and {Izquierdo}, Andr{\'e}s F. and {Kanagawa}, Kazuhiro and {Lesur}, Geoffroy and {Lodato}, Giuseppe and {Loomis}, Ryan A. and {Orihara}, Ryuta and {Pinte}, Christophe and {Price}, Daniel J. and {Stadler}, Jochen and {Teague}, Richard and {Fernandez}, Gaylor Wafflard- and {Winter}, Andrew J. and {W{\"o}lfer}, Lisa and {Yen}, Hsi-Wei and {Yoshida}, Tomohiro C. and {Zawadzki}, Brianna},
        title = "{exoALMA. XV. Interpreting the Height of CO Emission Layer}",
      journal = {\apjl},
         year = 2025,
        month = may,
       volume = {984},
       number = {1},
          eid = {L20},
        pages = {L20},
          doi = {10.3847/2041-8213/adc42e},
archivePrefix = {arXiv},
       eprint = {2504.20012},
 primaryClass = {astro-ph.EP},
       adsurl = {https://ui.adsabs.harvard.edu/abs/2025ApJ...984L..20R}
}

@ARTICLE{Trapman2023,
       author = {{Trapman}, Leon and {Rosotti}, Giovanni and {Zhang}, Ke and {Tabone}, Beno{\^\i}t},
        title = "{How Large Is a Disk-What Do Protoplanetary Disk Gas Sizes Really Mean?}",
      journal = {\apj},
         year = 2023,
        month = sep,
       volume = {954},
       number = {1},
          eid = {41},
        pages = {41},
          doi = {10.3847/1538-4357/ace7d1},
archivePrefix = {arXiv},
       eprint = {2307.07600},
 primaryClass = {astro-ph.EP},
       adsurl = {https://ui.adsabs.harvard.edu/abs/2023ApJ...954...41T}
}

@software{optool,
       author = {{Dominik}, Carsten and {Min}, Michiel and {Tazaki}, Ryo},
        title = "{OpTool: Command-line driven tool for creating complex dust opacities}",
 howpublished = {Astrophysics Source Code Library, record ascl:2104.010},
         year = 2021,
        month = apr,
          eid = {ascl:2104.010},
       adsurl = {https://ui.adsabs.harvard.edu/abs/2021ascl.soft04010D}
}

@ARTICLE{Woitke2016,
       author = {{Woitke}, P. and {Min}, M. and {Pinte}, C. and {Thi}, W.-F. and {Kamp}, I. and {Rab}, C. and {Anthonioz}, F. and {Antonellini}, S. and {Baldovin-Saavedra}, C. and {Carmona}, A. and {Dominik}, C. and {Dionatos}, O. and {Greaves}, J. and {G{\"u}del}, M. and {Ilee}, J.~D. and {Liebhart}, A. and {M{\'e}nard}, F. and {Rigon}, L. and {Waters}, L.~B.~F.~M. and {Aresu}, G. and {Meijerink}, R. and {Spaans}, M.},
        title = "{Consistent dust and gas models for protoplanetary disks. I. Disk shape, dust settling, opacities, and PAHs}",
      journal = {\aap},
         year = 2016,
        month = feb,
       volume = {586},
          eid = {A103},
        pages = {A103},
          doi = {10.1051/0004-6361/201526538},
archivePrefix = {arXiv},
       eprint = {1511.03431},
 primaryClass = {astro-ph.EP},
       adsurl = {https://ui.adsabs.harvard.edu/abs/2016A&A...586A.103W}
}

@software{radmc3d,
       author = {{Dullemond}, C.~P. and {Juhasz}, A. and {Pohl}, A. and {Sereshti}, F. and {Shetty}, R. and {Peters}, T. and {Commercon}, B. and {Flock}, M.},
        title = "{RADMC-3D: A multi-purpose radiative transfer tool}",
 howpublished = {Astrophysics Source Code Library, record ascl:1202.015},
         year = 2012,
        month = feb,
          eid = {ascl:1202.015},
       adsurl = {https://ui.adsabs.harvard.edu/abs/2012ascl.soft02015D}
}

@ARTICLE{Kimmig2024,
       author = {{Kimmig}, C.~N. and {Dullemond}, C.~P.},
        title = "{Warped disk evolution in grid-based simulations}",
      journal = {\aap},
         year = 2024,
        month = sep,
       volume = {689},
          eid = {A45},
        pages = {A45},
          doi = {10.1051/0004-6361/202348660},
archivePrefix = {arXiv},
       eprint = {2406.02754},
 primaryClass = {astro-ph.SR},
       adsurl = {https://ui.adsabs.harvard.edu/abs/2024A&A...689A..45K}
}

@ARTICLE{Chiang1997,
       author = {{Chiang}, E.~I. and {Goldreich}, P.},
        title = "{Spectral Energy Distributions of T Tauri Stars with Passive Circumstellar Disks}",
      journal = {\apj},
         year = 1997,
        month = nov,
       volume = {490},
       number = {1},
        pages = {368-376},
          doi = {10.1086/304869},
archivePrefix = {arXiv},
       eprint = {astro-ph/9706042},
 primaryClass = {astro-ph},
       adsurl = {https://ui.adsabs.harvard.edu/abs/1997ApJ...490..368C}
}

@ARTICLE{Rosenfeld2014,
       author = {{Rosenfeld}, Katherine A. and {Chiang}, Eugene and {Andrews}, Sean M.},
        title = "{Fast Radial Flows in Transition Disk Holes}",
      journal = {\apj},
         year = 2014,
        month = feb,
       volume = {782},
       number = {2},
          eid = {62},
        pages = {62},
          doi = {10.1088/0004-637X/782/2/62},
archivePrefix = {arXiv},
       eprint = {1312.3817},
 primaryClass = {astro-ph.SR},
       adsurl = {https://ui.adsabs.harvard.edu/abs/2014ApJ...782...62R}
}

@ARTICLE{Aizawa2026,
       author = {{Aizawa}, Masataka and {Orihara}, Ryuta},
        title = "{On the Interpretation of Velocity Residuals in Protoplanetary Disks}",
      journal = {\aj},
         year = 2026,
        month = jan,
       volume = {171},
       number = {1},
          eid = {6},
        pages = {6},
          doi = {10.3847/1538-3881/ae1a75},
archivePrefix = {arXiv},
       eprint = {2507.23287},
 primaryClass = {astro-ph.EP},
       adsurl = {https://ui.adsabs.harvard.edu/abs/2026AJ....171....6A}
}

@ARTICLE{Zuleta2024,
       author = {{Zuleta}, A. and {Birnstiel}, T. and {Teague}, R.},
        title = "{Kinematical signatures: Distinguishing between warps and radial flows}",
      journal = {\aap},
         year = 2024,
        month = dec,
       volume = {692},
          eid = {A56},
        pages = {A56},
          doi = {10.1051/0004-6361/202451145},
archivePrefix = {arXiv},
       eprint = {2410.14457},
 primaryClass = {astro-ph.EP},
       adsurl = {https://ui.adsabs.harvard.edu/abs/2024A&A...692A..56Z}
}

@ARTICLE{Garg2021,
       author = {{Garg}, H. and {Pinte}, C. and {Christiaens}, V. and {Price}, D.~J. and {Lazendic}, J.~S. and {Boehler}, Y. and {Casassus}, S. and {Marino}, S. and {Perez}, S. and {Zuleta}, A.},
        title = "{Non-Keplerian spirals, a gas-pressure dust trap, and an eccentric gas cavity in the circumbinary disc around HD 142527}",
      journal = {\mnras},
         year = 2021,
        month = jun,
       volume = {504},
       number = {1},
        pages = {782-791},
          doi = {10.1093/mnras/stab800},
archivePrefix = {arXiv},
       eprint = {2010.15310},
 primaryClass = {astro-ph.EP},
       adsurl = {https://ui.adsabs.harvard.edu/abs/2021MNRAS.504..782G}
}

@ARTICLE{Perez2018,
       author = {{P{\'e}rez}, Laura M. and {Benisty}, Myriam and {Andrews}, Sean M. and {Isella}, Andrea and {Dullemond}, Cornelis P. and {Huang}, Jane and {Kurtovic}, Nicol{\'a}s T. and {Guzm{\'a}n}, Viviana V. and {Zhu}, Zhaohuan and {Birnstiel}, Tilman and {Zhang}, Shangjia and {Carpenter}, John M. and {Wilner}, David J. and {Ricci}, Luca and {Bai}, Xue-Ning and {Weaver}, Erik and {{\"O}berg}, Karin I.},
        title = "{The Disk Substructures at High Angular Resolution Project (DSHARP). X. Multiple Rings, a Misaligned Inner Disk, and a Bright Arc in the Disk around the T Tauri star HD 143006}",
      journal = {\apjl},
         year = 2018,
        month = dec,
       volume = {869},
       number = {2},
          eid = {L50},
        pages = {L50},
          doi = {10.3847/2041-8213/aaf745},
archivePrefix = {arXiv},
       eprint = {1812.04049},
 primaryClass = {astro-ph.SR},
       adsurl = {https://ui.adsabs.harvard.edu/abs/2018ApJ...869L..50P}
}

@ARTICLE{Mayama2018,
       author = {{Mayama}, Satoshi and {Akiyama}, Eiji and {Pani{\'c}}, Olja and {Miley}, James and {Tsukagoshi}, Takashi and {Muto}, Takayuki and {Dong}, Ruobing and {de Leon}, Jerome and {Mizuki}, Toshiyuki and {Oh}, Daehyeon and {Hashimoto}, Jun and {Sai}, Jinshi and {Currie}, Thayne and {Takami}, Michihiro and {Grady}, Carol A. and {Hayashi}, Masahiko and {Tamura}, Motohide and {Inutsuka}, Shu-ichiro},
        title = "{ALMA Reveals a Misaligned Inner Gas Disk inside the Large Cavity of a Transitional Disk}",
      journal = {\apjl},
         year = 2018,
        month = nov,
       volume = {868},
       number = {1},
          eid = {L3},
        pages = {L3},
          doi = {10.3847/2041-8213/aae88b},
archivePrefix = {arXiv},
       eprint = {1810.06941},
 primaryClass = {astro-ph.EP},
       adsurl = {https://ui.adsabs.harvard.edu/abs/2018ApJ...868L...3M}
}

@ARTICLE{Walsh2017,
       author = {{Walsh}, Catherine and {Daley}, Cail and {Facchini}, Stefano and {Juh{\'a}sz}, Attila},
        title = "{CO emission tracing a warp or radial flow within {\ensuremath{\lesssim}}100 au in the HD 100546 protoplanetary disk}",
      journal = {\aap},
         year = 2017,
        month = nov,
       volume = {607},
          eid = {A114},
        pages = {A114},
          doi = {10.1051/0004-6361/201731334},
archivePrefix = {arXiv},
       eprint = {1710.00703},
 primaryClass = {astro-ph.EP},
       adsurl = {https://ui.adsabs.harvard.edu/abs/2017A&A...607A.114W}
}

@ARTICLE{Teague2025,
       author = {{Teague}, Richard and {Benisty}, Myriam and {Facchini}, Stefano and {Fukagawa}, Misato and {Pinte}, Christophe and {Andrews}, Sean M. and {Bae}, Jaehan and {Barraza-Alfaro}, Marcelo and {Cataldi}, Gianni and {Cuello}, Nicol{\'a}s and {Curone}, Pietro and {Czekala}, Ian and {Fasano}, Daniele and {Flock}, Mario and {Galloway-Sprietsma}, Maria and {Garg}, Himanshi and {Hall}, Cassandra and {Hammond}, Iain and {Hilder}, Thomas and {Huang}, Jane and {Ilee}, John D. and {Izquierdo}, Andr{\'e}s F. and {Kanagawa}, Kazuhiro and {Lesur}, Geoffroy and {Lodato}, Giuseppe and {Longarini}, Cristiano and {Loomis}, Ryan A. and {Masset}, Fr{\'e}d{\'e}ric and {Menard}, Francois and {Orihara}, Ryuta and {Price}, Daniel J. and {Rosotti}, Giovanni and {Stadler}, Jochen and {Testi}, Leonardo and {Yen}, Hsi-Wei and {Wafflard-Fernandez}, Gaylor and {Wilner}, David J. and {Winter}, Andrew J. and {W{\"o}lfer}, Lisa and {Yoshida}, Tomohiro C. and {Zawadzki}, Brianna},
        title = "{exoALMA. I. Science Goals, Project Design, and Data Products}",
      journal = {\apjl},
         year = 2025,
        month = may,
       volume = {984},
       number = {1},
          eid = {L6},
        pages = {L6},
          doi = {10.3847/2041-8213/adc43b},
archivePrefix = {arXiv},
       eprint = {2504.18688},
 primaryClass = {astro-ph.EP},
       adsurl = {https://ui.adsabs.harvard.edu/abs/2025ApJ...984L...6T}
}

@ARTICLE{Winter2025,
       author = {{Winter}, Andrew J. and {Benisty}, Myriam and {Izquierdo}, Andr{\'e}s F. and {Lodato}, Giuseppe and {Teague}, Richard and {Kimmig}, Carolin N. and {Andrews}, Sean M. and {Bae}, Jaehan and {Barraza-Alfaro}, Marcelo and {Cuello}, Nicol{\'a}s and {Curone}, Pietro and {Czekala}, Ian and {Facchini}, Stefano and {Fasano}, Daniele and {Hall}, Cassandra and {Hardiman}, Caitlyn and {Hilder}, Thomas and {Ilee}, John D. and {Fukagawa}, Misato and {Longarini}, Cristiano and {M{\'e}nard}, Fran{\c{c}}ois and {Orihara}, Ryuta and {Pinte}, Christophe and {Price}, Daniel J. and {Rosotti}, Giovanni and {Stadler}, Jochen and {Wilner}, David J. and {W{\"o}lfer}, Lisa and {Yen}, Hsi-Wei and {Yoshida}, Tomohiro C. and {Zawadzki}, Brianna},
        title = "{exoALMA. XVIII. Interpreting Large-scale Kinematic Structures as Moderate Warping}",
      journal = {\apjl},
         year = 2025,
        month = sep,
       volume = {990},
       number = {1},
          eid = {L10},
        pages = {L10},
          doi = {10.3847/2041-8213/adf113},
archivePrefix = {arXiv},
       eprint = {2507.11669},
 primaryClass = {astro-ph.EP},
       adsurl = {https://ui.adsabs.harvard.edu/abs/2025ApJ...990L..10W}
}

@ARTICLE{Held2025,
       author = {{Held}, Loren E. and {Ogilvie}, Gordon I.},
        title = "{Disc breaking and parametric instability in warped accretion discs}",
      journal = {arXiv e-prints},
         year = 2025,
        month = nov,
          eid = {arXiv:2511.18655},
        pages = {arXiv:2511.18655},
          doi = {10.48550/arXiv.2511.18655},
archivePrefix = {arXiv},
       eprint = {2511.18655},
 primaryClass = {astro-ph.SR},
       adsurl = {https://ui.adsabs.harvard.edu/abs/2025arXiv251118655E}
}

@ARTICLE{Fairbairn2023,
       author = {{Fairbairn}, Callum W. and {Ogilvie}, Gordon I.},
        title = "{Parametric instability in warped astrophysical discs: growth, saturation, and feedback}",
      journal = {\mnras},
         year = 2023,
        month = mar,
       volume = {520},
       number = {1},
        pages = {1022-1043},
          doi = {10.1093/mnras/stad211},
archivePrefix = {arXiv},
       eprint = {2301.07176},
 primaryClass = {astro-ph.SR},
       adsurl = {https://ui.adsabs.harvard.edu/abs/2023MNRAS.520.1022F}
}

@ARTICLE{Deng2021,
       author = {{Deng}, Hongping and {Ogilvie}, Gordon I. and {Mayer}, Lucio},
        title = "{Parametric instability in a free-evolving warped protoplanetary disc}",
      journal = {\mnras},
         year = 2021,
        month = jan,
       volume = {500},
       number = {3},
        pages = {4248-4256},
          doi = {10.1093/mnras/staa3504},
archivePrefix = {arXiv},
       eprint = {2010.00862},
 primaryClass = {astro-ph.EP},
       adsurl = {https://ui.adsabs.harvard.edu/abs/2021MNRAS.500.4248D}
}

@ARTICLE{Gammie2000,
       author = {{Gammie}, Charles F. and {Goodman}, Jeremy and {Ogilvie}, Gordon I.},
        title = "{Linear and non-linear theory of a parametric instability of hydrodynamic warps in Keplerian discs}",
      journal = {\mnras},
         year = 2000,
        month = nov,
       volume = {318},
       number = {4},
        pages = {1005-1016},
          doi = {10.1046/j.1365-8711.2000.03669.x},
archivePrefix = {arXiv},
       eprint = {astro-ph/0001539},
 primaryClass = {astro-ph},
       adsurl = {https://ui.adsabs.harvard.edu/abs/2000MNRAS.318.1005G}
}

@ARTICLE{Ogilvie2013-2,
       author = {{Ogilvie}, Gordon I. and {Latter}, Henrik N.},
        title = "{Hydrodynamic instability in warped astrophysical discs}",
      journal = {\mnras},
         year = 2013,
        month = aug,
       volume = {433},
       number = {3},
        pages = {2420-2435},
          doi = {10.1093/mnras/stt917},
archivePrefix = {arXiv},
       eprint = {1303.0264},
 primaryClass = {astro-ph.SR},
       adsurl = {https://ui.adsabs.harvard.edu/abs/2013MNRAS.433.2420O}
}

@ARTICLE{Ogilvie2013,
       author = {{Ogilvie}, Gordon I. and {Latter}, Henrik N.},
        title = "{Local and global dynamics of warped astrophysical discs}",
      journal = {\mnras},
         year = 2013,
        month = aug,
       volume = {433},
       number = {3},
        pages = {2403-2419},
          doi = {10.1093/mnras/stt916},
archivePrefix = {arXiv},
       eprint = {1303.0263},
 primaryClass = {astro-ph.SR},
       adsurl = {https://ui.adsabs.harvard.edu/abs/2013MNRAS.433.2403O}
}

@ARTICLE{Dullemond2022,
       author = {{Dullemond}, C.~P. and {Kimmig}, C.~N. and {Zanazzi}, J.~J.},
        title = "{On the equations of warped disc dynamics}",
      journal = {\mnras},
         year = 2022,
        month = apr,
       volume = {511},
       number = {2},
        pages = {2925-2947},
          doi = {10.1093/mnras/stab2791},
archivePrefix = {arXiv},
       eprint = {2109.12035},
 primaryClass = {astro-ph.SR},
       adsurl = {https://ui.adsabs.harvard.edu/abs/2022MNRAS.511.2925D}
}

@ARTICLE{Martin2019,
       author = {{Martin}, Rebecca G. and {Lubow}, Stephen H. and {Pringle}, J.~E. and {Franchini}, Alessia and {Zhu}, Zhaohuan and {Lepp}, Stephen and {Nealon}, Rebecca and {Nixon}, C.~J. and {Vallet}, David},
        title = "{Generalized Warped Disk Equations}",
      journal = {\apj},
         year = 2019,
        month = apr,
       volume = {875},
       number = {1},
          eid = {5},
        pages = {5},
          doi = {10.3847/1538-4357/ab0bb7},
archivePrefix = {arXiv},
       eprint = {1902.11073},
 primaryClass = {astro-ph.HE},
       adsurl = {https://ui.adsabs.harvard.edu/abs/2019ApJ...875....5M}
}

@ARTICLE{Lubow2000,
       author = {{Lubow}, S.~H. and {Ogilvie}, G.~I.},
        title = "{On the Tilting of Protostellar Disks by Resonant Tidal Effects}",
      journal = {\apj},
         year = 2000,
        month = jul,
       volume = {538},
       number = {1},
        pages = {326-340},
          doi = {10.1086/309101},
archivePrefix = {arXiv},
       eprint = {astro-ph/0003028},
 primaryClass = {astro-ph},
       adsurl = {https://ui.adsabs.harvard.edu/abs/2000ApJ...538..326L}
}

@ARTICLE{Ogilvie1999,
       author = {{Ogilvie}, G.~I.},
        title = "{The non-linear fluid dynamics of a warped accretion disc}",
      journal = {\mnras},
         year = 1999,
        month = apr,
       volume = {304},
       number = {3},
        pages = {557-578},
          doi = {10.1046/j.1365-8711.1999.02340.x},
archivePrefix = {arXiv},
       eprint = {astro-ph/9812073},
 primaryClass = {astro-ph},
       adsurl = {https://ui.adsabs.harvard.edu/abs/1999MNRAS.304..557O}
}

@ARTICLE{Pringle1992,
       author = {{Pringle}, J.~E.},
        title = "{A simple approach to the evolution of twisted accretion discs}",
      journal = {\mnras},
         year = 1992,
        month = oct,
       volume = {258},
       number = {4},
        pages = {811-818},
          doi = {10.1093/mnras/258.4.811},
       adsurl = {https://ui.adsabs.harvard.edu/abs/1992MNRAS.258..811P}
}

@ARTICLE{Papaloizou1983,
       author = {{Papaloizou}, J.~C.~B. and {Pringle}, J.~E.},
        title = "{The time-dependence of non-planar accretion discs}",
      journal = {\mnras},
         year = 1983,
        month = mar,
       volume = {202},
        pages = {1181-1194},
          doi = {10.1093/mnras/202.4.1181},
       adsurl = {https://ui.adsabs.harvard.edu/abs/1983MNRAS.202.1181P}
}

@ARTICLE{Shakura1973,
       author = {{Shakura}, N.~I. and {Sunyaev}, R.~A.},
        title = "{Black holes in binary systems. Observational appearance.}",
      journal = {\aap},
         year = 1973,
        month = jan,
       volume = {24},
        pages = {337-355},
       adsurl = {https://ui.adsabs.harvard.edu/abs/1973A&A....24..337S}
}

@ARTICLE{Romanova2021,
       author = {{Romanova}, M.~M. and {Koldoba}, A.~V. and {Ustyugova}, G.~V. and {Blinova}, A.~A. and {Lai}, D. and {Lovelace}, R.~V.~E.},
        title = "{3D MHD simulations of accretion on to stars with tilted magnetic and rotational axes}",
      journal = {\mnras},
         year = 2021,
        month = sep,
       volume = {506},
       number = {1},
        pages = {372-384},
          doi = {10.1093/mnras/stab1724},
archivePrefix = {arXiv},
       eprint = {2012.10826},
 primaryClass = {astro-ph.SR},
       adsurl = {https://ui.adsabs.harvard.edu/abs/2021MNRAS.506..372R}
}

@ARTICLE{Foucart2011,
       author = {{Foucart}, Francois and {Lai}, Dong},
        title = "{Evolution of spin direction of accreting magnetic protostars and spin-orbit misalignment in exoplanetary systems - II. Warped discs}",
      journal = {\mnras},
         year = 2011,
        month = apr,
       volume = {412},
       number = {4},
        pages = {2799-2815},
          doi = {10.1111/j.1365-2966.2010.18176.x},
archivePrefix = {arXiv},
       eprint = {1009.3233},
 primaryClass = {astro-ph.EP},
       adsurl = {https://ui.adsabs.harvard.edu/abs/2011MNRAS.412.2799F}
}

@ARTICLE{Kuffmeier2024,
       author = {{Kuffmeier}, M. and {Pineda}, J.~E. and {Segura-Cox}, D. and {Haugb{\o}lle}, T.},
        title = "{Constraints on the primordial misalignment of star-disk systems}",
      journal = {\aap},
         year = 2024,
        month = oct,
       volume = {690},
          eid = {A297},
        pages = {A297},
          doi = {10.1051/0004-6361/202450410},
archivePrefix = {arXiv},
       eprint = {2405.12670},
 primaryClass = {astro-ph.SR},
       adsurl = {https://ui.adsabs.harvard.edu/abs/2024A&A...690A.297K}
}

@ARTICLE{Kuffmeier2021,
       author = {{Kuffmeier}, M. and {Dullemond}, C.~P. and {Reissl}, S. and {Goicovic}, F.~G.},
        title = "{Misaligned disks induced by infall}",
      journal = {\aap},
         year = 2021,
        month = dec,
       volume = {656},
          eid = {A161},
        pages = {A161},
          doi = {10.1051/0004-6361/202039614},
archivePrefix = {arXiv},
       eprint = {2110.04309},
 primaryClass = {astro-ph.SR},
       adsurl = {https://ui.adsabs.harvard.edu/abs/2021A&A...656A.161K}
}

@ARTICLE{Dullemond2019,
       author = {{Dullemond}, C.~P. and {K{\"u}ffmeier}, M. and {Goicovic}, F. and {Fukagawa}, M. and {Oehl}, V. and {Kramer}, M.},
        title = "{Cloudlet capture by transitional disk and FU Orionis stars}",
      journal = {\aap},
         year = 2019,
        month = aug,
       volume = {628},
          eid = {A20},
        pages = {A20},
          doi = {10.1051/0004-6361/201832632},
archivePrefix = {arXiv},
       eprint = {1911.05158},
 primaryClass = {astro-ph.EP},
       adsurl = {https://ui.adsabs.harvard.edu/abs/2019A&A...628A..20D}
}

@ARTICLE{Cuello2019,
       author = {{Cuello}, Nicol{\'a}s and {Dipierro}, Giovanni and {Mentiplay}, Daniel and {Price}, Daniel J. and {Pinte}, Christophe and {Cuadra}, Jorge and {Laibe}, Guillaume and {M{\'e}nard}, Fran{\c{c}}ois and {Poblete}, Pedro P. and {Montesinos}, Mat{\'\i}as},
        title = "{Flybys in protoplanetary discs: I. Gas and dust dynamics}",
      journal = {\mnras},
         year = 2019,
        month = mar,
       volume = {483},
       number = {3},
        pages = {4114-4139},
          doi = {10.1093/mnras/sty3325},
archivePrefix = {arXiv},
       eprint = {1812.00961},
 primaryClass = {astro-ph.EP},
       adsurl = {https://ui.adsabs.harvard.edu/abs/2019MNRAS.483.4114C}
}

@ARTICLE{Clarke1993,
       author = {{Clarke}, C.~J. and {Pringle}, J.~E.},
        title = "{Accretion disc response to a stellar fly-by}",
      journal = {\mnras},
         year = 1993,
        month = mar,
       volume = {261},
       number = {1},
        pages = {190-202},
          doi = {10.1093/mnras/261.1.190},
       adsurl = {https://ui.adsabs.harvard.edu/abs/1993MNRAS.261..190C}
}

@ARTICLE{Rabago2023,
       author = {{Rabago}, Ian and {Zhu}, Zhaohuan and {Martin}, Rebecca G. and {Lubow}, Stephen H.},
        title = "{Grid-based simulations of polar circumbinary discs: polar alignment and vortex formation}",
      journal = {\mnras},
         year = 2023,
        month = apr,
       volume = {520},
       number = {2},
        pages = {2138-2147},
          doi = {10.1093/mnras/stad242},
archivePrefix = {arXiv},
       eprint = {2301.07796},
 primaryClass = {astro-ph.EP},
       adsurl = {https://ui.adsabs.harvard.edu/abs/2023MNRAS.520.2138R}
}

@ARTICLE{Deng2022,
       author = {{Deng}, Hongping and {Ogilvie}, Gordon I.},
        title = "{Non-linear behaviour of warped discs around a central object with a quadrupole moment}",
      journal = {\mnras},
         year = 2022,
        month = jun,
       volume = {512},
       number = {4},
        pages = {6078-6092},
          doi = {10.1093/mnras/stac858},
archivePrefix = {arXiv},
       eprint = {2203.13582},
 primaryClass = {astro-ph.EP},
       adsurl = {https://ui.adsabs.harvard.edu/abs/2022MNRAS.512.6078D}
}

@ARTICLE{Facchini2018,
       author = {{Facchini}, Stefano and {Juh{\'a}sz}, Attila and {Lodato}, Giuseppe},
        title = "{Signatures of broken protoplanetary discs in scattered light and in sub-millimetre observations}",
      journal = {\mnras},
         year = 2018,
        month = feb,
       volume = {473},
       number = {4},
        pages = {4459-4475},
          doi = {10.1093/mnras/stx2523},
archivePrefix = {arXiv},
       eprint = {1709.08369},
 primaryClass = {astro-ph.SR},
       adsurl = {https://ui.adsabs.harvard.edu/abs/2018MNRAS.473.4459F}
}

@ARTICLE{Facchini2013,
       author = {{Facchini}, Stefano and {Lodato}, Giuseppe and {Price}, Daniel J.},
        title = "{Wave-like warp propagation in circumbinary discs - I. Analytic theory and numerical simulations}",
      journal = {\mnras},
         year = 2013,
        month = aug,
       volume = {433},
       number = {3},
        pages = {2142-2156},
          doi = {10.1093/mnras/stt877},
archivePrefix = {arXiv},
       eprint = {1306.4331},
 primaryClass = {astro-ph.SR},
       adsurl = {https://ui.adsabs.harvard.edu/abs/2013MNRAS.433.2142F}
}

@ARTICLE{Papaloizou1995,
       author = {{Papaloizou}, J.~C.~B. and {Lin}, D.~N.~C.},
        title = "{On the Dynamics of Warped Accretion Disks}",
      journal = {\apj},
         year = 1995,
        month = jan,
       volume = {438},
        pages = {841},
          doi = {10.1086/175127},
       adsurl = {https://ui.adsabs.harvard.edu/abs/1995ApJ...438..841P}
}

@ARTICLE{Lodato2013,
       author = {{Lodato}, Giuseppe and {Facchini}, Stefano},
        title = "{Wave-like warp propagation in circumbinary discs - II. Application to KH 15D}",
      journal = {\mnras},
         year = 2013,
        month = aug,
       volume = {433},
       number = {3},
        pages = {2157-2164},
          doi = {10.1093/mnras/stt878},
archivePrefix = {arXiv},
       eprint = {1306.4333},
 primaryClass = {astro-ph.SR},
       adsurl = {https://ui.adsabs.harvard.edu/abs/2013MNRAS.433.2157L}
}

@ARTICLE{Foucart2013,
       author = {{Foucart}, Francois and {Lai}, Dong},
        title = "{Assembly of Protoplanetary Disks and Inclinations of Circumbinary Planets}",
      journal = {\apj},
         year = 2013,
        month = feb,
       volume = {764},
       number = {1},
          eid = {106},
        pages = {106},
          doi = {10.1088/0004-637X/764/1/106},
archivePrefix = {arXiv},
       eprint = {1211.3721},
 primaryClass = {astro-ph.EP},
       adsurl = {https://ui.adsabs.harvard.edu/abs/2013ApJ...764..106F}
}

@ARTICLE{Kraus2020,
       author = {{Kraus}, Stefan and {Kreplin}, Alexander and {Young}, Alison K. and {Bate}, Matthew R. and {Monnier}, John D. and {Harries}, Tim J. and {Avenhaus}, Henning and {Kluska}, Jacques and {Laws}, Anna S.~E. and {Rich}, Evan A. and {Willson}, Matthew and {Aarnio}, Alicia N. and {Adams}, Fred C. and {Andrews}, Sean M. and {Anugu}, Narsireddy and {Bae}, Jaehan and {ten Brummelaar}, Theo and {Calvet}, Nuria and {Cur{\'e}}, Michel and {Davies}, Claire L. and {Ennis}, Jacob and {Espaillat}, Catherine and {Gardner}, Tyler and {Hartmann}, Lee and {Hinkley}, Sasha and {Labdon}, Aaron and {Lanthermann}, Cyprien and {LeBouquin}, Jean-Baptiste and {Schaefer}, Gail H. and {Setterholm}, Benjamin R. and {Wilner}, David and {Zhu}, Zhaohuan},
        title = "{A triple-star system with a misaligned and warped circumstellar disk shaped by disk tearing}",
      journal = {Science},
         year = 2020,
        month = sep,
       volume = {369},
       number = {6508},
        pages = {1233-1238},
          doi = {10.1126/science.aba4633},
archivePrefix = {arXiv},
       eprint = {2004.01204},
 primaryClass = {astro-ph.SR},
       adsurl = {https://ui.adsabs.harvard.edu/abs/2020Sci...369.1233K}
}

@ARTICLE{Kimmig2025,
       author = {{Kimmig}, Carolin N. and {Villenave}, Marion},
        title = "{Asymmetric signatures of warps in edge-on disks}",
      journal = {\aap},
         year = 2025,
        month = jun,
       volume = {698},
          eid = {A146},
        pages = {A146},
          doi = {10.1051/0004-6361/202453313},
archivePrefix = {arXiv},
       eprint = {2504.05399},
 primaryClass = {astro-ph.SR},
       adsurl = {https://ui.adsabs.harvard.edu/abs/2025A&A...698A.146K}
}

@ARTICLE{Bohn2022,
       author = {{Bohn}, A.~J. and {Benisty}, M. and {Perraut}, K. and {van der Marel}, N. and {W{\"o}lfer}, L. and {van Dishoeck}, E.~F. and {Facchini}, S. and {Manara}, C.~F. and {Teague}, R. and {Francis}, L. and {Berger}, J.-P. and {Garcia-Lopez}, R. and {Ginski}, C. and {Henning}, T. and {Kenworthy}, M. and {Kraus}, S. and {M{\'e}nard}, F. and {M{\'e}rand}, A. and {P{\'e}rez}, L.~M.},
        title = "{Probing inner and outer disk misalignments in transition disks. Constraints from VLTI/GRAVITY and ALMA observations}",
      journal = {\aap},
         year = 2022,
        month = feb,
       volume = {658},
          eid = {A183},
        pages = {A183},
          doi = {10.1051/0004-6361/202142070},
archivePrefix = {arXiv},
       eprint = {2112.00123},
 primaryClass = {astro-ph.EP},
       adsurl = {https://ui.adsabs.harvard.edu/abs/2022A&A...658A.183B}
}

@ARTICLE{Villenave2024,
       author = {{Villenave}, Marion and {Stapelfeldt}, Karl R. and {Duch{\^e}ne}, Gaspard and {M{\'e}nard}, Fran{\c{c}}ois and {Wolff}, Schuyler G. and {Perrin}, Marshall D. and {Pinte}, Christophe and {Tazaki}, Ryo and {Padgett}, Deborah L.},
        title = "{JWST Imaging of Edge-on Protoplanetary Disks. II. Appearance of Edge-on Disks with a Tilted Inner Region: Case Study of IRAS04302+2247}",
      journal = {\apj},
         year = 2024,
        month = jan,
       volume = {961},
       number = {1},
          eid = {95},
        pages = {95},
          doi = {10.3847/1538-4357/ad0c4b},
archivePrefix = {arXiv},
       eprint = {2311.07668},
 primaryClass = {astro-ph.SR},
       adsurl = {https://ui.adsabs.harvard.edu/abs/2024ApJ...961...95V}
}

@ARTICLE{Garufi2022,
       author = {{Garufi}, A. and {Dominik}, C. and {Ginski}, C. and {Benisty}, M. and {van Holstein}, R.~G. and {Henning}, Th. and {Pawellek}, N. and {Pinte}, C. and {Avenhaus}, H. and {Facchini}, S. and {Galicher}, R. and {Gratton}, R. and {M{\'e}nard}, F. and {Muro-Arena}, G. and {Milli}, J. and {Stolker}, T. and {Vigan}, A. and {Villenave}, M. and {Moulin}, T. and {Origne}, A. and {Rigal}, F. and {Sauvage}, J.-F. and {Weber}, L.},
        title = "{A SPHERE survey of self-shadowed planet-forming disks}",
      journal = {\aap},
         year = 2022,
        month = feb,
       volume = {658},
          eid = {A137},
        pages = {A137},
          doi = {10.1051/0004-6361/202141692},
archivePrefix = {arXiv},
       eprint = {2111.07856},
 primaryClass = {astro-ph.GA},
       adsurl = {https://ui.adsabs.harvard.edu/abs/2022A&A...658A.137G}
}

@ARTICLE{Benisty2018,
       author = {{Benisty}, M. and {Juh{\'a}sz}, A. and {Facchini}, S. and {Pinilla}, P. and {de Boer}, J. and {P{\'e}rez}, L.~M. and {Keppler}, M. and {Muro-Arena}, G. and {Villenave}, M. and {Andrews}, S. and {Dominik}, C. and {Dullemond}, C.~P. and {Gallenne}, A. and {Garufi}, A. and {Ginski}, C. and {Isella}, A.},
        title = "{Shadows and asymmetries in the T Tauri disk HD 143006: evidence for a misaligned inner disk}",
      journal = {\aap},
         year = 2018,
        month = nov,
       volume = {619},
          eid = {A171},
        pages = {A171},
          doi = {10.1051/0004-6361/201833913},
archivePrefix = {arXiv},
       eprint = {1809.01082},
 primaryClass = {astro-ph.EP},
       adsurl = {https://ui.adsabs.harvard.edu/abs/2018A&A...619A.171B}
}

@ARTICLE{Benisty2017,
       author = {{Benisty}, M. and {Stolker}, T. and {Pohl}, A. and {de Boer}, J. and {Lesur}, G. and {Dominik}, C. and {Dullemond}, C.~P. and {Langlois}, M. and {Min}, M. and {Wagner}, K. and {Henning}, T. and {Juhasz}, A. and {Pinilla}, P. and {Facchini}, S. and {Apai}, D. and {van Boekel}, R. and {Garufi}, A. and {Ginski}, C. and {M{\'e}nard}, F. and {Pinte}, C. and {Quanz}, S.~P. and {Zurlo}, A. and {Boccaletti}, A. and {Bonnefoy}, M. and {Beuzit}, J.~L. and {Chauvin}, G. and {Cudel}, M. and {Desidera}, S. and {Feldt}, M. and {Fontanive}, C. and {Gratton}, R. and {Kasper}, M. and {Lagrange}, A.-M. and {LeCoroller}, H. and {Mouillet}, D. and {Mesa}, D. and {Sissa}, E. and {Vigan}, A. and {Antichi}, J. and {Buey}, T. and {Fusco}, T. and {Gisler}, D. and {Llored}, M. and {Magnard}, Y. and {Moeller-Nilsson}, O. and {Pragt}, J. and {Roelfsema}, R. and {Sauvage}, J.-F. and {Wildi}, F.},
        title = "{Shadows and spirals in the protoplanetary disk HD 100453}",
      journal = {\aap},
         year = 2017,
        month = jan,
       volume = {597},
          eid = {A42},
        pages = {A42},
          doi = {10.1051/0004-6361/201629798},
archivePrefix = {arXiv},
       eprint = {1610.10089},
 primaryClass = {astro-ph.EP},
       adsurl = {https://ui.adsabs.harvard.edu/abs/2017A&A...597A..42B}
}

@ARTICLE{Muro-Arena2020,
       author = {{Muro-Arena}, G.~A. and {Benisty}, M. and {Ginski}, C. and {Dominik}, C. and {Facchini}, S. and {Villenave}, M. and {van Boekel}, R. and {Chauvin}, G. and {Garufi}, A. and {Henning}, T. and {Janson}, M. and {Keppler}, M. and {Matter}, A. and {M{\'e}nard}, F. and {Stolker}, T. and {Zurlo}, A. and {Blanchard}, P. and {Maurel}, D. and {Moeller-Nilsson}, O. and {Petit}, C. and {Roux}, A. and {Sevin}, A. and {Wildi}, F.},
        title = "{Shadowing and multiple rings in the protoplanetary disk of HD 139614}",
      journal = {\aap},
         year = 2020,
        month = mar,
       volume = {635},
          eid = {A121},
        pages = {A121},
          doi = {10.1051/0004-6361/201936509},
archivePrefix = {arXiv},
       eprint = {1911.09612},
 primaryClass = {astro-ph.EP},
       adsurl = {https://ui.adsabs.harvard.edu/abs/2020A&A...635A.121M}
}

@ARTICLE{Marino2015,
       author = {{Marino}, S. and {Perez}, S. and {Casassus}, S.},
        title = "{Shadows Cast by a Warp in the HD 142527 Protoplanetary Disk}",
      journal = {\apjl},
         year = 2015,
        month = jan,
       volume = {798},
       number = {2},
          eid = {L44},
        pages = {L44},
          doi = {10.1088/2041-8205/798/2/L44},
archivePrefix = {arXiv},
       eprint = {1412.4632},
 primaryClass = {astro-ph.EP},
       adsurl = {https://ui.adsabs.harvard.edu/abs/2015ApJ...798L..44M}
}

@ARTICLE{Casassus2019,
       author = {{Casassus}, Simon and {P{\'e}rez}, Sebasti{\'a}n},
        title = "{Kinematic Detections of Protoplanets: A Doppler Flip in the Disk of HD 100546}",
      journal = {\apjl},
         year = 2019,
        month = oct,
       volume = {883},
       number = {2},
          eid = {L41},
        pages = {L41},
          doi = {10.3847/2041-8213/ab4425},
archivePrefix = {arXiv},
       eprint = {1906.06302},
 primaryClass = {astro-ph.EP},
       adsurl = {https://ui.adsabs.harvard.edu/abs/2019ApJ...883L..41C}
}

@ARTICLE{Young2022,
       author = {{Young}, Alison K. and {Alexander}, Richard and {Rosotti}, Giovanni and {Pinte}, Christophe},
        title = "{Characteristics of small protoplanetary disc warps in kinematic observations}",
      journal = {\mnras},
         year = 2022,
        month = jun,
       volume = {513},
       number = {1},
        pages = {487-502},
          doi = {10.1093/mnras/stac840},
archivePrefix = {arXiv},
       eprint = {2203.12348},
 primaryClass = {astro-ph.EP},
       adsurl = {https://ui.adsabs.harvard.edu/abs/2022MNRAS.513..487Y}
}

@INPROCEEDINGS{Pinte2023,
       author = {{Pinte}, C. and {Teague}, R. and {Flaherty}, K. and {Hall}, C. and {Facchini}, S. and {Casassus}, S.},
        title = "{Kinematic Structures in Planet-Forming Disks}",
    booktitle = {Protostars and Planets VII},
         year = 2023,
       editor = {{Inutsuka}, S. and {Aikawa}, Y. and {Muto}, T. and {Tomida}, K. and {Tamura}, M.},
       series = {Astronomical Society of the Pacific Conference Series},
       volume = {534},
        month = jul,
        pages = {645},
          doi = {10.48550/arXiv.2203.09528},
archivePrefix = {arXiv},
       eprint = {2203.09528},
 primaryClass = {astro-ph.EP},
       adsurl = {https://ui.adsabs.harvard.edu/abs/2023ASPC..534..645P}
}

@ARTICLE{Juhasz2017,
       author = {{Juh{\'a}sz}, Attila and {Facchini}, Stefano},
        title = "{Observational signatures of linear warps in circumbinary discs}",
      journal = {\mnras},
         year = 2017,
        month = apr,
       volume = {466},
       number = {4},
        pages = {4053-4073},
          doi = {10.1093/mnras/stw3389},
archivePrefix = {arXiv},
       eprint = {1701.02611},
 primaryClass = {astro-ph.EP},
       adsurl = {https://ui.adsabs.harvard.edu/abs/2017MNRAS.466.4053J}
}

\onecolumn 
\begin{appendix}

\section{Radiative transfer: temperature structure and emission height}

\begin{figure*}[ht!]
    \sidecaption
    \includegraphics[width=12cm]{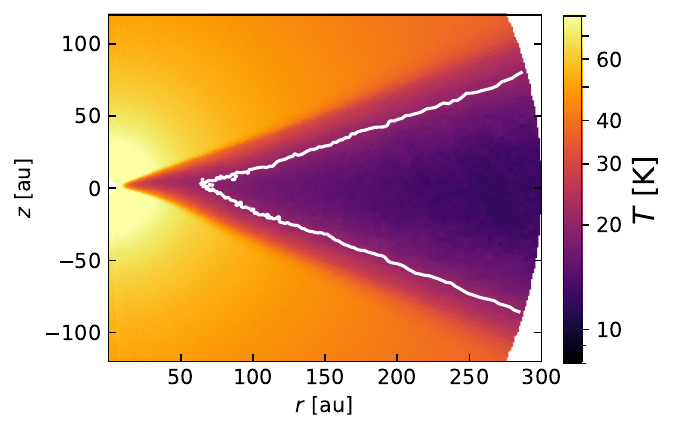}
    \caption{Dust temperature structure as computed with RADMC-3D of the warped and twisted disk with $\beta_0=5^\circ$ and $\gamma_0=540^\circ$ (fiducial setup). The white contour shows the condensation temperature of $20\,\mathrm{K}$.}
    \label{fig:temperature}
\end{figure*}

In Figure~\ref{fig:temperature}, we show a cross-section of the temperature structure in our fiducial setup. The white line indicates the location where the temperature falls below $20\,\mathrm{K}$, which is the threshold we assume for freeze-out of $^{12}\mathrm{CO}$ \citep{Williams2014}.

\begin{figure*}[ht!]
    \sidecaption
    \includegraphics[width=12cm]{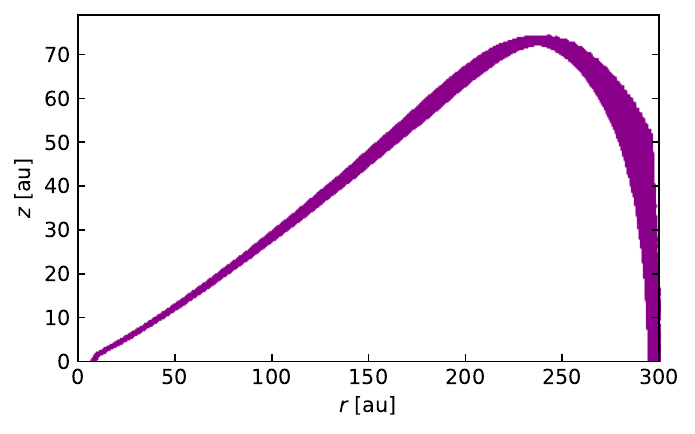}
    \caption{Emission height ($\tau=1$ surface) of a planar (unwarped) disk evaluated with RADMC-3D. We evaluate each pixel of the image individually and do not average over azimuth, which causes the thickness of the line.}
    \label{fig:emissionheight}
\end{figure*}

We additionally estimate the emission height of $^{12}\mathrm{CO}$-line emission in our models in the test case of a planar (unwarped) disk with the parameters described in Sect.~\ref{sec:methods-diskmodel}. Fig.~\ref{fig:emissionheight} shows the surface, where the optical depth of the line emission is $\tau=1$.

\section{Examples of limiting cases} \label{sec:appendix-limits}

In this section, we showcase in more detail some edge cases where the method of interpreting residuals in the kinematics as deviations in inclination and position angles to trace warps \citep[see][]{Winter2025} starts to fail. This occurs especially for the worst case model, where we do not include any knowledge of emission height and pressure correction in the disk, but simply subtract the Keplerian velocity at the midplane of the disk. We consider very small warp tilts, higher inclinations of the outer disk, reduced spectral resolution, and incorrectly assumed disk aspect ratios. We note, that for these models, we assume that we know the stellar mass perfectly. However, the method is not very sensitive in errors on the stellar mass, as shown in Sect.~\ref{sec:uncertainties}.

\subsection{Very small warp tilt}

\begin{figure*}[ht!]
    \sidecaption
    \includegraphics[width=12cm]{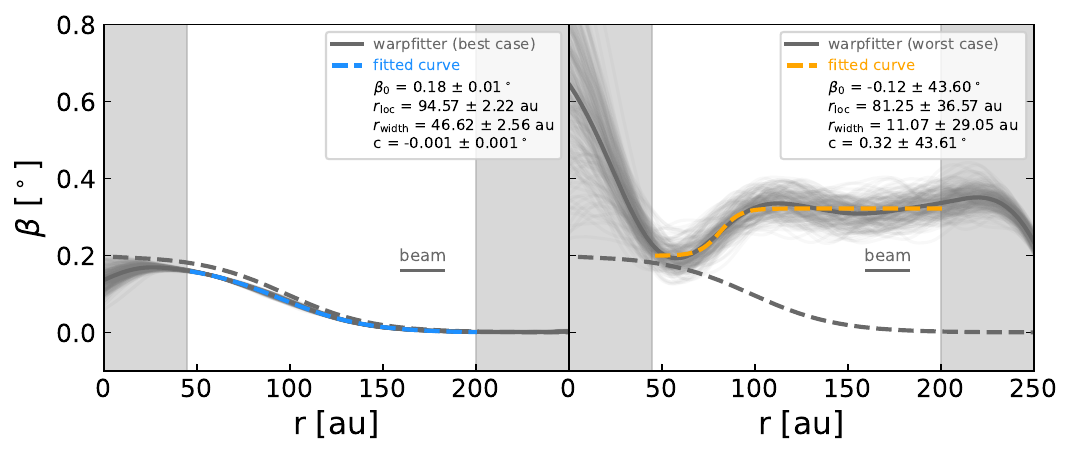}
    \caption{Fits of the warp shape fitted using \texttt{warpfitter} for the model of a twisted disk with $\beta_\mathrm{0}=0.2^\circ$ and $\gamma_\mathrm{0}=540^\circ$, similar to Fig.~\ref{fig:fits-10deg}.}
    \label{fig:fits-betamax02}
\end{figure*}

Fig.~\ref{fig:fits-betamax02} shows the fitted warp shape $\beta(r)$ for the model of a twisted disk with $\beta_\mathrm{0}=0.2^\circ$ and $\gamma_\mathrm{0}=540^\circ$. The best case model is able to trace the warp shape very well. On the other hand, the worst case model is not able to capture the true warp structure (dashed line). Here, the outer disk appears to have an offset in inclination, likely the result of subtracting the simple midplane Keplerian velocity, which becomes more inaccurate for the outer region, where the disk becomes thicker. The fit of the curve then traces the wrong warp shape, leading to a negative warp maximum $\beta_\mathrm{0}$, as the curve is flipped.

\subsection{Higher inclination of the outer disk}

\begin{figure*}[ht!]
    \sidecaption
    \includegraphics[width=12cm]{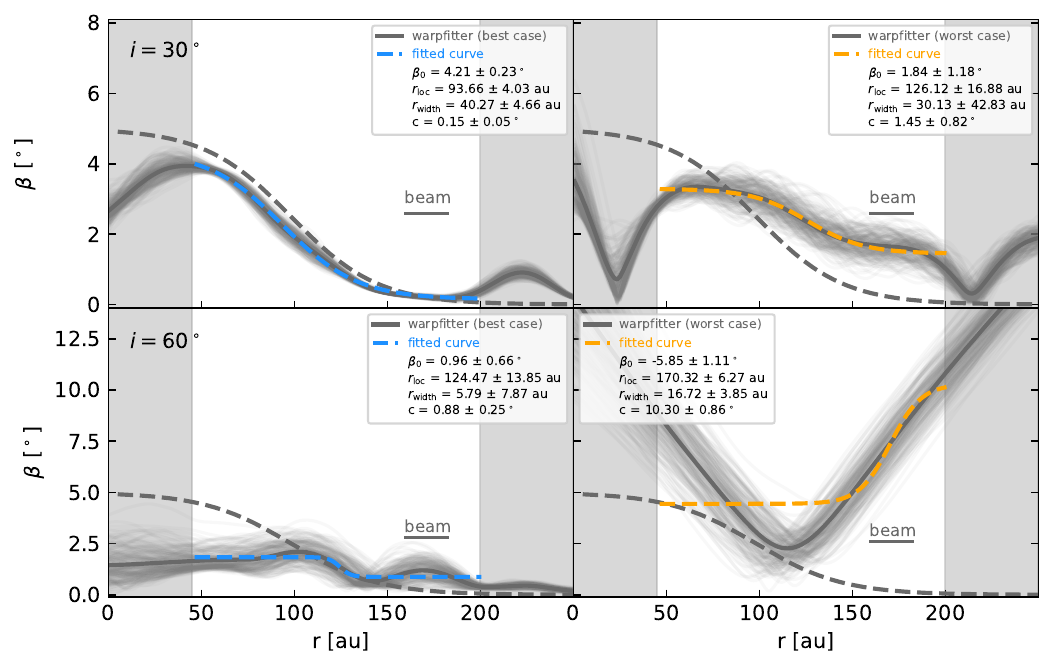}
    \caption{Fits of the warp shape for a twisted disk with $\beta_\mathrm{0}=5^\circ$ and $\gamma_\mathrm{0}=540^\circ$ and an inclination of $30^\circ$. Similar to Fig.~\ref{fig:fits-10deg}.}
    \label{fig:fits-30and60deg}
\end{figure*}

\begin{figure*}[ht!]
    \sidecaption
    \includegraphics[width=12cm]{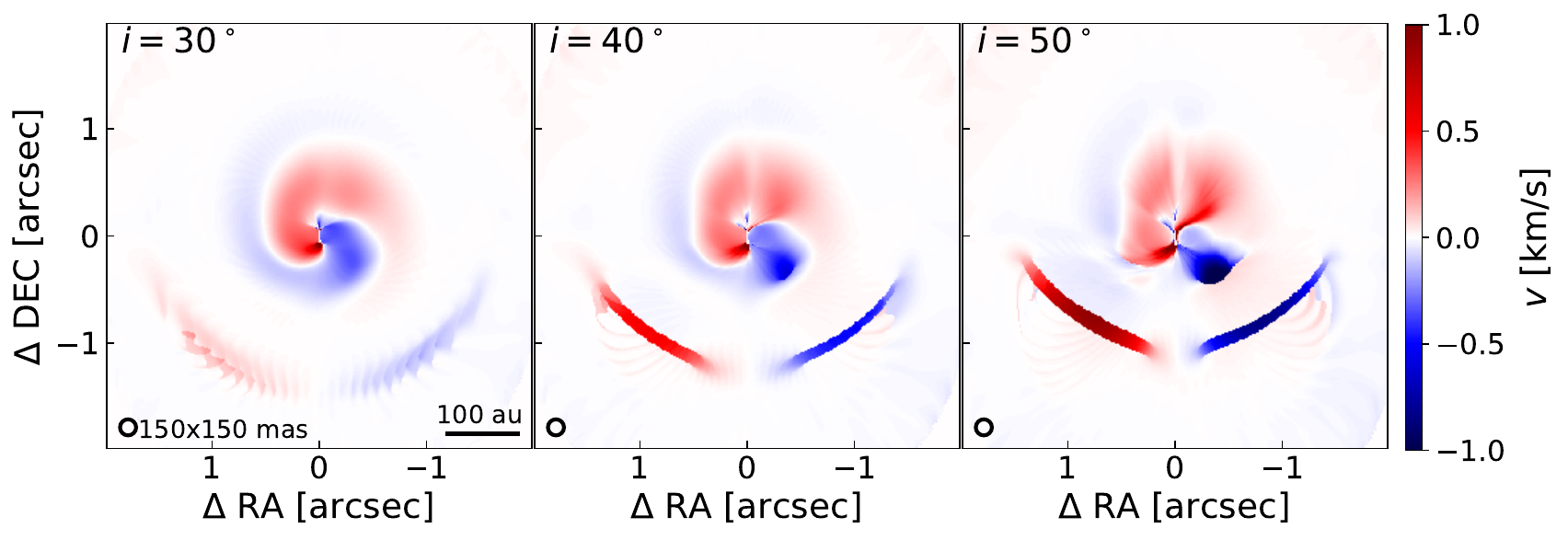}
    \caption{Residual maps of the twisted disk with $\beta_\mathrm{0}=5^\circ$ and $\gamma_\mathrm{0}=540^\circ$ for different inclinations. The maps were created using the best case model, where disk thickness and pressure corrections are perfectly known.}
    \label{fig:kinematic-maps-inclination}
\end{figure*}

For disks with a higher inclination, it is more difficult to fit the warp shape. We present two examples in Fig.~\ref{fig:fits-30and60deg}, specifically $i=30^\circ$ and $60^\circ$. For an inclination of $30^\circ$, the warp shape can be traced well in the best case model. However, a warp maximum occurs in the outer region, which is caused by the back side emission of the disk, as we show in Fig.~\ref{fig:kinematic-maps-inclination}, first panel. This is the reason we exclude the outer disk region in the curve fit. In observational analysis, this would be cropped, but we decided to show it in this plot to highlight this effect. The worst case model is not able to trace the warp shape well due to offset residuals in the outer disk regions, as discussed in the main text.
For a disk inclined by $60^\circ$ (Fig.~\ref{fig:fits-30and60deg}, bottom row), both the best and worst case models are not able to retrieve the warp shape of the disk.

\subsection{Varying the spectral resolution}

\begin{figure}[ht!]
    \sidecaption
    \includegraphics[width=10cm]{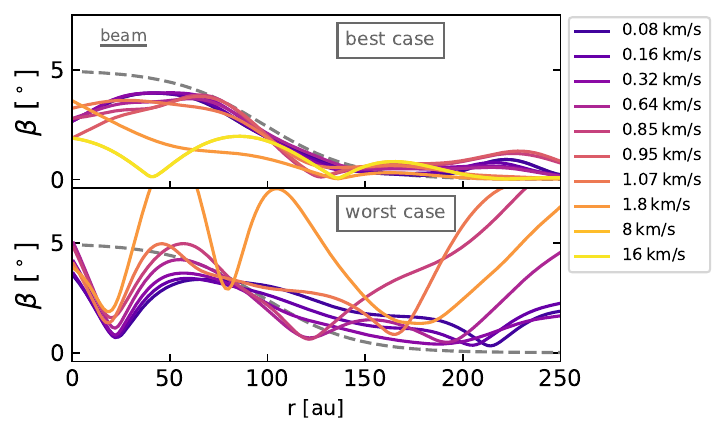}
    \caption{Warp profiles for a twisted disk with $\beta_\mathrm{0}=5^\circ$ and $\gamma_\mathrm{0}=540^\circ$, where the outer disk is inclined by $30^\circ$ for different spectral resolutions. The top panel shows the profiles in the best case scenario, while the bottom panel shows the worst case scenario.}
    \label{fig:spectralres-incl30}
\end{figure}

Fig.~\ref{fig:spectralres-incl30} shows the best fits by \texttt{warpfitter} for a warped and twisted disk with with $\beta_\mathrm{0}=5^\circ$ and $\gamma_\mathrm{0}=540^\circ$ for different spectral resolutions. To obtain the lower spectral resolution, we binned the fiducial resolution of ${0.08\,\mathrm{km/s}}$.
We see that the best case scenario starts to fail at resolutions $\geq 1.8\,\mathrm{km/s}$ (orange line). The worst case scenario already starts to fail for lower spectral resolutions $\geq 0.85\,\mathrm{km/s}$, as the outer region of the disk obtains an apparent warp tilt increase. This likely occurs as the worst case model does not account for the emission height and pressure correction, which seem to become increasingly important for lower spectral resolutions.
Nontheless, many existing observations have a spectral resolution higher than $0.8\,\mathrm{km/s}$, and our result implies that it is possible to extract a warp shape in these observations.

\subsection{Incorrectly assumed disk aspect ratio} \label{sec:appendix-aspectratio}

The best-case scenario assumes that the vertical structure of the disk and the corresponding pressure correction are perfectly known when computing the unwarped reference model. In observations, these quantities are subject to uncertainties. To investigate how an incorrectly assumed vertical structure affects the recovered warp, we repeat the analysis with different aspect ratios for the warped disk and the planar disk used to compute the kinematic residuals.

\begin{figure*}[ht]
    \centering
    \begin{minipage}[c]{0.48\textwidth}
        \centering
        \includegraphics[width=\linewidth]{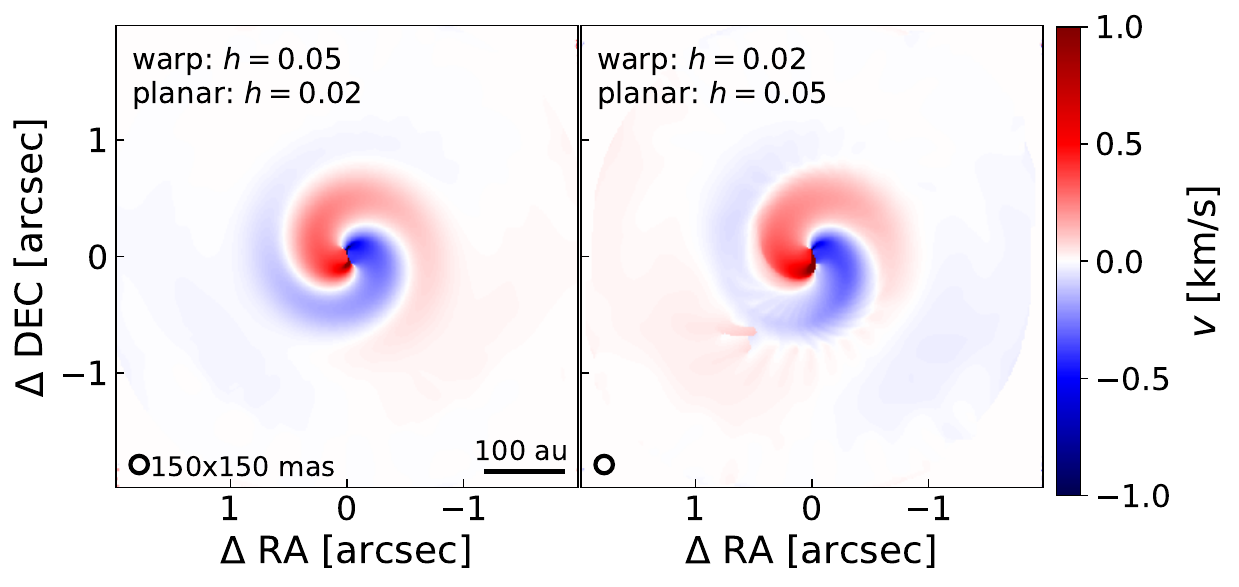}
        \textbf{(a)}\par\smallskip
    \end{minipage}
    \hfill
    \begin{minipage}[c]{0.48\textwidth}
        \centering
        \includegraphics[width=\linewidth]{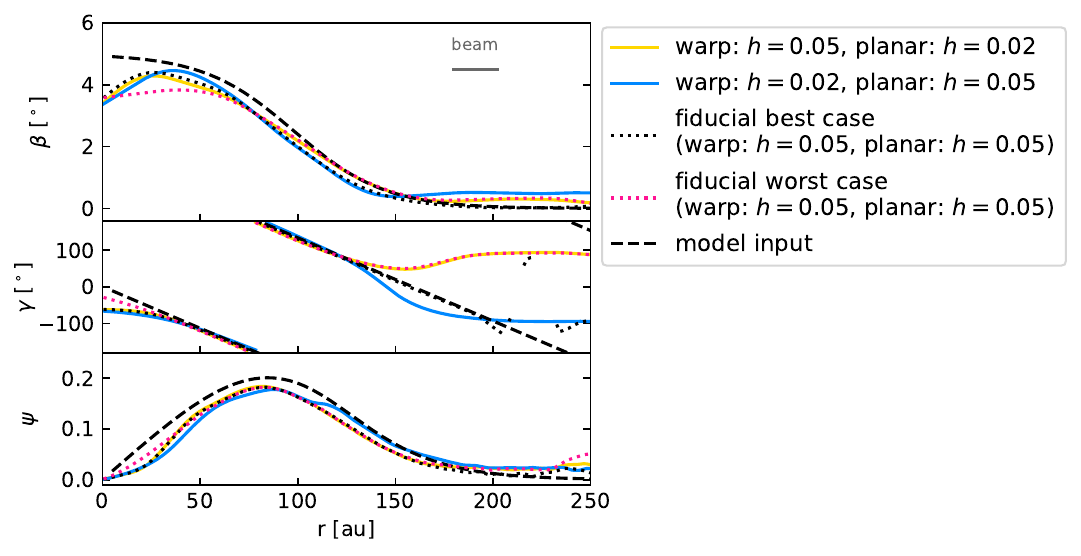}
        \textbf{(b)}\par\smallskip
    \end{minipage}

    \caption{
        Effect of an incorrectly assumed disk aspect ratio
        for an outer disk inclination of $10^\circ$.
        (a) Kinematic residual maps obtained by subtracting
        planar reference models with different aspect ratios.
        (b) Corresponding recovered tilt, twist, and warp
        amplitude profiles, compared with the input profiles
        and the fiducial best- and worst-case results.
    }
    \label{fig:aspect_ratio_10}
\end{figure*}

\begin{figure*}[ht]
    \centering
    \begin{minipage}[c]{0.48\textwidth}
        \centering
        \includegraphics[width=\linewidth]{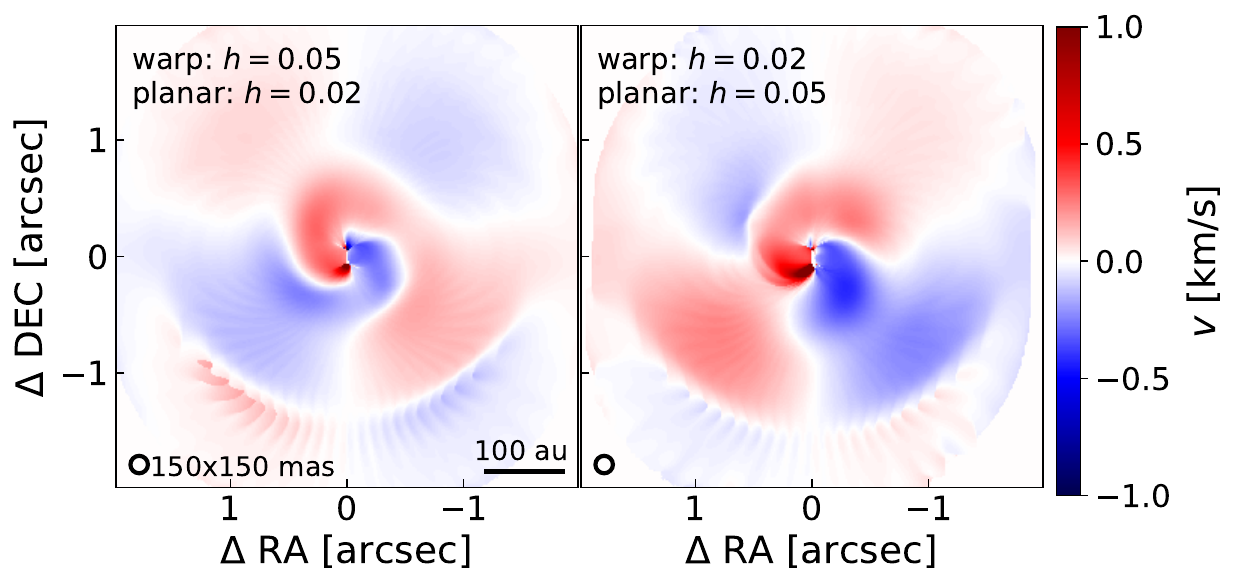}
        \textbf{(a)}\par\smallskip
    \end{minipage}
    \hfill
    \begin{minipage}[c]{0.48\textwidth}
        \centering
        \includegraphics[width=\linewidth]{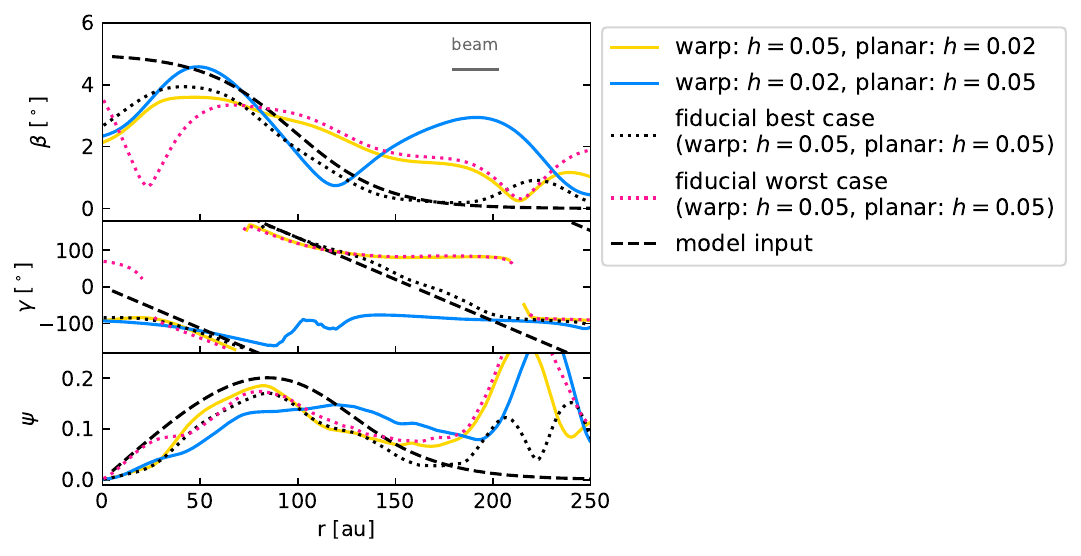}
        \textbf{(b)}\par\smallskip
    \end{minipage}
    \caption{
        Same as Fig.~\ref{fig:aspect_ratio_10}
        for an outer disk inclination of $30^\circ$.
    }
    \label{fig:aspect_ratio_30}
\end{figure*}

We consider two combinations: a warped disk with an aspect ratio of $h_0 = 0.05$ and a planar reference disk with $h_0 = 0.02$, and the reverse combination. Both warped models have the fiducial warp parameters $\beta_0 = 5^\circ$ and $\gamma_0 = 540^\circ$. We compute the residuals by subtracting the synthetic velocity map of the planar disk from that of the warped disk, and apply the same fitting procedure as in the main text. Figs.~\ref{fig:aspect_ratio_10} and \ref{fig:aspect_ratio_30} show the residual maps (panel a) and recovered profiles (panel b) for outer disk inclinations of $10^\circ$ and $30^\circ$, respectively. For comparison, we include the input warp profiles and the best and worst case results for the fiducial model with $h_0 = 0.05$.

At an inclination of $10^\circ$, the recovered tilt profiles retain the overall shape of the input warp for both combinations of aspect ratios. The differences between these profiles and the fiducial best and worst case results are relatively small. At $30^\circ$, the differences become larger, particularly in the outer disk. Focusing on the inner regions, where the warp transition is located, we find that the warp shape is recovered as well as or better than in the worst case scenario.

The recovered warp amplitude profiles $\psi$ are also similar for the different models at $10^\circ$, comparable to those of the fiducial best case. At $30^\circ$, the recovered profiles differ more strongly. The main peak around the warp transition is still recovered, although its amplitude differs between the models. The additional peak in the outer disk is likely related to backside emission. Its strength and shape may also be affected by the mismatch in disk thickness and the corresponding differences in the velocity field between the warped disk and the planar reference model.

\section{Investigation of the back side contamination}

\begin{figure}[ht!]
    \sidecaption
    \includegraphics[width=12cm]{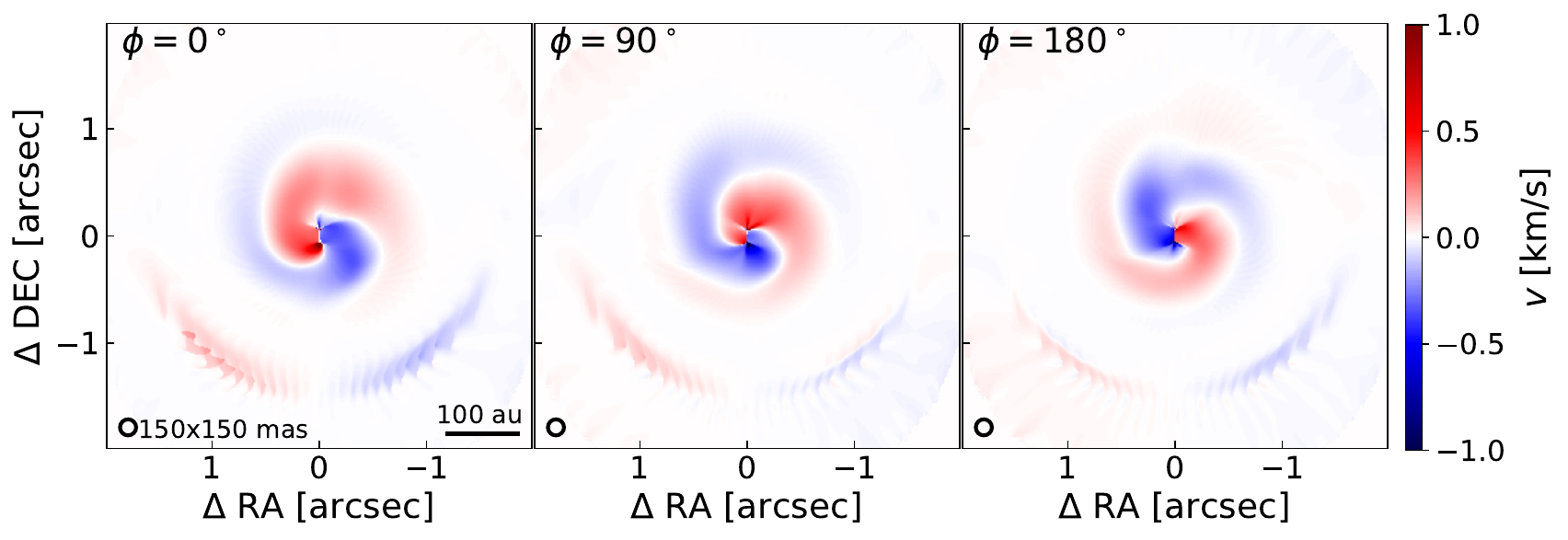}
    \caption{Kinematic residuals of a twisted disk as in Fig.~\ref{fig:kinematic-maps-wa5} for different azimuthal orientations of the twisted disk.}
    \label{fig:maps-phis}
\end{figure}

In this section, we take a closer look at the emission from the back side of the disk that occurs in disks with higher inclination. In a first attempt to assess if there are signatures characteristic to warps, we perform ray-tracing models of line emission where we rotate the azimuthal orientation of the twisted disk, shown in Fig.~\ref{fig:maps-phis}. To compute the residuals, we use the best case scenario. We find that the feature resulting from the back side emission does not qualitatively change for different orientations.

This back side feature occurs at higher inclination because a spatially more extended region of the back surface is visible. It is clearly an issue of subtracting a slightly incorrect model to obtain the residuals. If the back side is visible to a region that is slightly tilted with respect to the outer region, the emission differs from the unwarped disk model, resulting in the residuals.

\section{Radial flows and sloshing motions: kinematic maps}

\subsection{Radial flows}

\begin{figure}[ht!]
    \sidecaption
    \includegraphics[width=12cm]{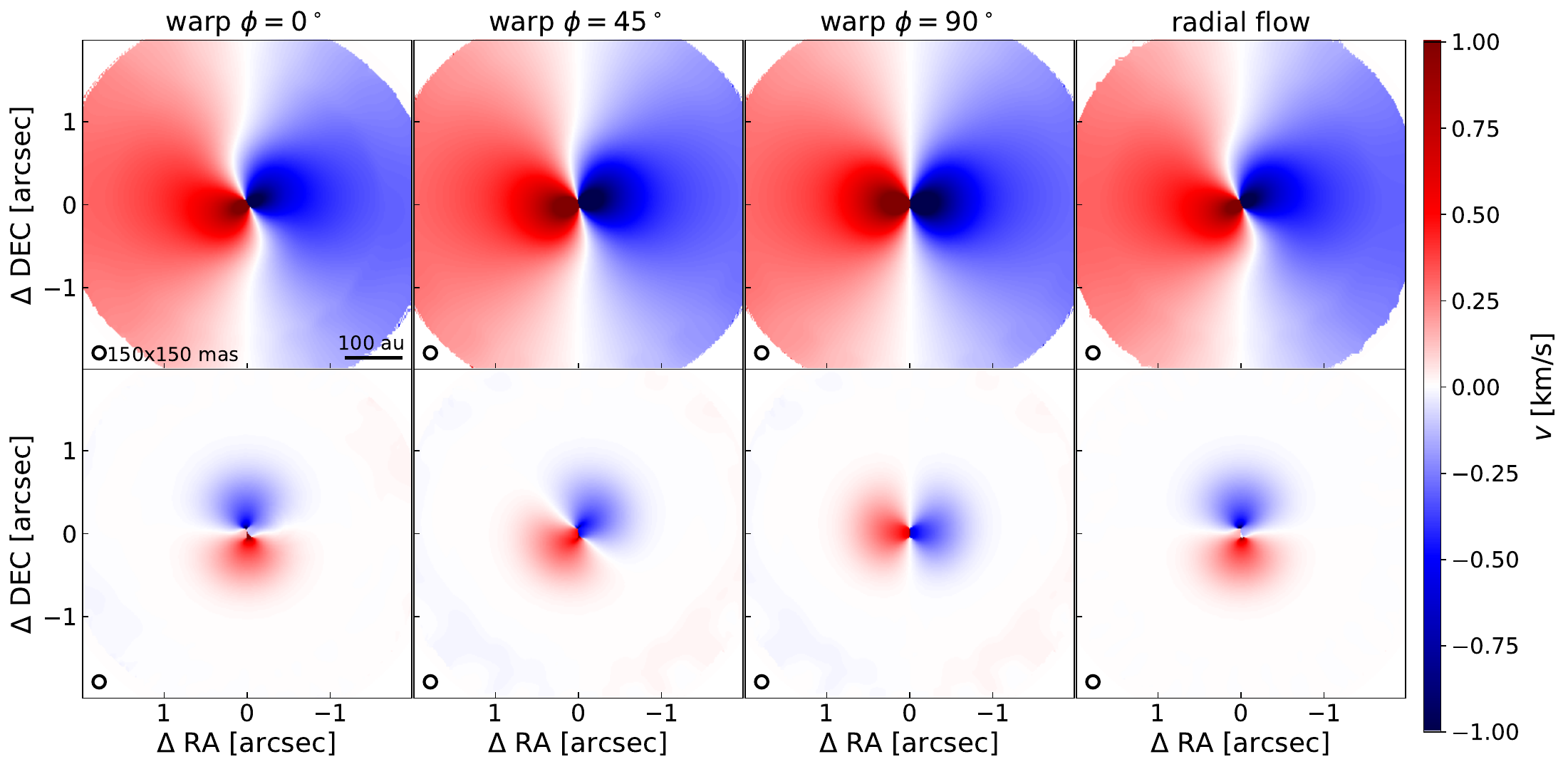}
    \caption{Kinematic maps of a warped disk in different viewing orientations (left three columns), and the radial flow model (right column). The residuals from an unwarped Keplerian model are shown in the bottom row.}
    \label{fig:maps-radialflow-phi}
\end{figure}

Fig.~\ref{fig:maps-radialflow-phi} shows a warped disk in different azimuthal orientations of the warp with respect to the direction in which the disk is inclined. The left most column is the view where the inner region is tilted perpendicular to the observed inclination of the outer disk. In the second column, this region is tilted in a direction of $45^\circ$ with respect to the disk inclination, and the third column the region is tilted in the same direction. The appearance of the disk in the kinematic map differs in all three cases and the residuals from a Keplerian model (bottom row) is rotated depending on the warp orientation.

The radial flow (right column) does not change its appearance for different orientations of the disk with respect to the observer, as it is an axisymmetric setup. This means that for an axisymmetric radial flow of the form given in Eq.~\ref{eq:raidal-flow}, the residuals will always be a bipolar structure in the direction of inclination, or in other words perpendicular to the position angle axis.

\begin{figure*}[ht!]
    \sidecaption
    \includegraphics[width=12cm]{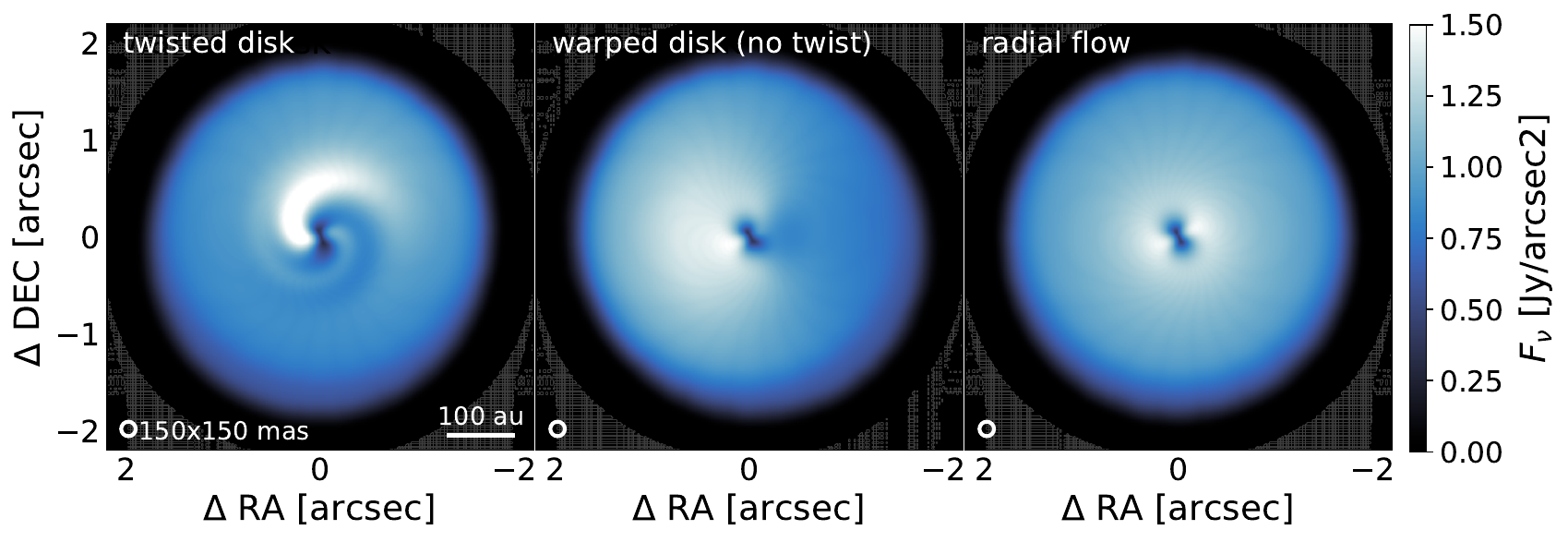}
    \caption{Peak intensity maps along the spectral axis (Moment~8) of a warped and twisted disk (first panel), a warped, but untwisted disk (second panel), and an unwarped disk that contains a radial flow (third panel). In all panels, the (outer) disk is synthetically observed at $10^\circ$ inclination and the images are deprojected.}
    \label{fig:mom0-peak}
\end{figure*}

For a comparison to Fig.~\ref{fig:mom0-maps}, we also add the peak intensity maps in Fig.~\ref{fig:mom0-peak}. The asymmetries in the warped and twisted case are even more obvious here.

\subsection{Hydrodynamic effects}

\begin{figure}[ht!]
    \sidecaption
    \includegraphics[width=12cm]{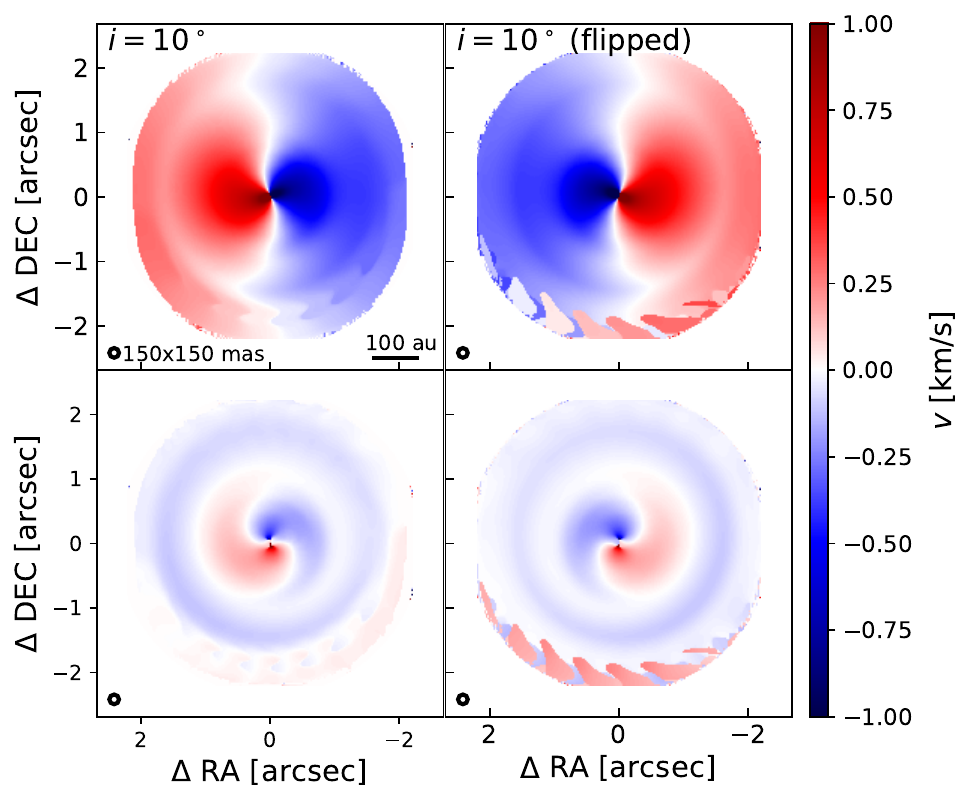}
    \caption{Kinematic maps of the hydrodynamic simulation presented in Sect.~\ref{sec:sloshing} at ${t=11686\,\mathrm{yr}}$ (100 orbits at ${r_0=5.2\,\mathrm{au}}$). The bottom row shows the residuals from an unwarped Keplerian model. This corresponds to the best case model. The left column is the disk viewed from the top surface of the simulated disk, the right column is the bottom disk surface, both viewed at an inclination of $10^\circ$.}
    \label{fig:kinematic-maps-hydro-flipped}
\end{figure}

The kinematic maps of the hydrodynamic simulation, shown in Fig.~\ref{fig:kinematic-maps-hydro-flipped}, show that the apparent twist changes its direction when looking at the disk from the top versus from the bottom. This aligns with the hypothesis that the apparent twist is caused by the sloshing motions.
We note that in the flipped case, the disk rotation also changes direction (from counter-clockwise to clockwise). Thus, use also flip the unwarped disk model to compute the residuals. Therefore, the rotational direction does not influence the resulting warp parameters.
It is also interesting to note that the spiral pattern induced by the sloshing motions can not exceed $180^\circ$, or in other words half of the disk, as it is caused by projection effects of the velocities.

\end{appendix}

\begin{appendix}





\FloatBarrier 
\clearpage

\end{appendix}
\end{document}